\documentclass[12pt]{article}
\usepackage{mathtools,amssymb,mathrsfs,microtype,tikz}
\usepackage[
 colorlinks=true,
 linkcolor=blue,
 urlcolor=blue,
 citecolor=red,
 pdfstartview=FitV,
 linktocpage
 ]{hyperref}
\usepackage[backend=bibtex,style=numeric-comp,sorting=none,maxbibnames=99, minbibnames=99]{biblatex}
\newtheorem{definition}{Definition}[section]
\newtheorem{conjecture}{Conjecture}[section]
\newtheorem{proposition}{Proposition}[section]
\newtheorem{remark}{Remark}[section]
\newtheorem{example}{Example}[section]

\makeatletter\@addtoreset{equation}{section}\makeatother

\DeclareMathOperator{\sign}{sign}
\DeclareMathOperator{\diag}{diag}
\DeclareMathOperator{\Aut}{Aut}
\renewcommand{\title}[1]{\vbox{\center\LARGE{#1}}\vspace{5mm}}
\renewcommand{\author}[1]{\vbox{\center\large#1}\vspace{5mm}}
\newcommand{\address}[1]{\vbox{\center\em#1}}

\begin{document}
 
\begin{titlepage}
\begin{center}
\vspace{5mm}
%\hfill {\tt HU-EP-09/40}\\
\hfill {\tt }\\
\vspace{8mm}

\title{\makebox[\textwidth]{Symmetries of Abelian Chern-Simons Theories and Arithmetic}}
\vspace{10mm}
Diego Delmastro,${}^{ab}$\footnote{\href{mailto:ddelmastro@perimeterinstitute.ca}
{\tt ddelmastro@perimeterinstitute.ca}}
Jaume Gomis${}^{a}$\footnote{\href{mailto:jgomis@perimeterinstitute.ca}
{\tt jgomis@perimeterinstitute.ca}}
\vskip 7mm
\address{
${}^a$Perimeter Institute for Theoretical Physics,\\
Waterloo, Ontario, N2L 2Y5, Canada}
\address{
${}^b$ Department of Physics, University of Waterloo,\\ Waterloo, ON N2L 3G1, Canada}
\end{center}

\vspace{5mm}
\abstract{
We determine the unitary and anti-unitary Lagrangian and quantum symmetries of arbitrary abelian Chern-Simons theories. The symmetries depend sensitively on the arithmetic properties (e.g.~prime factorization) of the matrix of Chern-Simons levels, revealing interesting connections with number theory. We give a complete characterization of the symmetries of abelian topological field theories and along the way find many theories that are non-trivially time-reversal invariant by virtue of a quantum symmetry, including $U(1)_k$ Chern-Simons theory and $(\mathbb Z_k)_\ell$ gauge theories. For example, we prove that $U(1)_k$ Chern-Simons theory is time-reversal invariant if and only if $-1$ is a quadratic residue modulo $k$, which happens if and only if all the prime factors of $k$ are Pythagorean (i.e.,~of the form $4n+1$), or Pythagorean with a single additional factor of $2$. Many distinct non-abelian finite symmetry groups are found.
}
\vfill\eject

\vspace{20pt}

{\hypersetup{linkcolor=black}
\tableofcontents
\thispagestyle{empty}
}

\end{titlepage}

%\maketitle

\section{Introduction and Summary}
\setcounter{footnote}{0} 

Symmetries play a pivotal role in our description of nature. In classical physics symmetries generate solutions of the equations of motion and in quantum mechanics symmetries imply selection rules and constrain physical observables. 't Hooft anomalies for global symmetries, being renormalization-group invariant, provide powerful nonperturbative constraints on the dynamics. By a classic result of Wigner, symmetries in quantum mechanics are implemented in the Hilbert space either by unitary or anti-unitary operators, and the corresponding transformations are linear and anti-linear, respectively.

Invariance of the classical action under a transformation $g$ imposes nontrivial constraints on the correlation functions of the theory. These are encapsulated in Ward identities. Invariance of the action under a transformation $g$ is a sufficient condition for $g$ to be a symmetry. However, this is not necessary. A transformation $g$ that does not leave the action $S$ invariant
\begin{equation}
g\cdot S\neq S
\end{equation}
is nevertheless a symmetry of the quantum theory if it obeys the Ward identities
\begin{equation}
\langle g\cdot \mathcal O_{1}\cdots g\cdot \mathcal O_m\rangle=
\begin{cases}
\langle \mathcal O_{1}\cdots \mathcal O_m\rangle\qquad &g~\text{unitary}\\[+4pt]
\langle \mathcal O_{1}\cdots \mathcal O_m\rangle^*\qquad &g~\text{anti-unitary}
\end{cases}\,,
\label{Ward}
\end{equation}
where $*$ implements complex conjugation. We shall refer to such non-Lagrangian symmetries as \emph{quantum symmetries}. Naturally, determining whether a theory has a quantum symmetry is nontrivial. In this work we characterize all the symmetries, quantum or otherwise, of abelian Chern-Simons theories.

Chern-Simons theories are ubiquitous in physics and mathematics. They arise as the emergent infrared description of gapped, quantum phases of matter such as the integer and fractional quantum Hall effect, quantum spin liquids and analogs of topological insulators and superconductors (see e.g~\cite{Wen:2004ym,Fradkin:1991nr}). Chern-Simons theories capture the nonperturbative infrared dynamics of $2+1$ dimensional gauge theories with massless fermions~\cite{Gomis:2017ixy,Cordova:2017vab,Bashmakov:2018wts,Benini:2018umh, Choi:2018ohn,Choi:2018tuh,Cordova:2018qvg}, and describe the low-energy dynamics of domain walls connecting vacua of $3+1$ dimensional gauge theories~\cite{Acharya:2001dz,Gaiotto:2017yup,Gaiotto:2017tne,Choi:2018tuh}. Chern-Simons theory, a topological quantum field theory (TQFT), has also found beautiful and profound applications in mathematics, starting with Witten's work~\cite{Witten:1988hf} on the topological invariants of knots and three-manifolds. 
 
In this paper we give a complete description of all the unitary and anti-unitary symmetries of abelian Chern-Simons theories, the simplest incarnation being $U(1)_k$ Chern-Simons theory, described by the Lagrangian
\begin{equation}
\mathcal L=\frac{k}{4\pi}a\,\mathrm da\,,
\end{equation}
where $a$ is a $U(1)$ gauge field and the coupling constant is quantized, $k\in \mathbb Z$. More generally, an arbitrary abelian TQFT can be described by a collection of such fields coupled via an integral symmetric matrix $K$
with Lagrangian
\begin{equation}
\mathcal L=\frac{1}{4\pi}a^tK\mathrm da\,,
\end{equation}
where $a^t=(a_1,\ldots,a_n)$. These theories have been studied intensely and enjoy a myriad of applications. In spite of this, we unearth a rich structure of symmetries in these theories, which depends on the arithmetic properties of the Chern-Simons levels $K$, revealing interesting connections with number theory.

Symmetries in topological phases of matter have been at the forefront of recent developments at the intersection of condensed matter, particle physics, and mathematics. These gapped phases are encoded by emergent TQFTs. Gapped phases with no topological order (no nontrivial anyons) and protected by symmetries describe SPT phases (see e.g.~\cite{PhysRevB.83.075103,PhysRevB.83.075102,PhysRevB.83.035107,PhysRevB.84.165139,PhysRevB.87.155114}) while phases with topological order (with nontrivial anyons) and enriched by symmetries give rise to the so-called SET phases (see e.g.~\cite{PhysRevB.87.104406,PhysRevB.93.155121,PhysRevB.87.165107,Barkeshli:2014cna,PhysRevX.6.041068}). Symmetries and 't Hooft anomalies of TQFTs have recently played a key role in understanding the nonperturbative infrared dynamics of gauge theories~\cite{Gomis:2017ixy,Cordova:2017vab,Bashmakov:2018wts,Benini:2018umh, Choi:2018ohn,Choi:2018tuh,Cordova:2018qvg}. Despite a lot of work, little is concretely known about the symmetries of TQFTs. Here we tackle this problem for abelian TQFTs.

\medskip

For the reader's convenience we summarize here a sample of our main results:

\begin{itemize}

\item $U(1)_k$ is a time-reversal invariant spin TQFT,\footnote{If $k$ is odd, $U(1)_k$ is a spin TQFT. For $k$ even it is bosonic but can be turned into a spin TQFT by tensoring with a transparent fermion $\{\boldsymbol 1,\psi\}$. See section~\ref{sec:TQFT} for details.} that is, it admits an anti-unitary symmetry, if and only if $-1$ is a quadratic residue modulo $k$ (cf.~proposition~\ref{th:time_reversal}). Equivalently:
\begin{equation}
U(1)_{+k}\ \longleftrightarrow\ U(1)_{-k}\qquad\Longleftrightarrow\qquad q^2=-1\mod k\text{ ~~~for some $q\in\mathbb Z$.}
\end{equation}
Therefore, $U(1)_k$ Chern-Simons theory is time-reversal invariant if and only if
\begin{equation}
k\in \mathbb T:=\{k\in \mathbb Z\,|\, kp-q^2=1\quad \text{for some $p,q\in\mathbb Z$}\}\,.
\end{equation}
This result can also be stated as $U(1)_k$ being dual to $U(1)_{-k}$ when $k\in\mathbb T$, which we denote by $U(1)_{+k}\ \longleftrightarrow\ U(1)_{-k}$.
The integer $k$ is in $\mathbb T$ if and only if all its prime factors are Pythagorean (i.e.,~congruent to $1$ modulo $4$), or Pythagorean with a single factor of $2$. Any time-reversal symmetry is of order $4$, except for $k=1,2$, when it is of order $2$ (cf.~proposition~\ref{pr:order_4}).

The set of time-reversal invariant $U(1)_k$ Chern-Simons theories includes the subset $k\in \mathbb P:=\{k\in \mathbb Z\,|\, kp^2-q^2=1\quad \text{for some $p,q\in\mathbb Z$}\}\subset \mathbb T$. The set $\mathbb P$ corresponds to those values of the level for which the (negative) Pell equation is solvable, which was shown by Witten~\cite{witten:unpublished,Benini:2018reh} to lead to time-reversal invariance. 

We prove that the time-reversal symmetry is a quantum symmetry if and only if $k\in \mathbb T\setminus \mathbb P$ (cf.~proposition~\ref{lm:lagrangian_pell}). By studying the time-reversal invariance of $U(1)_k\times U(1)_{k'}$ we obtain an interesting number-theoretic conjecture, to wit, $k\in \mathbb T$ if and only if there exist some $k'\in \mathbb P$ such that $kk'\in \mathbb P$. We argue that this conjecture follows from a well-known conjecture by Hardy-Littlewood (cf.~conjecture~\ref{cj:pell}).

\item All the unitary symmetries of $U(1)_k$ are of order $2$, and the number of such symmetries depends on the number of distinct prime factors of $k$, usually denoted by $\omega(k)$. More precisely, the group of unitary symmetries of $U(1)_k$ is (cf.~proposition~\ref{pr:full_group_u1})
\begin{equation}
 (\mathbb Z_2)^{\varpi(k)}\,, \qquad \varpi(k):=\begin{cases} \omega(k) & k\ \text{odd}\\ \omega(k/2) & k\ \text{even.}
 \end{cases}
 \label{defff}
\end{equation}
When $U(1)_k$ with $k$ even is upgraded to an spin TQFT by considering $U(1)_k\times \{\bf{1},\psi\}$, an additional factor of $\mathbb Z_2$ appears when $k$ is a multiple of $8$. All but one factor of $\mathbb Z_2$ in~\eqref{defff}, which corresponds to charge conjugation, are quantum symmetries. When $k\in\mathbb T$, the total group of symmetries is the central product of its unitary subgroup and $\mathbb Z_4$.

\item The unitary and anti-unitary symmetries of $U(1)^n$ Chern-Simons theory with matrix of levels $K$ correspond to the integral-valued matrices $Q$, invertible modulo $K$, that solve (cf.~proposition~\ref{pr:group_symmetries_K})
\begin{equation} 
\begin{aligned}
\text{unitary:}\qquad & Q^tK^{-1}Q-K^{-1}=P\\
\text{anti-unitary:}\qquad & Q^tK^{-1}Q+K^{-1}=P
\end{aligned}
\end{equation}
for some integral-valued matrix $P$. While the first equation always admits solutions, the second one need not, and only when there is a solution is the theory time-reversal invariant. The group of symmetries is finite and generically non-abelian. A given symmetry is quantum if and only if $P\neq 0$ for all the $Q$'s that implement it.

\item Two abelian Chern-Simons theories described by matrices $K_1,K_2$ (not necessarily of the same dimension) are dual if and only if there exist suitable matrices $Q,P$ such that
\begin{equation}
Q^tK_1^{-1}Q-K_2^{-1}=P
\end{equation}
(see section~\ref{sec:dualities} for the precise formulation and the conditions on $Q,P$). In this sense, the unitary symmetries of $K$ correspond to the self-dualities $K\leftrightarrow K$, and the anti-unitary symmetries to dualities $K\leftrightarrow-K$.

\item The twisted gauge theory $(\mathbb Z_{k_1})_{k_2}$ (also known as $\mathbb Z_{k_1}$ Dijkgraaf-Witten theory~\cite{Dijkgraaf:1989pz} when $k_2$ is even, and which can be realized by the $U(1)^2$ Chern-Simons theory with $K=\begin{pmatrix}0&k_1\\k_1&k_2\end{pmatrix}$ with $k_2\in[0,2k_1)$) is conjectured to be time-reversal invariant if and only if $k_2$ is proportional to $\mu(k_1)$ (cf.~conjecture~\ref{cj:DW_time_rev})
\begin{equation}
k_2\propto \mu(k_1)
\end{equation}
where $\mu(n)$ equals $n$ divided by all its Pythagorean prime factors (e.g. $\mu(10)=\frac{2\times 5}{5}=2$). The conjecture has been verified for $k_1\in[0,200]$ and all $k_2$. We compute the explicit group of unitary and anti-unitary symmetries of $(\mathbb Z_{k_1})_{k_2}$ for small values of the levels; see table~\ref{tab:DW_examples} for a sample. The time-reversal symmetry of $(\mathbb Z_{k_1})_{k_2}$ implies in particular a duality between abelian TQFTs
\begin{equation}
 (\mathbb Z_{k_1})_{+k_2} \ \longleftrightarrow \ (\mathbb Z_{k_1})_{-k_2} \qquad\Longleftrightarrow\qquad k_2\propto \mu(k_1)\,.
\end{equation}
The theory $(\mathbb Z_{k})_{0}$ has conjecturally $2^{\omega(k)}\phi(k)$ unitary transformations and as many anti-unitary ones (where $\phi(k)$ is the Euler totient function, which counts the number of integers $q\in[1,k)$ relatively prime to $k$). Among these symmetries, there is a unitary $\mathbb Z_2$ subgroup which is Lagrangian, and four anti-unitary Lagrangian symmetries (except for $k=2$, which only has two). For $k>2$ the group of symmetries is non-abelian (see~\ref{cj:Z_k_symmetry} for the explicit conjecture), while for $k=2$, the group of symmetries is $\mathbb Z_2^2$, with a $\mathbb Z_2$ unitary subgroup.
 
\begin{table}
\begin{equation*}
\begin{array}{r|c|c|c|c}
k & \Aut((\mathbb Z_k)_0)& \Aut_U((\mathbb Z_k)_0)& \Aut((\mathbb Z_k)_{\mu(k)})& \Aut_U((\mathbb Z_k)_{\mu(k)})\\ \hline
2&\mathbb Z_2^2&\mathbb Z_2&\mathbb Z_2&0\\
3&D_8&\mathbb Z^2_2&D_8&\mathbb Z_2^2\\
4&D_8&\mathbb Z^2_2&D_8&\mathbb Z_2^2\\
5&\mathbb Z_4\circ D_8&D_8&\mathbb Z_4&\mathbb Z_2\\
6&\mathbb Z_2\times D_8&\mathbb Z^3_2&D_8&\mathbb Z_2^2\\
7&\mathbb Z_3\rtimes D_8&D_{12}&\mathbb Z_3\rtimes D_8&D_{12}\\
8&\mathbb Z_2\times D_8&\mathbb Z^3_2&\mathbb Z_2\times D_8&\mathbb Z_2^3\\
9&\mathbb Z_3\rtimes D_8&D_{12}&\mathbb Z_3\rtimes D_8&D_{12}\\
10&\mathbb Z_2\times \mathbb Z_4\circ D_8&\mathbb Z_2 \times D_8&\mathbb Z_4&\mathbb Z_2\\
11&\mathbb Z_5\rtimes D_8&D_{20}&\mathbb Z_5\rtimes D_8&D_{20}\\
12&\mathbb Z_2^2\wr \mathbb Z_2&\mathbb Z^4_2&\mathbb Z_2^2\wr \mathbb Z_2&\mathbb Z_2^4\\
\end{array}
\end{equation*}
\caption{The group of symmetries of $(\mathbb Z_{k_1})_{k_2}$, denoted by $\Aut(\,\star\,)$, and its unitary subgroup $\Aut_U(\,\star\,)$, for $k_1\in[0,12]$ and $k_2=0,\mu(k_1)$. For $k_2\not\propto \mu(k_1)$ there are no anti-unitary symmetries. See table~\ref{tab:Z_k_symmetry_full} for the group of symmetries up to $k_1=27$. (See Appendix~\ref{ap:grupo} for basic definitions).}
\label{tab:DW_examples}
\end{table} 

\item The so-called ``minimal abelian TQFT'' $\mathcal A^{N,t}$ is proven to be time-reversal invariant invariant if and only if $t$ is proportional to $\mu(N)$ (cf.~subsection~\ref{subsec:mimimal})
\begin{equation}
t\propto \mu(N)\,.
\end{equation}
These minimal theories have $N$ anyons with a $\mathbb Z_N$ fusion algebra, and their spin depends on the integer $t$.
\end{itemize}

TQFTs can also have a one-form symmetry group~\cite{Kapustin:2014gua,Gaiotto:2014kfa} on top of the usual (zero-form) symmetry group that we study in this paper. The Wilson lines describing the worldline of anyons transform in representations of this group. The one-form symmetries of abelian Chern-Simons theories are well understood (see e.g.~\cite{Hsin:2018vcg}). Given an abelian TQFT with an abelian Chern-Simons representation, the one-form symmetry group is $\mathbb Z_{k_1}\times \mathbb Z_{k_2}\ldots \times \mathbb Z_{k_n}$, where $\{k_i\}$ are the Smith invariants of $K$ (cf.~section~\ref{sec:K}). Interestingly, given a QFT with a zero-form symmetry group and a one-form symmetry group, these can combine into a nontrivial extension known as a 2-group (see e.g.~\cite{Kapustin:2013uxa,Benini:2018reh}). When a theory has a 2-group symmetry, the zero-form and one-form symmetries do not factorize; rather, they are mixed non-trivially. However, it is known that abelian TQFTs have a trivial 2-group of symmetries~\cite{Benini:2018reh,hsin:unpublished,Lee2018}: the zero-form and one-form symmetries factorize, and since the one-form symmetries are completely understood, what remains are the zero-form symmetries, which is the problem we address in this paper. Furthermore, since the 2-group in an abelian TQFT is trivial, the zero-form and one-form 't Hooft anomalies are well defined and can be classified using cohomology and cobordism groups~\cite{Kapustin:2014tfa,Kapustin2015,Guo:2018vij,Yonekura:2018ufj,Wen:2013oza,Freed:2014iua}, and ``anomaly indicators" detecting the 't Hooft anomalies (see e.g.~\cite{PhysRevLett.119.136801,Tachikawa:2016cha}) can be investigated. These anomaly indicators -- which are the partition function evaluated on the generators of the corresponding cobordism groups, and expressed in terms of the modular data of the TQFT (see below) -- are only known for a handful of symmetry groups.

The plan for the remainder of the paper is as follows. In section~\ref{sec:TQFT} we describe the general paradigm of symmetries in topological quantum field theories, and the simplifications that occur for abelian TQFTs. In section~\ref{sec:U(1)} we completely describe all the symmetries for the most characteristic abelian system: $U(1)_k$ Chern-Simons theory. In section~\ref{sec:K-matrix} we generalize the analysis to arbitrary abelian TQFTs, by realizing them as $U(1)^n$ Chern-Simons theories. We prove several results, and make a number of conjectures. In section~\ref{sec:examples} we work out a couple dozen examples in some detail, so as to illustrate the general formalism. Finally, we summarize definitions and notations in Appendix~\ref{app:not}
and leave some proofs and further results to Appendix~\ref{app:further}.

\section{TQFTs and Symmetry}
\label{sec:TQFT}

Before delving into the study of the symmetries of abelian Chern-Simons theories we describe how symmetries are realized in a TQFT in $2+1$ dimensions. We informally review the data defining a TQFT and how, in an abelian TQFT, it is completely fixed in terms of most elementary data, to wit, the anyon fusion algebra and the anyon spins. We then proceed with the physical and mathematical characterization of a symmetry in a TQFT. More details and mathematical elaborations can be found in the literature~\cite{Moore:1988qv,Moore:1989vd,Turaev:1994xb,Kitaev:2006lla,Barkeshli:2014cna,1742-5468-2013-09-P09016}. 

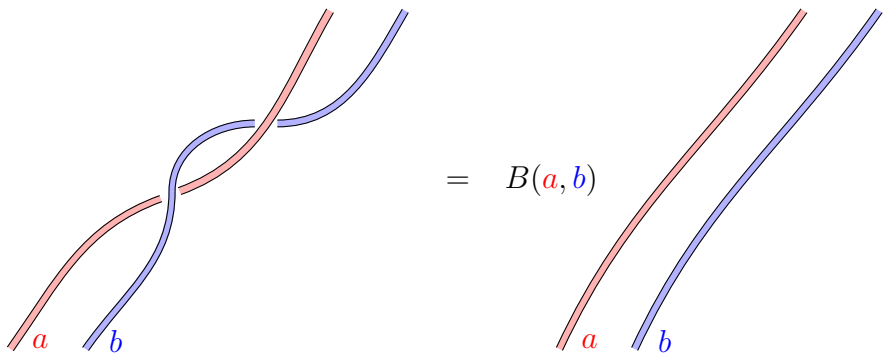
\begin{figure}[!h]
\centering
\begin{tikzpicture}

\draw [line width=1mm,red!30] (0,0) to[out=45+10,in=180+20] (2,2);
\draw [line width=1mm,red!30] (2.25,2.1) to[out=20,in=180+60] (4.25,4.5);

\draw (0-.04,.025) to[out=45+10,in=180+20] (2-.015,2+.04);
\draw (0+.04,-.025) to[out=45+10,in=180+20] (2+.015,2-.04);

\draw (2.25-.02,2.1+.045) to[out=20,in=180+60] (4.25-.045,4.5+.025);
\draw (2.25+.02,2.1-.045) to[out=20,in=180+60] (4.25+.045,4.5-.025);

\draw [line width=1mm,blue!30] (1,0) to[out=45+10,in=-90] (2.15,2.1);
\draw [line width=1mm,blue!30] (2.15,2.1) to[out=90,in=180] (3.25,3);
\draw [line width=1mm,blue!30] (3.55,3) to[out=0,in=180+60] (5.25,4.5);
\node at (.4,.1) {$\color{red}a$};
\node at (1.4,.1) {$\color{blue}b$};

\draw (1-.04,0+.02) to[out=45+10,in=-90] (2.15-.04,2.1);
\draw (1+.04,0-.02) to[out=45+10,in=-90] (2.15+.04,2.1);

\draw (2.15-.04,2.1) to[out=90,in=180] (3.25,3+.045);
\draw (2.15+.04,2.1) to[out=90,in=180] (3.25,3-.045);

\draw (3.55,3+.045) to[out=0,in=180+60] (5.25-.04,4.5+.02);
\draw (3.55,3-.045) to[out=0,in=180+60] (5.25+.04,4.5-.025);

\node at (6.8,2.25) {$=\hspace{10pt}B({\color{red}a},{\color{blue}b})$};

\begin{scope}[shift={(7.3,0)}]
\draw [line width=1mm,red!30] (0,0) to[out=45+20,in=-90-45+10] (3.25,4.5);
\draw [line width=1mm,blue!30] (1,0) to[out=45+20,in=-90-45+10] (4.25,4.5);

\draw (0-.04,0+.02) to[out=45+20,in=-90-45+10] (3.25-.04,4.5+.025);
\draw (0+.04,0-.02) to[out=45+20,in=-90-45+10] (3.25+.04,4.5-.025);

\draw (1-.04,0+.02) to[out=45+20,in=-90-45+10] (4.25-.04,4.5+.025);
\draw (1+.04,0-.02) to[out=45+20,in=-90-45+10] (4.25+.04,4.5-.025);

\node at (.4,.1) {$\color{red}a$};
\node at (1.4,.1) {$\color{blue}b$};
\end{scope}

\end{tikzpicture}
\caption{Braiding of anyons: if at least one of $a,b$ is abelian, then the two lines may be unbraided, a process that generates a phase $B(a,b)\in U(1)$.}
\label{fig:braiding}
\end{figure}

A TQFT can be understood as a finite collection of anyons -- particles with fractional statistics -- belonging to an anyon set $\mathcal A$ endowed with the following additional data:
 \begin{itemize}

\item \textbf{Fusion:} A commutative, associative product $\times\colon\mathcal A\times\mathcal A\to\mathcal A$ describing the fusion of anyons $a,b\in \mathcal A$ (see figure~\ref{fig:fusion}):
\begin{equation}
 a \times b=\sum_{c\in \mathcal A} N_{ab}{}^c\ c\,,
 \label{fusion} \end{equation}
where $N_{ab}{}^c\in \mathbb Z_{\geq 0}$ are the so-called \emph{fusion coefficients}. We denote the trivial anyon by $\boldsymbol 1$.

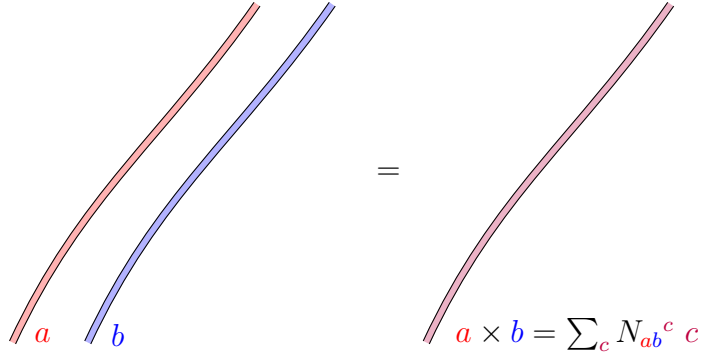
\begin{figure}[!h]
\centering
\begin{tikzpicture}

\draw [line width=1mm,red!30] (0,0) to[out=45+20,in=-90-45+10] (3.25,4.5);
\draw [line width=1mm,blue!30] (1,0) to[out=45+20,in=-90-45+10] (4.25,4.5);

\draw (0-.04,0+.02) to[out=45+20,in=-90-45+10] (3.25-.04,4.5+.025);
\draw (0+.04,0-.02) to[out=45+20,in=-90-45+10] (3.25+.04,4.5-.025);

\draw (1-.04,0+.02) to[out=45+20,in=-90-45+10] (4.25-.04,4.5+.025);
\draw (1+.04,0-.02) to[out=45+20,in=-90-45+10] (4.25+.04,4.5-.025);

\node at (.4,.1) {$\color{red}a$};
\node at (1.4,.1) {$\color{blue}b$};

\node at (5,2.25) {$=$};

\begin{scope}[shift={(5.5,0)}]
\draw [line width=1mm,purple!30] (0,0) to[out=45+20,in=-90-45+10] (3.25,4.5);

\draw (0-.04,0+.02) to[out=45+20,in=-90-45+10] (3.25-.04,4.5+.025);
\draw (0+.04,0-.02) to[out=45+20,in=-90-45+10] (3.25+.04,4.5-.025);

%\node at (6,2.25) {$\displaystyle{\color{red}a}\times{\color{blue}b}=\sum_{\color{purple}c}N_{{\color{red}a}{\color{blue}b}}{}^{\color{purple}c} \ {\color{purple}c}$};

\node at (2,.1) {${\color{red}a}\times{\color{blue}b}=\sum_{\color{purple}c}N_{{\color{red}a}{\color{blue}b}}{}^{\color{purple}c} \ {\color{purple}c}$};
\end{scope}

\end{tikzpicture}
\caption{Fusion of anyons: two unbraided lines with labels $a,b$ can be replaced by one with label $\sum_c N_{ab}{}^c\ c$.}
\label{fig:fusion}
\end{figure}

\item \textbf{Topological spin:} A map $\theta\colon \mathcal A\to U(1)$. The topological spin determines the anyonic character of an anyon. One usually writes $\theta(a)=:\exp(2\pi i h_a)$, where $h_a\colon\mathcal A\to \mathbb Q/\mathbb Z$ is the spin of $a$. The topological spin controls the framing anomaly of a knot (the dependence of observables on the choice of the homotopy class of a normal vector field, see figure~\ref{fig:spin}).

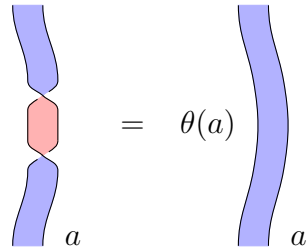
\begin{figure}[!h]
\centering
\begin{tikzpicture}

\filldraw[blue!30] (-.2,0) to[out=90,in=-90] (0,1) to[out=90,in=180+45] (.2,1.2) to[out=-45,in=90] (.4,1) to[out=-90,in=90] (.4-.2,0) -- cycle;
\draw (-.2,0) to[out=90,in=-90] (0,1) to[out=90,in=180+35] (.13,1.15);% to[out=-45,in=90] (.4,1) -- (.4,0);

\filldraw[red!30] (.2,1.2) to[out=90+45,in=-90] (0,1.4) -- (0,1.8) to[out=90,in=180+45] (.2,2) to[out=-45,in=90] (.4,1.8) -- (.4,1.4) to[out=-90,in=45] cycle;
\draw (.2,1.2) to[out=90+45,in=-90] (0,1.4) -- (0,1.8) to[out=90,in=180+35] (.13,1.95);
\draw (.2,2) to[out=-45,in=90] (.4,1.8) -- (.4,1.4);% to[out=-90,in=45] cycle;
\draw (.4,1.4) to[out=-90,in=35] (.28,1.25);

\filldraw[blue!30] (.4-.2,3.2) to[out=-90,in=90] (.4,2.2) to[out=-90,in=45] (.2,2) to[out=90+45,in=-90] (0,2.2) to[out=90,in=-90] (-.2,3.2) -- cycle;
\draw (.4-.2,3.2) to[out=-90,in=90] (.4,2.2) to[out=-90,in=35] (.28,2.05);
\draw (.21,1.99) to[out=90+45,in=-90] (0,2.2) to[out=90,in=-90] (-.2,3.2);

\draw (.19,1.21) to[out=-45,in=90] (.4,1) to[out=-90,in=90] (.4-.2,0);

\node at (.6,.1) {$a$};

\node at (2,1.6) {$=\hspace{10pt}\theta(a)$};

\begin{scope}[shift={(3,0)}]
\filldraw[blue!30] (-.2,0) to[out=90,in=-100] (0,1) to[out=80,in=-80] (0,2.2) to[out=100,in=-90] (-.2,3.2) -- (.4-.2,3.2) to[out=-90,in=100] (.4,2.2) to[out=-80,in=80] (.4,1) to[out=-100,in=90] (.4-.2,0) -- cycle;
\draw (-.2,0) to[out=90,in=-100] (0,1) to[out=80,in=-80] (0,2.2) to[out=100,in=-90] (-.2,3.2);
\draw (.4-.2,3.2) to[out=-90,in=100] (.4,2.2) to[out=-80,in=80] (.4,1) to[out=-100,in=90] (.4-.2,0);
\node at (.6,.1) {$a$};
\end{scope}

\end{tikzpicture}
\caption{Topological spin: anyons are to be thought of as ribbons rather than knots. Observables depend on the twisting thereof, through their spin.}
\label{fig:spin}
\end{figure}

\item \textbf{$\boldsymbol S$- and $\boldsymbol T$-matrices:} A representation of the modular group. The $S$-matrix determines the braiding phase $B\colon \mathcal A\times \mathcal A\rightarrow U(1)$ between anyons (see figure~\ref{fig:braiding})
\begin{equation}
B(a,b)=\frac{S_{ab}}{S_{\boldsymbol 1b}}\,,
\end{equation}
while $T_{ab}=\theta_a e^{-2\pi i c/24}\delta_{ab}$, where $c$ is the \emph{chiral central charge} of the TQFT, which controls the framing anomaly (the dependence of observables on the 2-framing of the manifold).

\item \textbf{$\boldsymbol F$- and $\boldsymbol R$-symbols:} The associator and braiding isomorphism, encoding the fusion of multiple anyons and their half-braiding. This data is defined modulo local, redundant isomorphisms (gauge transformations) $U$ defined on fusion vector spaces. The gauge-transformed data, which we denote by $UF$ and $UR$, is physically equivalent to $F$ and $R$, and define the same TQFT.
\end{itemize}

This data is subject to nontrivial consistency conditions, known as the Moore-Seiberg relations, which include the hexagon and pentagon relations involving the $F$- and $R$-symbols. These relations imply that some of the data above is actually redundant; for example, the topological spin $\theta$ is a gauge invariant combination of the $F$- and $R$-symbols. The TQFT data defines a modular tensor category. This data can be used to compute an arbitrary correlation function of the TQFT (cf. (\ref{Ward})).
 
An anyon $a$ is said to be \emph{abelian} if the fusion of $a$ with an arbitrary anyon $b$ contains a single anyon $c=c(a,b)$, i.e.
\begin{equation}
a\times b=c \qquad\qquad \forall b \in \mathcal A\,.
\end{equation}
In terms of the fusion coefficients (\ref{fusion}), $a$ is abelian if for any $b$ the sum $\sum_{c\in \mathcal A}N_{ab}{}^c$ equals $1$. An abelian anyon $a\in\mathcal A$ has a unique inverse $\bar a\in\mathcal A$ such that $a\times \bar a=\boldsymbol 1$, and therefore abelian anyons form a finite abelian group, the one-form symmetry group of the TQFT \cite{Gaiotto:2014kfa}.

An abelian TQFT is a TQFT in which all anyons in $\mathcal A$ are abelian. Therefore, in an abelian TQFT the anyon fusion algebra defines a finite abelian group, which we also denote by $\mathcal A$. Remarkably, an abelian TQFT is completely determined by the group $\mathcal A$ encoding the fusion of anyons, and by the topological spin $\theta\colon \mathcal A\rightarrow U(1)$ of the anyons, which is a quadratic, homogeneous function on $\mathcal A$ \cite{Belov:2005ze,Stirling:2008bq,Kapustin:2010hk,Lee2018}.\footnote{$\theta$ is a quadratic function if the symmetric form in~\eqref{braid} is bilinear, i.e.~$B(a\times b,c)=B(a,c)B(b,c)$. Homogeneity means that $\theta(a^n)=\theta(a)^{n^2}$ for any $n\in \mathbb Z$, which implies that $\theta(\boldsymbol1)=1$.}
The entire TQFT data can be reconstructed from $\mathcal A$ and such a $\theta$.\footnote{The central charge is determined by $(\mathcal A, \theta)$ only modulo $8$. This indeterminacy can be understood as coming from the fact that one may always tensor by an even unimodular lattice, which has no lines, but may add central charge; the minimal such lattice is $E_8$, which has signature $8$. Some more refined observables (see e.g.~\cite{Belov:2005ze,Kapustin:2010hk}) are sensitive to the actual value of $c$, and not only to it modulo $8$. If we are interested in such observables, the TQFT data should be taken as $(\mathcal A,\theta,c)$ rather than just $(\mathcal A,\theta)$. This will not play a major role in this work.} The braiding phase of the abelian TQFT with fusion $\mathcal A$ and spin $\theta$ takes the form
\begin{equation}
B(a,b)=\frac{\theta(a\times b)}{\theta(a)\theta(b)}\,,\qquad a,b\in \mathcal A\,,
\label{braid}
\end{equation}
while the corresponding $S$-matrix is 
\begin{equation}
S(a,b)=\frac{B(a,b)}{\sqrt{|\mathcal A|}}\,.
\end{equation}
Importantly, given $(\mathcal A,\theta)$ there is a unique equivalence class of $F$ and $R$ symbols, and therefore a unique TQFT with that $(\mathcal A,\theta)$. Summarizing, in an abelian TQFT the entire theory is completely fixed in terms of $(\mathcal A,\theta)$. This statement is \emph{not} true in a generic non-abelian TQFT, which is what makes the abelian case more tractable.

The discussion above applies as stated for a bosonic TQFT, a theory that does not require specifying a spin structure on the three-manifold where it is defined. Many interesting TQFTs, including abelian Chern-Simons theories, do require a choice of a spin structure to be defined. Such TQFTs are known as \emph{spin} TQFTs. In a spin TQFT there is a distinguished abelian anyon $\psi$ with topological spin $\theta(\psi)=-1$ and trivial braiding with all other anyons. This implies that $\psi$ squares to the trivial anyon, i.e. $\psi\times \psi=\boldsymbol1$, and that $\theta(a\times \psi)=-\theta(a)$ for all $a\in \mathcal A$. In other words, a spin TQFT has a local (spin $1/2$) fermion, which endows the data above with a $\mathbb Z_2$-grading (i.e., anyons come in pairs $\{a,a\times\psi\}$).

Any abelian TQFT, bosonic or spin, admits a representation as an abelian Chern-Simons theory~\cite{Belov:2005ze,Kapustin:2010hk,Nikulin_1980,Cano:2013ooa,Lee2018}, and is completely determined by $(\mathcal A,\theta)$. Therefore, in spite that a complete and universally accepted axiomatization of a spin TQFT from a categorical point of view is lacking, the abelian Chern-Simons realization of the TQFT and its datum $(\mathcal A,\theta)$ suffice to determine the symmetries of spin abelian TQFTs (we also provide path integral arguments to exhibit the symmetries of abelian Chern-Simons theories that do not rely on the precise categorical characterization of spin TQFTs). 

The symmetries of a TQFT are, by definition, the automorphisms of its data~\cite{Barkeshli:2014cna}. An automorphism $g$ of a TQFT is a permutation of the anyons $g\colon \mathcal A\to \mathcal A$
\begin{equation}
a\mapsto g(a)
\end{equation}
that preserves the fusion algebra $\mathcal A$
\begin{equation}
g(a\times b)=g(a) \times g(b)\qquad \Longleftrightarrow\qquad N_{g(a)g(b)}{}^{g(c)}=N_{ab}{}^c\,.
\end{equation}
If the symmetry of the TQFT is unitary it must preserve the data modulo gauge transformations
\begin{equation}
\theta(g(a))=\theta(a)\,,\qquad S_{g(a)g(b)}=S_{ab}\,,\qquad g\cdot F=UF\,\qquad g\cdot R=UR\,,
\end{equation}
while if the symmetry is anti-unitary it preserves the data modulo gauge transformations, up to complex conjugation 
\begin{equation}
\theta(g(a))=\theta(a)^*\,\qquad S_{g(a)g(b)}=S^*_{ab}\,\qquad g\cdot F=UF^*\,,\qquad g\cdot R=UR^*\,.
\end{equation}
Despite this explicit characterization, little is known about the actual symmetries of TQFTs. By contrast, the one-form symmetries of a TQFT are completely understood; they are determined by the abelian anyons and their fusion. Henceforth, when we discuss symmetries we refer to usual (zero-form) symmetries.

As reviewed above, in an abelian TQFT the entire data is completely determined by the abelian group $\mathcal A$ encoding the fusion algebra and the topological spin $\theta$. A necessary condition for the transformation $g$ to a symmetry of an abelian TQFT is that $g\colon \mathcal A\to \mathcal A$ is an automorphism of the finite group $\mathcal A$
\begin{equation}
 g(a\times b)=g(a)\times g(b)\,.
\end{equation}
The set of automorphisms of $\mathcal A$, denoted by $\Aut(\mathcal A)$, is a finite, generically nonabelian group. An automorphism $g$ of $\mathcal A$ lifts to a unitary symmetry of the abelian TQFT if and only if
\begin{equation}
\theta(g(a))=\theta(a)\,,
\label{unitspin}
\end{equation}
and to an anti-unitary symmetry if and only if
\begin{equation}
\theta(g(a))=\theta(a)^*\,.
\label{aunitspin}
\end{equation}
If such an automorphism $g$ exists, it is guaranteed that the entire data of the abelian TQFT is preserved and $g$ is a symmetry. In other words, the group of symmetries of an abelian TQFT is the subgroup of $\Aut(\mathcal A)$ that preserves the topological spins (up to complex conjugation for anti-unitary symmetries). We introduce the following notation for this group:
\begin{definition}\label{def:aut}
\normalfont Given an abelian TQFT, we let $\Aut(\mathcal A,\theta)\subseteq\Aut(\mathcal A)$ denote the group of all symmetries, and $\Aut_U(\mathcal A,\theta)\subseteq\Aut(\mathcal A,\theta)$ the subgroup of unitary symmetries.
\end{definition}

The main goal of this work is to study the object $\Aut(\mathcal A,\theta)$. We determine it explicitly in the case of $U(1)_k$, and give a complete characterization thereof for arbitrary abelian theories. We will also work out a few illustrative examples in some detail.

\section{$U(1)_k$ Chern-Simons}
\label{sec:U(1)}
We begin by reviewing Chern-Simons theory with gauge group $U(1)$. The generalization to the gauge group $U(1)^n$ is the content of section~\ref{sec:K-matrix}.

The Lagrangian of $U(1)_k$ Chern-Simons theory is 
\begin{equation}
\mathcal L=\frac{k}{4\pi}a\, \mathrm da\,,
\label{lagrangian}
\end{equation}
where $a$ is a $U(1)$ gauge gauge field and the coupling $k\in \mathbb Z$ is quantized. Being topological, the theory can be defined on an arbitrary (oriented, framed) three-manifold, perhaps with a choice of spin structure depending on the parity of $k$. The equations of motion are
\begin{equation} \label{eq:EOM}
\mathrm da=0
\end{equation}
and the classical field configurations are flat connections.

The gauge invariant operators in this theory are the Wilson lines 
\begin{equation}
W_\alpha(\gamma):=\exp\!\left[i \alpha\int_\gamma a\right],\qquad \alpha\in\mathbb Z\,.
\end{equation}
Physically, $W_\alpha$ describes the worldline of an anyon $\alpha$ with topological spin
\begin{equation}\label{eq:spin}
\theta(\alpha)= e^{2\pi i h_\alpha}\,,\qquad h_\alpha=\frac{\alpha^2}{2k}\,.
\end{equation}
The spin of an anyon $h_\alpha$ is only well-defined modulo an integer, because it cannot be distinguished from an anyon enriched with a soft $a$-photon, which has spin $h=1$. If we introduce a background electromagnetic field, the anyon $\alpha$ is seen to carry a fractional charge given by $\alpha/k$, as follows from the coupling $\frac{1}{2\pi} A\,\mathrm da$.

The anyon fusion algebra is determined by the OPE of the corresponding Wilson lines: $\alpha \times \beta=\alpha+\beta$. The braiding phase acquired by an anyon $\alpha$ circumnavigating around an anyon $\beta$ is 
\begin{equation} \label{eq:braiding}
B(\alpha,\beta)=\frac{\theta(\alpha\times \beta)}{\theta(\alpha)\theta(\beta)}=\mathrm e^{2\pi i\, \frac{\alpha\beta}{k}}\,.
\end{equation}
%This is measured by the correlation function $\langle e^{i n\int_{C_1} a} e^{i m\int_{C_2} a}\rangle$, where $C_1$ and $C_2$ are curves which link. {\bf (add figure)}

It follows from~\eqref{eq:braiding} and~\eqref{eq:spin} that the anyon $\alpha=k$ has trivial braiding with respect to all other anyons, and has spin $h=0\mod 1$ for $k$ even and spin $h=1/2\mod1$ for $k$ odd. Therefore $U(1)_k$ is a spin TQFT for odd $k$, and a bosonic TQFT for even $k$. The former describes, for example, the fractional quantum Hall fluid at filling fraction $\nu=1/k$, where the anyon $\alpha=k$ represents the microscopic electron.

Since the anyons $\alpha$ and $\alpha+k$ have indistinguishable braiding properties, and identical spins for $k$ even, and spins that differ by $1/2$ for $k$ odd, the lines of $U(1)_k$ are subject to an equivalence relation: anyons related by a transparent bosonic anyon are to be identified. A bosonic theory can be made into a spin theory by tensoring with the trivial spin TQFT of a transparent fermion $\{\boldsymbol1,\psi\}$. We will often follow the convention of leaving this factor implicit when discussing spin TQFTs.

Summarizing, the anyon set and the fusion algebra of $U(1)_k$ is:
\begin{itemize}
\item $U(1)_k$, $k$ even: the theory has $k$ anyons labeled by $\alpha\in \{0,1,\ldots,k-1\}$ and a $\mathcal A\cong\mathbb Z_k$ fusion algebra
\begin{equation}
\alpha\times \beta=\alpha+\beta\mod k\,.
\end{equation}
The theory is bosonic and can be defined on an arbitrary three-manifold. 

\item $U(1)_k$, $k$ odd: the theory has $2k$ anyons labeled by $\alpha\in \{0,1,\ldots,2k-1\}$ and a $\mathcal A\cong\mathbb Z_{2k}$ fusion algebra
\begin{equation}
\alpha\times \beta=\alpha+\beta\mod 2k\,.
\end{equation}
It is a spin TQFT, as signalled by the presence of the transparent fermion $\alpha=k$.

\item $U(1)_k\times \{\boldsymbol 1,\psi\}$, $k$ even: the theory has $2k$ anyons labeled by the pair $(\alpha,\beta)$, where $\alpha\in \{0,1,\ldots,k-1\}$ and $\beta\in \{0,1\}$, and the fusion algebra is $\mathcal A\cong\mathbb Z_k\times \mathbb Z_2$
 \begin{equation}
(\alpha,\beta)\times (\alpha',\beta')=(\alpha+\alpha'\mod k,\ \beta+\beta'\mod2)\,.
\end{equation}
It is a spin TQFT by virtue of the tensoring with $\{\boldsymbol1,\psi\}$, where $\psi$ is represented by the Wilson line with charges $(0,1)$.
\end{itemize}

We now proceed to determine the full set of symmetries of $U(1)_k$ Chern-Simons theory.

\subsection{Symmetries of $U(1)_k$}
 
 We start with the manifest Lagrangian symmetries. $U(1)_k$ with $k>2$ has a $\mathbb Z_2$ unitary Lagrangian symmetry $\mathsf C: a\mapsto -a$, charge conjugation, under which $\mathcal L\mapsto\mathcal L$, and that acts on the anyons as 
\begin{equation}
\mathsf C\colon \alpha\mapsto -\alpha\,. 
\label{chargeconj}
\end{equation}
The operation $\mathsf C$ is not a symmetry of $U(1)_1$ and $U(1)_2$ because charge conjugation acts trivially on all the lines, since $1=-1\mod2$.

Time-reversal is an anti-unitary transformation 
\begin{equation}
\mathsf T\colon~\begin{cases}
a_0(x^0)\mapsto +a_0(-x^0)\\
a_i(x^0)\mapsto - a_i(-x^0)
\end{cases}\,
\label{timere}
\end{equation}
which acts on the Wilson lines as $\mathsf T\colon W_\alpha(\gamma)\mapsto W_\alpha(\mathsf T\gamma)$, where $\mathsf T\gamma$ denotes the time-reflected image of the curve $\gamma$. While $\mathsf T$ is a symmetry of the equations of motion~\eqref{eq:EOM}, it does not leave the action invariant, i.e.~$\mathcal L\mapsto -\mathcal L$. This transformation is not a quantum symmetry either since it does not obey the corresponding Ward identity~\eqref{Ward}. Therefore, if $\mathsf T$ is to be a symmetry of $U(1)_k$, it must act non-trivially on the anyon labels:
\begin{equation}
\mathsf T\colon W_\alpha(\gamma)\mapsto W_{\mathsf T(\alpha)}(\mathsf T\gamma)
\end{equation}
for some $\mathsf T\colon\mathcal A\to\mathcal A$.

In order to study the quantum symmetries of $U(1)_k$ Chern-Simons theory we first need to understand the automorphisms of its fusion algebra $\mathcal A$. Indeed, as explained in section~\ref{sec:TQFT}, a transformation $g$ is a symmetry of a TQFT if it is an automorphism of its data $(\mathcal A,\theta)$ which requires, first and foremost, that $g\in\Aut(\mathcal A)$. As usual, any element of $\Aut(\mathcal A)$ is completely determined by its action on the generators of $\mathcal A$. With this in mind, the automorphisms of the fusion algebra $\mathcal A$ of $U(1)_k$ Chern-Simons theory are as follows:
\begin{itemize}
\item $U(1)_k$, $k$ even. The most general endomorphism of $\mathcal A\cong\mathbb Z_k$ acts as $g: \alpha\mapsto q\alpha\mod k$, where $q:=g(1)\in \mathcal A$ and $\alpha\in \{0,1\,\ldots, k-1\}$. This lifts to an automorphism of $\mathbb Z_k$ if and only if $g$ maps a generator of $\mathbb Z_k$ into a generator of $\mathbb Z_k$. This requires $q$ to be relatively prime to $k$, i.e.~$\gcd(q,k)=1$: 
\begin{equation}\label{homoeven}
g\colon\alpha\mapsto q\alpha\mod k\,,\qquad\gcd(q,k)=1\,.
\end{equation}
The number of automorphisms (and of generators) of $\mathbb Z_k$ is the number of totatives of $k$: the number of integers $1\leq q\leq k$ such that $\gcd(q,k)=1$. This number is counted by the Euler totient function $\phi(k)$. The automorphism group $\Aut(\mathbb Z_k)$ is the multiplicative group of integers modulo $k$, an abelian group often denoted as $\mathbb Z_k^\times$.

\item $U(1)_k$, $k$ odd. The most general endomorphism of $\mathcal A\cong\mathbb Z_{2k}$ acts as $g\colon \alpha \mapsto q\alpha\mod 2k$, where $q:=g(1)\in \mathcal A$ and $\alpha\in \{0,1,\ldots, 2k-1\}$. It is an automorphism if and only if $q$ is coprime to $2k$:
\begin{equation}\label{homoodd}
g\colon\alpha\mapsto q\alpha\mod 2k\,,\qquad\gcd(q,2k)=1\,. 
\end{equation}
The automorphisms automatically preserve the transparent fermion ($\alpha=k$) since $qk=k \mod 2k$ for $q$ odd. The number of automorphisms of $\mathbb Z_{2k}$ is the Euler totient function $\phi(2k)=\phi(k)$, the last equality by virtue of $k$ being odd. The automorphism group is $\Aut(\mathbb Z_{2k})=\mathbb Z_{2k}^\times$.
 
\item $U(1)_k\times \{\boldsymbol1,\psi\}$, $k$ even. The most general endomorphism of $\mathcal A\cong\mathbb Z_{k}\times \mathbb Z_{2}$ acts as 
\begin{equation} \label{actkevenspin}
g\colon\begin{pmatrix} \alpha\\ \beta\end{pmatrix}\mapsto \begin{pmatrix}a &b\\c&d\end{pmatrix}
 \begin{pmatrix}\alpha\\ \beta \end{pmatrix}\begin{matrix}\mod k\\\mod 2\end{matrix}\,,\qquad \begin{matrix}
a,b\in\mathbb Z_k\\
c,d\in\mathbb Z_2
 \end{matrix}
\end{equation}
where $\alpha\in \{0,1,\ldots,k-1\}$ and $\beta\in \{0,1\}$. Such a map is an automorphism if and only if it is invertible $(\mathrm{mod}\ k,\mathrm{mod}\ 2)$. The automorphism group of $\mathbb Z_k\times\mathbb Z_2$ does not admit as straightforward a description as in the previous cases, but its order is known: $4\phi(k)$ if $k/2$ is even, and $6\phi(k)$ if $k/2$ is odd~\cite{semusstructure,sommer2013automorphism}. The automorphism group is generically non-abelian.

Locality of the TQFT requires that the automorphism $g$ preserves the transparent fermion, $g(\psi)=\psi$, that is, it fixes the anyon $(0,1)$. This implies that the candidate symmetries of $U(1)_k\times \{\boldsymbol1,\psi\}$ with $k$ even are the automorphisms of $\mathbb Z_{k}\times \mathbb Z_{2}$ with $b=0$ and $d=1$. In order for the transformation to be invertible, one must have $\gcd(a,k)=1$ or, if $k/2$ is odd, $\gcd(a,k/2)=1$. The number of such transformations is $2\phi(k)$ and $3\phi(k)$ for $k/2$ even and odd, respectively.

\end{itemize}

This immediately shows that $U(1)_1$ and $U(1)_2$ have no symmetries since $\Aut(\mathbb Z_2)$ is trivial, and indeed charge conjugation $\mathsf C$ acts trivially in these theories.

We have thus characterised all the automorphisms of $\mathcal A$. These are the candidate transformations to be a symmetry of the TQFT. They uplift to symmetries if they respect the topological spin of the lines (up to complex conjugation for anti-unitary symmetries). We turn to this question next.

\subsubsection{Anti-unitary Symmetries}

We start by studying the anti-unitary symmetries of $U(1)_k$ Chern-Simons theory. We already established that the canonical time-reversal 
transformation~\eqref{timere} is not a symmetry of $U(1)_k$. Since the TQFT data of $U(1)_k$ Chern-Simons theory is determined by the fusion algebra $\mathcal A$ and the topological spin $\theta$, an automorphism $\mathsf T\in\Aut(\mathcal A)$ will lead to an anti-unitary symmetry if and only if 
\begin{equation}\label{eq:eqntimerever}
\theta(\mathsf T(\alpha))=\theta(\mathsf T(\alpha))^*\quad\Longleftrightarrow\quad h_{\mathsf T(\alpha)}=-h_{\alpha}\mod 1\,.
\end{equation}

This condition is not satisfied by every automorphism of $\mathcal A$. More importantly, depending on the value of $k$, there will be cases where there are no automorphisms at all that satisfy~\eqref{eq:eqntimerever}. This is precisely what happens for even $k$, when we regard $U(1)_k$ as a bosonic theory\footnote{This result also follows from the fact that the central charge of $U(1)_k$ is not proportional to $4$.}:

\begin{proposition}\label{lm:bosonic}
The bosonic theory $ U(1)_k$, with $k$ even, is never time-reversal invariant.
\end{proposition}

\emph{Proof}. Consider the permutation $\mathsf T\colon\alpha\mapsto q\alpha$ for some $q\in[0,k)$. This operation satisfies $h_{\mathsf T(\alpha)}=-h_{\alpha}\mod 1$ if and only if
\begin{equation}
\frac{q^2\alpha^2}{2k}+\frac{\alpha^2}{2k}\in\mathbb Z\,.
\end{equation}

If we take, for example, the fundamental line $\alpha=1$, this requires $\frac{1+q^2}{2k}$ to be an integer. But $q$ must odd for $\mathsf T$ to be an automorphism, and so $1+q^2=2\mod4$, which means that $\frac{1+q^2}{2k}$ cannot be an integer.\hfill$\square$

\medskip

We therefore see that the theory $U(1)_k$ can only possibly be time-reversal invariant if we regard it as a spin TQFT. And even if we do so, there will still be some values of $k$ for which $U(1)_k$ admits no time-reversal permutation at all. To see this, define the following:

\begin{definition}
\normalfont We let $\mathbb T\subset \mathbb Z$ be the set of integers $k$ such that $-1$ is a quadratic residue modulo $k$, i.e. $q^2=-1\mod k$ for some $q\in \mathbb Z$. In other words,
\begin{equation}
\mathbb T:=\{k\in \mathbb Z\,|\, kp-q^2=1\quad\text{for some $p,q\in\mathbb Z$}\}\,.
\end{equation}
\end{definition}

With this, we prove that
\begin{proposition}\label{th:time_reversal}
The spin theory $U(1)_k$ is time-reversal invariant if and only if $k\in\mathbb T$.
\end{proposition}

\emph{Proof}. We begin with the case of odd $k$, that is, $U(1)_k$, where $\mathcal A\cong\mathbb Z_{2k}$. We shall look for the most general automorphism $\mathsf T\in\Aut(\mathcal A)$ that satisfies~\eqref{eq:eqntimerever}. Any such operation is of the form
\begin{equation}
\mathsf T(\alpha)=q\alpha,\qquad q:=\mathsf T(1)\in[0,2k)\,.
\end{equation}

If we impose that $h_{\mathsf T(1)}=-h_1\mod 1$, we get $1+q^2=2pk$ for some integer $p$. It is easy to show that this equation is solvable if and only if $k\in\mathbb T$. One direction is obvious; for the opposite direction, assume that $1+\tilde q^2=\tilde pk$. If $\tilde p$ is even, we are done; if it is odd, then we can set
\begin{equation}\label{eq:even_from_odd}
q:=\tilde q+k,\qquad p:=\tilde q+\frac{\tilde p+k}{2}
\end{equation}
which satisfy $1+q^2=2pk$, as required (note that $\tilde p+k$ is even, and so $p\in\mathbb Z$). 

Once we ensure the spin of the generator transforms properly under $\mathsf T$, it is easy to show that so do the rest of lines. Indeed,
\begin{equation}
h_{\mathsf T(\alpha)}=\frac{q^2\alpha^2}{2k}= \frac{(2pk-1)\alpha^2}{2k}=-\frac{\alpha^2}{2k}\mod 1\,,
\end{equation}
 where we have used that $1+q^2=2pk$.
 
Finally, it is also easy to show that \emph{any} integer $q$ that solves $1+q^2=2pk$ will be a time-reversal operation. Indeed, $1+q^2=2pk$ implies that any common factor to $k$ and $q$ must divide $1$, and so $\gcd(q,k)=1$, which means that $\alpha\mapsto q\alpha$ is invertible.

We now move on to the even $k$ case, that is, $U(1)_k\times \{\boldsymbol1,\psi\}$, where $\mathcal A\cong\mathbb Z_k\times\mathbb Z_2$, where the first factor is generated by the fundamental line $(1,0)$, and the second one by the transparent fermion $\psi=(0,1)$.

Any fusion endomorphism is fixed once we choose its action on the generators. In fact, the transparent fermion is the only spin $h=1/2$ line that braids trivially to all other lines (because $U(1)_k$ is bosonic), and thus the action of time-reversal on it is fixed to $\mathsf T(\psi)\equiv \psi$. Therefore, we only have freedom to choose how time-reversal acts on $(1,0)$. We write $\mathsf T(1,0):=(q_1,q_2)$ for a pair of integers $q_1,q_2$, where $q_1\in \{0,1,\ldots,k-1\}$ and $q_2\in \{0,1\}$.

Proposition~\ref{lm:bosonic} implies that $q_2=0$ is not possible. Therefore, the candidate anti-unitary transformation is $\mathsf T(1,0)\equiv (q,1)$ for some integer $q\in[0,k)$, and so the most general endomorphism is of the form
\begin{equation}
\mathsf T(\alpha,\beta)=(q\alpha,\alpha+\beta)\,.
\end{equation}

We now insist that the spin of $(1,0)$ is mapped into its negative under time-reversal. Imposing that $h_1=-h_{\mathsf T(1)\otimes\psi}\mod 1$ we get $1+q^2=(2p-1)k$ for some integer $p$. Once again, it is easy to show that this equation is solvable if and only if $k\in\mathbb T$. One direction is obvious; for the opposite direction, assume that $1+\tilde q^2=\tilde pk$. Then, upon reducing the equation modulo $4$, it becomes clear that $\tilde p$ has to be odd, and so we can write $\tilde p=2p-1$, as we wanted to show.

Once we ensure the spin of the generator transforms properly under $\mathsf T$, it is easy to show that so do the rest of lines. Indeed,
\begin{equation}
h_{\mathsf T(\alpha,\beta)}=\frac{q^2\alpha^2}{2k}+\frac12(\alpha+\beta)^2=\frac{(-1+k(2p-1))\alpha^2}{2k}+\frac12(\alpha+\beta)^2\,,
\end{equation}
where we have used $q^2=-1+k(2p-1)$. This is clearly equal to
\begin{equation}
h_{\mathsf T(\alpha,\beta)}=-\frac{\alpha^2}{2k}+\frac12\beta^2=-h_{(\alpha,\beta)}\mod1
\end{equation}
as required.

Finally, it is also easy to show that \emph{any} integer $q$ that solves $1+q^2=(2p-1)k$ will be a time-reversal operation. Indeed, and as before, this equation can only be satisfied if $\gcd(q,k)=\gcd(q,k/2)=1$, and so $(\alpha,\beta)\mapsto(q\alpha,\alpha+\beta)$ is invertible (i.e.~an automorphism of $\mathbb Z_k\times \mathbb Z_2$). \hfill$\square$

\medskip

As we can see, the set $\mathbb T\subset \mathbb Z$ plays a key role in the study of the time-reversal properties of $U(1)_k$ (and, as we shall see, of $U(1)^n$). We therefore make a few remarks about this set:
\begin{itemize}

\item The first few solutions are $k=1$, 2, 5, 10, 13, 17, 25, 26, 29, 34, 37, 41, 50, 53, 58, 61, 65, 73, 74, 82, 85, 89, $97,\dots$.

%\item It is easy to see that if $k\in \mathbb T$, then the associated $p$ is also in $\mathbb T$.

%\item The integers $k,q$ are coprime (because $kp-q^2=1$ implies that $\gcd(k,q)$ divides $1$).

%\item One has $kp\not\propto 4$ (inasmuch as $q^2+1$ never vanishes modulo $4$).

%\item $k\in \mathbb T$ if and only if $d\in \mathbb T$ for all $d|k$.

\item A given $k$ is in $\mathbb T$ if and only if it can be written as $k=a^2+b^2$ for relatively prime $a,b\in\mathbb Z$ (see e.g.~\cite{trove.nla.gov.au/work/6440368}, theorem 3.21).

\item Given the prime decomposition of $k$
\begin{equation}
k=2^e\left[\,\prod_{\pi=1\ \mathrm{mod}\ 4} \pi^\alpha\,\right]\left[\,\prod_{\pi=3\ \mathrm{mod}\ 4} \pi^\beta\,\right]\,,
\end{equation}
 $k\in \mathbb T$ if and only if $e\in\{0,1\}$ and $\beta_i\equiv0$ (see e.g.~\cite{trove.nla.gov.au/work/6440368}, theorem 3.20). In other words, $k\in \mathbb T$ if and only if all its prime factors are Pythagorean, or Pythagorean with a single factor of $2$. This implies, for example, that $\mathbb T^{\mathrm{even}}=2\mathbb T^{\mathrm{odd}}$.

\item The set $\mathbb T$ contains a special subset $\mathbb P$, defined as those integers $k$ for which the (negative) Pell equation is solvable:
\begin{equation}
\mathbb P:=\{k\in \mathbb Z\,|\, kp^2-q^2=1\quad\text{for some $p,q\in\mathbb Z$}\}\,.
\end{equation}
Unlike $\mathbb T$, the set $\mathbb P$ has no simple characterization in terms of the prime decomposition of $k$. See Appendix~\ref{app:further} for some mode details about Pell numbers.

\item The density of $\mathbb T$ is $\#\{k\in\mathbb T\,\mid\,k\le x\}\sim x/\sqrt{\log x}$. It is conjectured that around $57\%$ of the numbers in $\mathbb T$ are in $\mathbb P$~\cite{Stevenhagen1995,Bosma1996DensityCF}.

\end{itemize}

If $k\in\mathbb T$, there exists an integer $q\in[0,k)$ such that $q^2=-1\mod k$. We explain in the Appendix~\ref{app:further} how to construct $q$ explicitly.

\medskip

We now go back to the theory $U(1)_k$. We have the following:
\begin{proposition}\label{pr:order_4}
The time-reversal symmetry of $U(1)_k$ is an order-four operation (except for $k=1,2$, where it is of order two).
\end{proposition}

\emph{Proof}. We shall prove that $\mathsf T^2=\mathsf C$, where $\mathsf C\colon \alpha\mapsto-\alpha$ is the unitary $\mathbb Z_2$ charge conjugation symmetry~\eqref{chargeconj}. From this it follows that $\mathsf T^4=1$, and therefore $\mathsf T$ is an order-four operation (except for $k=1,2$, where $\mathsf C$ is trivial).\footnote{The symmetry algebra $\mathsf T^2=\mathsf C$ can in principle be extended by fermion parity $(-1)^F$, which does not act on the Wilson lines. The full symmetry algebra is, therefore, either $\mathbb Z_4\times\mathbb Z_2$ (corresponding to $\mathsf T^4=1$) or $\mathbb Z_8$ (corresponding to $\mathsf T^4=(-1)^F$). Figuring out which of these options is realized requires determining how $(-1)^F$ acts on these theories, a subject that is beyond the scope of this paper.}

Showing that $\mathsf T^2=\mathsf C$ is straightforward. If $k$ is odd, then
\begin{equation}%\label{eq:T2_C_odd}
\begin{aligned}
-\mathsf T^2(\alpha)=-q^2\alpha=(1-2pk)\alpha=\alpha\mod 2k\,.
\end{aligned}
\end{equation}
Similarly, if $k$ is even, then 
\begin{equation}%\label{eq:T2_C_even}
\begin{aligned}
-\mathsf T^2(\alpha,\beta)&=(-q^2\alpha,(q+1)\alpha+\beta)\\
&=((1-(2p-1)k)\alpha,(q+1)\alpha+\beta)\\
&=(\alpha,\beta)\mod (k,2)
\end{aligned}
\end{equation}
where we have used that $q$ is odd.\hfill$\square$

We see that if $k\in\mathbb T$, then there exists some anti-unitary operation $\mathsf T$ which satisfies a $\mathbb Z_4$ algebra. That being said, there will be, in general, more than one such permutations, and therefore the time-reversal transformation is not unique. We have the following result:
\begin{proposition}\label{pr:2omega}
If $U(1)_k$ is time-reversal invariant, there are $2^{\varpi(k)}$ different anti-unitary permutations, where $\varpi(k)$ denotes the number of distinct prime factors of $k$ for $k$ odd and of $k/2$ for $k$ even $($cf.~\eqref{defff}$)$.
\end{proposition}

\emph{Proof}. Indeed, there are as many permutations as there are solutions to $q^2=-1+(2p-1)k$ with $q\in[0,k)$ for $k$ even, and to $q^2=-1+2pk$ with $q\in[0,2k)$ for $k$ odd. We shall first show that this problem is equivalent to counting the solutions to $\tilde q^2=-1\mod k$:

\begin{itemize}
\item Consider the case with $k$ even. Then any solution to $\tilde q^2=-1+\tilde pk$ must necessarily have $\tilde p$ odd (for otherwise we reach a contradiction upon reducing the equation modulo $4$), and so we can write $(\tilde q,\tilde p)=(q,2p-1)$, which yields $q^2=-1+(2p-1)k$, as required.

\item We now consider the case with $k$ odd. We claim that the solutions to $q^2=-1+2pk$ with $q\in[0,2k)$ can be put in a bijection with solutions to $\tilde q^2=-1+\tilde pk$ with $\tilde q\in[0,k)$. First, assume we are given the set $\{\tilde q\in[0,k)\}$; we construct the set $\{q\in[0,2k)\}$ as follows: if $\tilde q$ is odd, then $\tilde p$ must be even, and so $(q,2p)=(\tilde q,\tilde p)$; on the other hand, if $\tilde q$ is even, then $\tilde p$ must be odd, and so $(q,2p)=(\tilde q+k,\tilde p+2\tilde q+k)$. Conversely, if we are given the set $\{q\in[0,2k)\}$, we write $(\tilde q,\tilde p)=(q,p)$ if $q\in[0,k)$, and $(\tilde q,\tilde p)=(q-k,p-2q+k)$ if $\tilde q\in[k,2k)$.
\end{itemize}

We thus see that we may reduce our problem to counting solutions to $q^2=-1\mod k$, both for $k$ even and odd. It is a well-known result that the number of solutions is precisely $2^{\varpi(k)}$, see for example theorem 6.3 in~\cite{Mollin:2008:FNT:1628707} (together with remark 6.2 therein). The intuition behind this result (and which can be generalised to any polynomial congruence) is the following. Any solution to $q^2=-1\mod k$ can be reconstructed uniquely from the solutions to $q_i^2=-1\mod \pi_i$, where $\pi_i$ are the prime factors of $k$. Each congruence $q_i^2=-1\mod \pi_i$ is solvable (because $\pi_i$ is Pythagorean), and it has two solutions $\pm q_i$ (and only two, as per Lagrange's theorem, except for $\pi=2$, where only solution is $q_i=1$, inasmuch as $1=-1\mod 2$). As there are $\varpi(k)$ congruences, each having two solutions, the total number of solutions is $2^{\varpi(k)}$, as claimed.\hfill$\square$

\medskip

%\begin{remark}\normalfont
%As a consistency check, we note that the number of time-reversal symmetries... blah blah (this can only be discussed after counting unitary symmetries too, so I'm deleting it for now).
%By definition, the symmetries of the system (the solutions $\{q\}$) form a subset of $\Aut(\mathbb Z_k)=\mathbb Z_k^*$, the multiplicative group of integers modulo $k$. For $k>2$, the solutions form a \emph{proper} subset, because $2^{\varpi(k)}<\varphi(k)$ (indeed, both sides are multiplicative and $\varphi(\pi^p)=\pi^{p-1}(\pi-1)>2^{\varpi(\pi^p)}=2$ for prime $\pi$). This was to be expected, because the automorphisms of the fusion algebra do not, in general, conserve the spin.
%\end{remark}

For completeness, we mention that one can prove that $k\in\mathbb T$ is sufficient for time-reversal invariance using a path integral argument, which is quite similar to one in~\cite{witten:unpublished,Benini:2018reh} where it was used to show time-reversal invariance for $k\in \mathbb P\subset\mathbb T$. The argument is straightforward but it does not prove that the condition $k\in\mathbb T$ is also necessary.

\begin{proposition}\label{lm:suff_lagran}
It follows from a path integral argument that when $k\in\mathbb T$ the theory $U(1)_k$ is time-reversal invariant as a spin TQFT.
\end{proposition}

\emph{Proof}. Take two arbitrary integers $m,n$ with $m$ is odd and $n$ even, and such that
\begin{equation}
mn-q^2=1
\end{equation}
for some integer $q$ (which can easily seen to be odd). We shall prove that $U(1)_m$ and $U(1)_n\times\{\boldsymbol1,\psi\}$ are both time-reversal invariant.

Take the Lagrangian of $U(1)_m\times U(1)_{-n}$
\begin{equation}
4\pi\mathcal L=m\,a\, \mathrm da-n\,b\, \mathrm db
\end{equation}
whose Wilson lines are of the form
\begin{equation}
\exp\left[i \alpha\int a+i \beta\int b\right],\qquad (\alpha,\beta)\in\mathbb Z_{2m}\times\mathbb Z_n\,.
\end{equation}
Under the $\mathrm{GL}_2(\mathbb Z)$ transformation
\begin{equation}
\mathsf T\colon\begin{pmatrix}a\\b\end{pmatrix}\mapsto \begin{pmatrix}q&-n\\m&-q\end{pmatrix}\begin{pmatrix}a\\b\end{pmatrix}\,,
\end{equation}
%(We note in passing that $\mathsf T^2=-1$.)
the Lagrangian becomes
\begin{equation}
\mathsf T\colon4\pi\mathcal L\mapsto-m\,a\, \mathrm da+n\,b\, \mathrm db\equiv -4\pi\mathcal L
\end{equation}
and the lines map according to
\begin{equation}\label{eq:map_lines_mn}
\mathsf T\colon(\alpha,\beta)\mapsto(q\alpha+m\beta,-n\alpha-q\beta)\,.
\end{equation}
We therefore see that $U(1)_m\times U(1)_{-n}\ \longleftrightarrow\ U(1)_{-m}\times U(1)_n$, i.e., the product is time-reversal invariant. The explicit duality map is given by~\eqref{eq:map_lines_mn}.

We now prove that $U(1)_m$ is time-reversal invariant. To this end, we note that the theory above contains a sub-group of lines of the form $(\alpha,0)$, which is isomorphic to $U(1)_m$, with isomorphism $\alpha\leftrightarrow(\alpha,0)$. Time-reversal restricts to a well-defined action on $U(1)_m$, because
\begin{equation}
\mathsf T\colon(\alpha,0)\mapsto(q\alpha,-n\alpha)\sim(q\alpha,0)
\end{equation}
where we have used the fact that $n$ is even.

We next prove that $U(1)_n\times\{\boldsymbol1,\psi\}$ is time-reversal invariant. To this end, we note that the theory above contains a sub-group of lines of the form $(0,\beta)$ and $(m,\beta)$, which is isomorphic to $U(1)_n\times\{\boldsymbol1,\psi\}$, with isomorphism $\beta\otimes \boldsymbol1\leftrightarrow (0,\beta)$ and $\beta\otimes\psi\leftrightarrow (m,\beta)$. Time-reversal restricts to a well-defined action on $U(1)_n\times\{\boldsymbol1,\psi\}$, because
\begin{equation}
\begin{aligned}
\mathsf T\colon(\,0\,,\beta)&\mapsto(m\beta,-q\beta)\\
\mathsf T\colon(m,\beta)&\mapsto(qm+m\beta,-nm-q\beta)\sim (m(1+\beta),-q\beta)
\end{aligned}
\end{equation}
where we have used the fact that $n$ is even and $q$ is odd. This completes the proof. \hfill$\square$

%\begin{remark}\label{rm:map_lines}\normalfont
As a consistency check, we note that the action of time-reversal on the lines of $U(1)_m$ is $\mathsf T(\alpha)=q\alpha$, and that on $U(1)_n\times\{\boldsymbol1, \psi\}$ is $\mathsf T(\beta\otimes\psi^\gamma)= q\beta\otimes\psi^{\beta+\gamma}$, with $\gamma=0,1$. This is precisely the same map we found in proposition~\ref{th:time_reversal}.
%\end{remark}

One can couple the theory $U(1)_k$ to electromagnetism by turning on a background $U(1)_B$ connection. If $k\in\mathbb T$, then time-reversal remains a symmetry in the presence of this background field, but at the cost of introducing a Chern-Simons counterterm for the electromagnetic field, with fractional coefficient. This means that there is a mixed $\mathsf T-U(1)_B$ 't Hooft anomaly,\footnote{We thank N. Seiberg for this comment.} and so the system can only be defined on the boundary of a $3+1$ manifold. Using the Lagrangian argument above, and following the same reasoning as in~\cite{10.1093/ptep/ptw083,Benini:2018reh}, it is easy to prove that the anomaly is given by a $3+1$ dimensional topological term $\theta=2\pi/k$ for $U(1)_B$.

\begin{remark}\normalfont
It is common that in theories that are symmetric under both time-reversal and charge conjugation, the operators $\mathsf T$ and $\mathsf{CT}$ constitute two separate $\mathbb Z_2$ symmetries, both of which represent suitable time-reversal operations. These two symmetries are independent: they have different anomalies, they may be affected by magnetic symmetries (if any), and may be interchanged under duality (see e.g.~\cite{Cordova:2017kue}). In our case, these two symmetries in fact combine into a single $\mathbb Z_4$ algebra, $\mathsf T^3=\mathsf{CT}$, and so they do not correspond to independent symmetries.
\end{remark}

\begin{remark}\normalfont
It is interesting to note that we obtained $k\in\mathbb T$ as a \emph{necessary} condition just by insisting that the fundamental line has a partner with opposite spin. In turns, this condition was also seen to be \emph{sufficient}, so one may wonder if a similar phenomenon may occur in other topological systems. In other words, given an arbitrary TQFT, does the matching of the spin of a single line guarantee that the theory is time-reversal invariant? Generically speaking, the answer is \emph{no}, as there are many examples where a specific pair of lines match but others do not. A much stronger test is the matching of \emph{all} the lines, that is, the condition that $\{h\}=\{-h\}\mod 1$ (with equality as multisets, that is, taking into account multiplicities). For example, one may we check that the set of spins matches for the theory $SU(N)_N$, for $N=1,5,13,17,\dots$ (both as a bosonic and a spin TQFTs), all of which happen to be Pythagorean primes. As suggestive as this may seem, the pairs of lines that have opposite spin do not in general have the same quantum dimension, so these theories are not time-reversal invariant. ($SU(N)_N/\mathbb Z_N$ is, however, time-reversal invariant for all $N$ \cite{Aharony:2016jvv})
\end{remark}

Upon turning on a background metric, the duality $U(1)_k\ \longleftrightarrow\ U(1)_{-k}$ no longer holds as written, because the two theories have a different framing anomaly, and so they couple to the background gravitational field differently. This can be interpreted as a mixed anomaly between time-reversal and gravity. To maintain the duality one must adjust gravitational Chern-Simons counterterms on both sides so that their central charges agree. In particular, one may use $U(1)_{\pm1}$ to add/subtract one unit of central charge, without otherwise changing the topological content of the theory. With this in mind, the precise duality reads
\begin{equation}
U(1)_{+k}\times U(1)_{-1}\ \longleftrightarrow\ U(1)_{-k}\times U(1)_{+1}\,.
\end{equation}
These theories can be represented by the matrices $K=\diag(\pm k,\mp1)$. In the bosonic case, we already included a factor $\{\boldsymbol1,\psi\}$ to make the theory into a spin theory; here we see that this factor also fixes the central charge, provided we identify $\{\boldsymbol1,\psi\}\equiv U(1)_{-\sign(k)}$. In the spin case, this factor also fixes the central charge, but leaves the spectrum of lines unaffected.

It is clear that without the factor of $U(1)_{\pm1}$, time-reversal cannot possibly be a Lagrangian symmetry of the $U(1)_k$ theory, because the only $\mathrm{GL}_1(\mathbb Z)$ transformations are $a\to \pm a$, neither of which maps $k\to -k$. More generally, the signature of the $K$-matrix is invariant under congruence ($\sign(K)\equiv \sign(G^tKG)$ for any $G \in\mathrm{GL}_n(\mathbb Z)$, as per the Sylvester law of inertia) and so time-reversal can only be a Lagrangian symmetry if the signature vanishes (inasmuch as the chiral central charge is odd under time-reversal). Once we fix the central charge, time-reversal may (but need not) become a Lagrangian symmetry. It is interesting to note that, in the case at hand, this happens only for a subset of $\mathbb T$: only for a specific set of values of $k$ is the Lagrangian time-reversal invariant. One can show that this is so if and only if $k\in\mathbb P$:
\begin{proposition}\label{lm:lagrangian_pell}
The Lagrangian of the theory $U(1)_k\times U(1)_{-1}$ is time-reversal invariant if and only if $k$ satisfies the negative Pell equation.
\end{proposition}

\emph{Proof}. The fact that this condition is necessary can be obtained by looking at the bottom-right component of the equation $K=-G^tKG$, where $K=\diag(k,-1)$. That this is also sufficient was originally shown in~\cite{witten:unpublished,Benini:2018reh}, and follows from the explicit change of variables
\begin{equation}
\mathsf T\colon\begin{pmatrix}a\\b\end{pmatrix}\mapsto \begin{pmatrix}q&p\\-kp&-q\end{pmatrix}\begin{pmatrix}a\\b\end{pmatrix},\qquad kp^2-q^2=1\,.
\end{equation}
Proposition~\ref{lm:double_pell} in Appendix~\ref{app:further} generalizes the construction to $K=\diag(k,k')$. \hfill$\square$

This means that if $k\in\mathbb T$ but it is not in $\mathbb P$, then $U(1)_k$ will be time-reversal invariant, but the invariance will not be a symmetry of the Lagrangian, not even if we include the factor of $U(1)_{\pm1}$. It is a quantum symmetry of $U(1)_k\times U(1)_{- 1}$. However, it is possible that in a different abelian Chern-Simons realization of the same TQFT data 
that the symmetry becomes Lagrangian. 

\begin{remark}\normalfont
As a physical application of proposition~\ref{th:time_reversal}, note that given an integer $\ell$ such that both $k$ and $k+\ell^2$ are in $\mathbb T$, the theory%\footnote{We thank D.~Gaiotto for a discussion regarding this point.}
\begin{equation}
U(1)_{k+\ell^2/2}+\Psi
\end{equation}
with $\Psi$ a Dirac fermions of charge $\ell$ is infrared time-reversal invariant for $m\neq0$. Indeed, integrating the fermions out we get
\begin{equation}
U(1)_{k}\ \longleftrightarrow\ U(1)_{-k}
\end{equation}
for $m\to-\infty$, and
\begin{equation}
U(1)_{k+\ell^2}\ \longleftrightarrow\ U(1)_{-(k+\ell^2)}
\end{equation}
for $m\to+\infty$. This suggests that the CFT at the massless point $m=0$ may be time-reversal invariant as well. These gauge theories, in spite of not being time-reversal invariant in the ultraviolet, have an emergent time-reversal symmetry across the entire infrared phase diagram. The first few solutions of $(k,k+\ell^2)\in\mathbb T\times\mathbb T$ are
\begin{equation}
\begin{aligned}
\ell&=0\colon\qquad k=1, 2, 5, 10, 13, 17, 26, 25, 29, \dots\\
\ell&=1\colon\qquad k=1, 25, 73, 145, 169, 193, 289, \dots\\
\ell&=2\colon\qquad k=1, 13, 37, 61, 85, 97, 109, 181, \dots\\
\ell&=3\colon\qquad k=1, 17, 25, 41, 65, 73, 97, 113, \dots\\& \text{etc}. 
\end{aligned}
\end{equation}
(For $\ell=0$ one gets an infrared emergent time-reversal symmetry in Maxwell-Chern-Simons theories). A similar phenomenon occurs in non-abelian theories. For example, using the Chern-Simons dualities $U(1)_k\leftrightarrow SU(k)_{-1}\leftrightarrow SO(k)_{-2}$ we observe that the theories $SU(k)_0$ and $SO(k)_0$ with two fundamental Dirac fermions, and $k\in\mathbb T$, are time-reversal invariant in their massive phases (necessarily also in their massless phase, because the UV theory is time-reversal invariant).

It is an interesting number-theoretic problem whether there exists, for a given $\ell\in\mathbb Z$, an infinite number of pairs with $(k,k+\ell^2)\in\mathbb T^2$. This is similar in spirit to the so-called Polignac conjecture, which states that there exists an infinite number of pairs of primes of the form $(\pi,\pi+n)$ for any $n\in2\mathbb N$ (recall that primes $\pi>2$ are in $\mathbb T$ iff they are Pythagorean). Assuming this conjecture with $\ell^2=n$ (which requires $\ell$ to be even), and noting that $\pi$ and $\pi+\ell^2$ are either both Pythagorean or neither is, suggests that indeed there exists an infinite number of pairs $(k,k+\ell^2)\in\mathbb T^2$, at least for $\ell$ even.
\end{remark}

\subsubsection{Unitary Symmetries}
 
We now move on to the unitary symmetries of $U(1)_k$. The principle is identical to the anti-unitary case, the only difference being a sign flip. By definition, an automorphism $\mathsf U\in\Aut(\mathcal A)$ is a unitary symmetry of $(\mathcal A,\theta)$ if and only if
\begin{equation}\label{eq:u1_k_unitary}
\theta(\mathsf U(\alpha))=\theta(\alpha)\quad\Longleftrightarrow\quad h_{\mathsf U(\alpha)}= h_{\alpha}\mod 1\,.
\end{equation}

As in the anti-unitary case, any permutation is fixed once we choose how the generators transform. The corresponding permutation will be a symmetry if it satisfies~\eqref{eq:u1_k_unitary}. But, unlike the case of anti-unitary symmetries, here the equation $h_{\mathsf U(\alpha)}= h_{\alpha}\mod 1$ always admits solutions: at least, the trivial permutation and charge conjugation $\mathsf C$ exist. These are transformations that leave the action of the theory invariant. We thus solve a more refined problem: the interesting automorphisms will be those that are neither trivial nor $\mathsf C$. Another difference with the anti-unitary case is that, in general, we will find non-trivial symmetries also in the bosonic case.

We begin with the following observation:
\begin{proposition}
All the unitary symmetries of $U(1)_k$ (as a bosonic TQFT if $k$ is even) are transformations of the form
\begin{equation}
\mathsf U\colon \alpha\mapsto q \alpha	
\end{equation}
for some integer $q$ that satisfies
\begin{equation}
q^2=1+2pk\,.
\label{unitk}
\end{equation}
Similarly, the unitary symmetries of $U(1)_k\times\{\boldsymbol1,\psi\}$ for $k$ even are of the form
\begin{equation}
\mathsf U\colon (\alpha,\beta)\mapsto (q \alpha,p \alpha+\beta)
\end{equation}
for some integer $q$ that satisfies
\begin{equation}
q^2=1+pk\,.
\label{unitkspin}
\end{equation}
The solutions $q=\pm 1$ $($with $p=0)$ always exist and corresponds to the trivial permutation, and charge conjugation $\mathsf C$~\eqref{chargeconj}, respectively. All other solutions correspond to quantum symmetries.
\end{proposition}

\emph{Proof}. The case of $U(1)_k$ (as a bosonic TQFT if $k$ is even) is essentially identical to the anti-unitary case. Let us therefore consider $U(1)_k\times\{\boldsymbol1,\psi\}$ with $k$ even. Any fusion endomorphism that fixes the transparent fermion is of the form
\begin{equation}
\mathsf U\colon (\alpha,\beta)\mapsto (q \alpha,c \alpha+\beta)
\end{equation}
for a pair of integers $c,q$. If $c$ is even, $\mathsf U$ does not mix the lines of $U(1)_k$ with the transparent fermion, and so this is a symmetry that was also present in the bosonic case. If $c$ is odd, the permutation does mix the lines, and so it is only a symmetry of the fermionic theory. In any case, requiring that the spin of the fundamental line is equal to the spin of its image under $\mathsf U$, we get
\begin{equation}\label{eq:unitary_q2_k}
%\frac{\alpha^2}{2k}+\frac{\beta^2}{2}=\frac{q^2\alpha^2}{2k}+\frac{p \alpha+\beta}2+\tilde p
q^2 = 1-(c^2 + 2\tilde p)k
\end{equation}
for some integer $\tilde p$. Letting $-p:=c^2+2\tilde p$ we get the expression in the proposition (note that $p$ and $c$ have the same parity, and therefore we can replace the latter by the former in the transformation $\mathsf U$). It is straightforward to check that if the spin of the fundamental line is invariant under $\mathsf U$, so is the spin of the rest of lines. Finally, it is easy to show that \emph{any} solution of~\eqref{eq:unitary_q2_k} corresponds to a permutation (i.e.~$q$ automatically has the appropriate coprimality with $k$ to define an automorphism).\hfill$\square$

\medskip

As in the anti-unitary case, all the unitary permutations have the same order: 
\begin{proposition}
All the unitary symmetries of $U(1)_k$ (either as a bosonic or as an spin TQFT) are of order-two.
\end{proposition}

\emph{Proof}. For $U(1)_k$ we have
\begin{equation}
\mathsf U^2\colon \alpha\mapsto q^2\alpha=\alpha+2pk\alpha%=\begin{cases}
%\alpha~ \mod k, ~~~k~\hbox{even}\\
%\alpha~ \mod 2k, ~~~k ~\hbox{odd}
%\end{cases}\,,
\end{equation}
which indeed equals $\alpha$. In the case of $U(1)_k\times\{\boldsymbol 1,\psi\}$, the argument is identical:
\begin{equation}
\mathsf U^2\colon (\alpha,\beta)\mapsto (q^2\alpha,p\alpha(q+1)+\beta)=(\alpha+pk\alpha,p\alpha(q+1)+\beta)%=(\alpha,\beta)~(\hbox{mod}~ k,\hbox{mod}~ 2)
\end{equation}
which, using the fact that $q$ is odd, yields $(\alpha,\beta)$, as claimed. \hfill$\square$

%note: if $p$ is even, $p(q+1)\equiv0\mod 2$ is obvious. If odd, then $q$ must be odd as well, for if we let $p=2a+1$ and $q=2b$, the equation $q^2=1+pk$ becomes $4b^2=1+k+2ak$, which is impossible (as $k$ is even).

\medskip

%As in lemma~\ref{lm:prime_factors}, we may always restrict any solution to the range $q\in[0,2k)$, inasmuch as if $(q_0,p_0)$ is a solution, then so is
%\begin{equation}
%(q,p)=(q_0+2kn,p_0-2q_0n-2kn^2)
%\end{equation}
%for any $n\in\mathbb Z$. For $k$ even, we may further reduce the range down to $q\in[0,k)$, because if $(q_0,p_0)$ is a solution, then so is
%\begin{equation}
%(q,p)=(q_0+kn,p_0-q_0n-\tfrac12kn^2)
%\end{equation}

Take the theory $U(1)_k$, without the factor of $\{\boldsymbol1,\psi\}$ for $k$ even. A slight modification of the argument in proposition~\ref{pr:2omega} proves that the number of solutions in the range $q\in[0,2k)$ for $k$ odd, and in the range $q\in[0,k)$ for $k$ even, is $2^{\varpi(k)}$, as in the anti-unitary case. Therefore, in order to have solutions other than $\mathsf U\in\{1,\mathsf C\}$, the level $k$ must not be a prime power or twice a prime power. Such non-trivial solutions will not be a symmetry of the classical Lagrangian, because $p\neq 0$. They correspond to quantum symmetries.

For $k$ even, one may also study the unitary symmetries of the theory as a spin TQFT, that is, of $U(1)_k\times\{\boldsymbol 1,\psi\}$. The symmetries of the bosonic theory are inherited in the fermionic theory, but new symmetries may appear -- those under which the transparent fermion mixes non-trivially. The automorphisms are given by the integers $q$ that satisfy $q^2=1+pk$, and whether the transparent fermion mixes is controlled by the parity of $p$. It is easy to show that the number of solutions is $2^{\varpi(k)}$ for $k=2 \mod 4$, and $2^{\varpi(k/2)+1}$ for $k=0 \mod 4$. Therefore, there is an enhancement of symmetry when going from the bosonic theory to the spin theory if and only if $k$ is a multiple of $8$: only in that case may the fermion mix. The additional transformation that appears when the theory is uplifted from bosonic to spin is generated by $q=k/2-1$ (with $p=k/4-1$). We summarise these claims as follows:
\begin{proposition}
All the unitary symmetries of $U(1)_k$ (both as a spin theory and as a bosonic theory in the case of $k$ even) are $\mathbb Z_2$-valued. There are $2^{\varpi(k)}$ permutations if $k$ is not a multiple of $8$. If $k=0\mod 8$, then there are $2^{\varpi(k)}$ permutations in the bosonic theory, and twice as many in the spin theory.
\end{proposition}

Needless to say, one may compose any non-trivial unitary symmetry with a given $\mathsf T$ to yield a different notion of time-reversal. Similarly, composing any two time-reversal operations results in a unitary symmetry, and composing two unitary symmetries leads to another unitary symmetry. 
%Indeed, if we have
%\begin{equation}
%2pk-q^2=1,\qquad 2\tilde pk+\tilde q^2=1
%\end{equation}
%then $q\tilde q$ satisfies
%\begin{equation}
%2(p+\tilde p-2p\tilde pk)k-(q\tilde q)^2=1
%\end{equation}
%and so it also represents a valid time-reversal permutation. Conversely, composing two anti-unitary operations we get a unitary one, $\mathsf T\tilde{\mathsf T}=\mathsf U$. Indeed, if we have
%\begin{equation}
%2pk-q^2=1,\qquad 2\tilde pk-\tilde q^2=1
%\end{equation}
%then $q\tilde q$ satisfies
%\begin{equation}
%2(p+\tilde p-2p\tilde pk)k+(q\tilde q)^2=1
%\end{equation}
%as expected. Finally, composing two unitary symmetries yields another unitary symmetry, for if we have
%\begin{equation}
%2pk+q^2=1,\qquad 2\tilde pk+\tilde q^2=1
%\end{equation}
%then $q\tilde q$ satisfies
%\begin{equation}
%2k (p+\tilde p-2p\tilde pk)+(q\tilde q)^2=1
%\end{equation}
In fact, a stronger result is true. Let $\{\mathsf T_i\}$ be all time-reversal symmetries, and $\{\mathsf U_i\}$ be unitary ones. Let $\mathsf U_0:=1$, pick some element of $\{\mathsf T_i\}$, and denote it by $\mathsf T_0$. Then any $\mathsf T_i$ can be obtained by acting with some $\mathsf U_i$ on $\mathsf T_0$. Indeed, it is easy to see that the sets
\begin{equation}
\{\mathsf T_i\}\qquad\text{and}\qquad \{\mathsf T_0\mathsf U_i\}
\end{equation}
contain the same number of elements (because $\mathsf T_0$ is invertible, so $\mathsf T_0\mathsf U_i\neq\mathsf T_0\mathsf U_j$ for $i\neq j$), and so they must be identical. Thus, perhaps after relabelling its elements, we have
\begin{equation}
\mathsf T_i\equiv \mathsf U_i\mathsf T_0
\end{equation}
and so one time-reversal permutation suffices to generate them all.

%Therefore, the algebra of symmetries is
%\begin{equation}
%\begin{aligned}
%\mathsf U_i\mathsf U_j&=f_{ijk}\mathsf U_k\\
%\mathsf T_i\mathsf U_j&=f_{ijk}\mathsf T_k\\
%\mathsf T_i\mathsf T_j&=-f_{ijk}\mathsf U_k
%\end{aligned}
%\end{equation}
%for some symmetric structure constants $f_{ijk}\in\{1,0\}$ (with $f_{iik}=\delta_{k0}$ and $f_{0ij}=\delta_{ij}$, etc.).
\medskip

Recalling definition~\ref{def:aut}, all these considerations can be put together to obtain the following:
\begin{proposition}\label{pr:full_group_u1}
The group of symmetries of $U(1)_k$ as a spin TQFT is
\begin{equation}
\begin{aligned}
\Aut(U(1)^{\mathrm{spin}}_k)&=\mathbb Z_4\times(\mathbb Z_2)^{\varpi(k)-1}\\%=\langle\mathsf T,\mathsf U_1,\dots, \mathsf U_{\varpi(k)-1}\rangle\\
\Aut_U(U(1)^{\mathrm{spin}}_k)&=(\mathbb Z_2)^{\varpi(k)}%=\langle\mathsf C,\mathsf U_1,\dots, \mathsf U_{\varpi(k)-1}\rangle
\end{aligned}
\end{equation}
if $k\in\mathbb T$, and
\begin{equation}
\Aut(U(1)^{\mathrm{spin}}_k)=\Aut_U(U(1)^{\mathrm{spin}}_k)=\begin{cases}(\mathbb Z_2)^{\varpi(k)+1} & k=0\mod 8\\ (\mathbb Z_2)^{\varpi(k)} & \mathrm{otherwise}\end{cases}
\end{equation}
otherwise. On the other hand, as a bosonic theory (with $k$ even), the group reads
\begin{equation}
\Aut(U(1)^{\mathrm{bosonic}}_k)=\Aut_U(U(1)^{\mathrm{bosonic}}_k)=(\mathbb Z_2)^{\varpi(k)}\,.%\begin{cases}(\mathbb Z_2)^{\varpi(k)} & k\equiv 0\mod 8\\ (\mathbb Z_2)^{\varpi(k)} & \mathrm{otherwise}\end{cases}
\end{equation}
\end{proposition}
 
\subsection{Minimal abelian TQFT}
\label{subsec:mimimal}

An important abelian theory that appears in the study of the one-form symmetries of three-dimensional TQFTs is the so-called ``minimal abelian TQFT''~\cite{Moore:1988qv,Barkeshli:2014cna,Bonderson:2007ci,Hsin:2018vcg}. This theory is denoted by $\mathcal A^{N,t}$ (also by $\mathbb Z_N^{(t)}$), with $N,t$ two integers, which must be coprime if we require the theory to be modular. The number of lines is $N$, which can be labelled as $s=1,2,\dots,N$. Fusion corresponds to addition modulo $N$, $s\times s'=(s+s'\mod N)$, i.e.~the fusion algebra is $\mathbb Z_N$. The spin of the line $s$ is $h_s=t\frac{s^2}{2N}$. For example, if $k$ is even, then $U(1)_k=\mathcal A^{k,1}$; if $k$ is odd, then $U(1)_k=\mathcal A^{2k,2}$ (which, indeed, is not modular, because the braiding matrix has a non-trivial kernel). All these theories admit an abelian Chern-Simons representation (e.g.~for $t=N-1$ the $K$-matrix is the Cartan matrix of $SU(N)$).

The analysis of the symmetries of $\mathcal A^{N,t}$ is essentially identical to that of $U(1)_k$ because the fusion algebra is also cyclic. For example, following the same reasoning as in the $1\times 1$ case, this theory is seen to be time-reversal invariant if and only if
\begin{equation}\label{eq:ANt}
2Np=t(1+q^2)
\end{equation}
is solvable for some integers $p,q$. It is easy to prove that this equation is solvable if and only if
\begin{equation}\label{eq:con_TN}
t\in\mu(N)\mathbb Z\,.
\end{equation}
Indeed, by reducing~\eqref{eq:ANt} modulo $\mu(N)$ we get $t(1+q^2)=0\mod\mu(N)$; but $(1+q^2)$ is never divisible by a prime of the form $4n+3$, and so $t$ itself mush vanish modulo $\mu(N)$, showing that $t\propto\mu(N)$ is necessary. Conversely, noting that $N/\mu(N)$ is always in $\mathbb T^{\mathrm{odd}}$, we know that there exists a pair of integers $\tilde p,q$ such that $2\tilde p N=\mu(N)(1+q^2)$; multiplying this equation by $t/\mu(N)$ and letting $p=\tilde pt/\mu(N)$ we find that $t\propto\mu(N)$ is also sufficient.

Alternatively, one may rewrite~\eqref{eq:con_TN} as a condition on $N$ instead of $t$, as follows:
\begin{equation}\label{eq:con_Tt}
N\in \bigcup_{d|t}d\,\mathbb T^{\mathrm{odd}}\,.
\end{equation}

Indeed, if $N\in d\,\mathbb T^{\mathrm{odd}}$ for some $d|t$, then there exists some $\tilde p,q$ such that $2(N/d)\tilde p=1+q^2$; multiplying this equation by $t/d$ and letting $p=\tilde pt/d$ shows that~\eqref{eq:ANt} is solvable. Conversely, if $N\notin d\,\mathbb T^{\mathrm{odd}}$ for any $d|t$ then, in particular, $N\notin t\,\mathbb T^{\mathrm{odd}}$ (and, if $t\in2\mathbb Z$, then $N\notin (t/2)\,\mathbb T^{\mathrm{odd}}$ either), and so equation~\eqref{eq:ANt} is not solvable (note that if $t$ is odd then $N$ must be odd as well).

%If $t$ is even, then $N\notin tT^o$ and $N\notin (t/2)T^o$, which means that $N\notin (t/2)T$, as required. If $t$ is odd, then $N$ must be odd too. Therefore, $N\notin tT^o$; furthermore, being odd, $2N\notin tT^o$ trivially. Together, $2N\notin tT$, as required.

If we further assume that $(N,t)=1$, the expression~\eqref{eq:con_Tt} can be simplified into
\begin{equation}
N\in \mathbb T^{\mathrm{odd}}\setminus \left(\bigcup_{\pi|p}\pi\,\mathbb Z\right)
\end{equation}
where $\pi$ are the Pythagorean prime factors of $p$.

As $\mathcal A^{N,t}$ has a single generator, its group of symmetries is abelian, and can be studied along the same lines as in the $U(1)_k$ case.

%\textbf{Question}: what is the number of solutions as a function of $N,t$? What are their order?

\section{$U(1)^n$ Chern-Simons theory}\label{sec:K-matrix}
 \label{sec:K}
We now move on to Chern-Simons theories that contain an arbitrary number of factors of $U(1)$. As a Lagrangian theory, the system is described by
\begin{equation}
\mathcal L=\frac{1}{4\pi}a^tK\mathrm da
\end{equation}
for a $U(1)^n$ gauge field $a^t=(a_1,a_2,\dots,a_n)$. The Lagrangian is metric independent and, although not manifestly so, gauge invariant provided the coefficient matrix $K\in\mathbb Z^{n\times n}$ is symmetric and integral-valued. Generically speaking, the theory depends on the orientation of spacetime and, if at least diagonal component of $K$ is odd, on the spin structure. The theory has central charge $c=\sign(K)$ (the signature of $K$), which controls the coupling to the Chern-Simons form for the background metric, via the framing anomaly. To keep matters simple, we shall often turn off this metric, and any other background field one may ultimately want to couple $a$ to.

The observables of the theory are the Wilson lines, modulo local bosonic operators. These lines are of the form
\begin{equation}
W_{\vec\alpha}(\gamma):=\exp\left[i\vec\alpha^t\int_\gamma a\right]\,,
\end{equation}
where $\vec\alpha\in\mathbb Z^n$ is the representation $U(1)^n\ni\theta\mapsto\mathrm e^{i\vec\alpha\cdot\theta}$. We shall call $\vec\alpha$ the \emph{charge} of $W_{\vec\alpha}$, and we will often denote the line itself by $\vec\alpha$.

These lines can be thought of as the worldlines of anyons, i.e., particles with fractional statistics. In particular, they have spin and may braid non-trivially. If a line $\vec\alpha$ braids around a line $\vec\beta$, their product picks up a phase $B(\vec\alpha,\vec\beta)\in U(1)$, where
\begin{equation}
B(\vec\alpha,\vec\beta):=\exp\left[2\pi i\,\vec\alpha^tK^{-1}\vec\beta\right]\,.
\end{equation}
Similarly, the topological spin of the line corresponds to half self-braiding,
\begin{equation}
\theta(\vec\alpha):=\exp\left[2\pi i\,h_{\vec\alpha}\right],\qquad h_{\vec\alpha}:=\frac12\vec\alpha^tK^{-1}\vec\alpha\,.
\end{equation}
The function $\theta$ is said to be a \emph{quadratic refinement} of the bilinear form $B$, because one has
\begin{equation}\label{eq:q_ref}
B(\vec\alpha,\vec\beta)\equiv\frac{\theta(\vec\alpha+\vec\beta)}{\theta(\vec\alpha)\theta(\vec\beta)}\,.
\end{equation}
This implies that the spin of the lines determines their braiding unambiguously; one need not keep track of the latter.

An operator is said to be \emph{local} if it braids trivially with any other line. In particular, any line with $\vec\alpha$ proportional to a column of $K$ satisfies $B(\vec\alpha,\vec\beta)\equiv1$ for any $\vec\beta$, and so it will be local. If, furthermore, the corresponding column has even diagonal element, then $h_{\vec\alpha}=0\mod 1$, and so the local line will be bosonic. As before, lines differing by such a local operator are identified, and so the degrees of freedom of the theory are in fact finite. More explicitly, we have the following:
\begin{itemize}
\item If all the diagonal components of $K$ are even, then all the local operators are bosonic, and we need not specify a spin structure to define the theory. It is a bosonic TQFT. Any two lines that are congruent modulo some linear combination (with integer coefficients) of the columns of $K$ are identified, which means that the lines live in the lattice $\mathbb Z^n/K\mathbb Z^n$. There are $|\det K|$ independent lines, which can be taken to be all the lattice points in the $n$-dimensional parallelepiped spanned by the columns of $K$.

\item If at least one diagonal component of $K$ is odd, the theory contains a local fermionic operator, which requires a choice of spin structure. The theory is a spin TQFT. Any two lines that are congruent modulo some linear combination (with integer coefficients) of the columns of $K$ are identified, except if they differ by a local fermion. This means that the lines live in the lattice $(\mathbb Z^n/K\mathbb Z^n)\times\mathbb Z_2$. There are $2|\det K|$ independent lines, which can be taken to be all the lattice points in the $n$-dimensional parallelepiped spanned by the columns of $K$, together with a $\mathbb Z_2$ label that specifies if the line carries a local fermion or not. Alternatively, a basis of lines can be taken to be all the lattice points in the $n$-dimensional parallelepiped spanned by the columns of $\tilde K$, where $\tilde K$ is the matrix given by doubling any one column of $K$ with odd diagonal component.

\end{itemize}

The spectrum of lines is given by the set $\mathcal A:=\mathbb Z^n/\!\sim$, where
\begin{equation}\label{eq:equiv_A}
\vec\alpha\sim\vec\beta\quad\Longleftrightarrow\quad \vec\alpha=\vec\beta+K\vec \gamma\,,
\end{equation}
where $\vec\gamma$ is any tuple of integers with
\begin{equation}
\sum_{K_{ii}\text{ odd}}\gamma_i=\text{even}\,.
\end{equation}

Reducing $\mathbb Z^n$ modulo $K$, instead of modulo $\sim$, would be tantamount to identifying the local fermion, if any, with the vacuum. In other words, we would forget about the information carried by such a line. This would not be correct: we need the $\mathbb Z_2$ label to signal the presence of $\psi$. This extra piece of information resolves the ambiguity in lifting the symmetric form $B$ into the quadratic form $\theta$. We shall nevertheless often refer to the equivalence $\sim$ as ``reduction modulo $K$'', in order to keep the notation as simple as possible.

Due to the abelian nature of the gauge fields, any pair of unbraided lines $\vec\alpha,\vec\beta$ can be brought together to form a line of charge $\vec\alpha+\vec\beta$. In other words, the fusion rules of the theory are
\begin{equation}
\vec\alpha\times\vec\beta:= (\vec\alpha+\vec\beta\mod K)\,.
\end{equation}

The theory described by a given matrix $K$ may have several symmetries. The main focus of this paper is to study the zero-form symmetries, but for completeness we mention that the one-form symmetry group can be obtained by bringing $K$ into its Smith normal form $K\to\diag(k_1,k_2,\dots,k_n)$, where $k_i$ is the greatest common divisor of all $i\times i$ minors of $K$. Given this canonical decomposition, the one-form symmetry group is
\begin{equation}
\bigoplus_{i=1}^n \mathbb Z_{k_i}\,.
\end{equation}

We now move on to the zero-form symmetries of the system. These are, by definition, the permutations of the lines that respect their topological properties. A unitary zero-form symmetry of the corresponding system is an automorphism $\mathsf U\colon\mathcal A\to\mathcal A$ that satisfies
\begin{equation}\label{eq:mtc_unitary}
\begin{aligned}
\mathsf U(a\times b)&=\mathsf U(a)\times\mathsf U(b)\\
\theta(\mathsf U(a))&=\theta(a)\\
B(\mathsf U(a),\mathsf U(b))&=B(a,b)
\end{aligned}
\end{equation}
for all $a,b\in\mathcal A$. Similarly, an anti-unitary zero-form symmetry is an automorphism $\mathsf T\colon\mathcal A\to\mathcal A$ that satisfies
\begin{equation}\label{eq:mtc_anti}
\begin{aligned}
\mathsf T(a\times b)&=\mathsf T(a)\times\mathsf T(b)\\
\theta(\mathsf T(a))&=\theta(a)^*\\
B(\mathsf T(a),\mathsf T(b))&=B(a,b)^*\,.
\end{aligned}
\end{equation}

Thanks to~\eqref{eq:q_ref}, the braiding is determined by the spin, and so the third condition is automatically guaranteed to hold if the first two do; we nevertheless find it convenient to keep track of the braiding matrix explicitly.

%More abstractly, a given abelian TQFT is uniquely specified by the data $(\mathcal A,\times,\theta,c)$. The pair $(\mathcal A,\times)$ (the \emph{fusion algebra}) is a finite abelian group, and $\theta\colon\mathcal A\to U(1)$ (the \emph{topological spin}) is a quadratic form on it; and $c$ (the \emph{central charge}) is an integer modulo $8$ (which, if $\theta$ is non-degenerate, is uniquely determined by $(\mathcal A,\theta)$, and so it need not be specified separately). This data determines a unique symmetric form (the \emph{braiding matrix}) $B\colon\mathcal A\times\mathcal A\to U(1)$, and a unique equivalence class of $F$- and $R$-symbols (modulo cocycles?). Two pairs of theories $(\mathcal A,\times,\theta,c)$ and $(\mathcal A',\times',\theta',c')$ are said to be equivalent if $\varphi\colon (\mathcal A,\times)\to (\mathcal A',\times')$ is an isomorphism and $\theta'=\theta\circ\varphi$ and $c\equiv c'\mod8$.

%A unitary symmetry of a theory is a (non-trivial) auto-equivalence, and an anti-unitary symmetry is an equivalence to its conjugate (where the conjugate of a theory is, by definition, the theory with data $(\mathcal A,\times,\theta^*,-c)$). If the theory admits such a symmetry, the existence of a coclycle satisfying $[F]=[F\circ\varphi]$ and $[R]=[R\circ\varphi]$ in the unitary case -- and $[F]=[F^*\circ\varphi]$ and $[R]=[R^*\circ\varphi]$ in the anti-unitary case -- is guaranteed to exist, because both sides are uniquely specified by the fusion algebra and topological spin.

We have denoted the anti-unitary symmetries by $\mathsf T$ because we will think of them as a time-reversal operation (or a reflection in the Euclidean setting). These symmetries do not always exist: only for some special matrices $K$ is the system independent of the orientation of spacetime. In particular, as the Lagrangian is odd under the reversal of orientation, we require $K$ and $-K$ to describe equivalent theories: the theories with matrices $K$ and $-K$ must be dual. 

A sufficient condition for the theories described by two matrices $K_1,K_2$ to be equivalent is that they are congruent, i.e., $\mathrm{GL}_n(\mathbb Z)$-equivalent: that there exists a unimodular matrix $G$ such that $K_1\equiv G^tK_2G$, as follows from the redefinition $a_2:=Ga_1$. The matrix $G$ is required to be unimodular because the change of variables has to be invertible and respect the normalisation of the gauge fields. We shall refer to these equivalences of theories as \emph{Lagrangian} (or classical) \emph{symmetries}, because they are manifest symmetries of the Lagrangian. As we shall show, one may have matrices $K_1,K_2$ that are not $\mathrm{GL}_n(\mathbb Z)$-equivalent, and yet the theories described by them are nevertheless equivalent. This latter notion of equivalence we refer to as a \emph{quantum symmetry}, or as a \emph{duality}. 
%These can always be understood as classical equivalences of $K\oplus \Sigma$, where $\Sigma$ is a unimodular matrix with even diagonal elements (and hence having no non-trivial lines); this relaxed notion of congruence is known as \emph{stable equivalence}.

Dualities of TQFTs are often valid only when the theory is regarded as a spin TQFT. In order to turn a bosonic theory into a spin TQFT, it suffices to tensor the theory by the trivial spin TQFT $U(1)_{\pm1}=\{\boldsymbol 1,\psi\}$, where $\boldsymbol1$ is a local boson and $\psi$ a local fermion. Tensoring a theory that is already spin by this trivial factor leaves the TQFT unaffected, inasmuch as we identify local fermions anyway (because they differ by a local boson: $\psi_1=(\psi_1\psi_2)\psi_2$).

If we turn on some background field that couples to a given TQFT, then one may need to adjust appropriate counterterms for it on both sides of the duality. The canonical example is the coupling to background gravity, which is controlled by the central charge of the theory (through the framing anomaly). In particular, the central charge -- being the signature of the $K$-matrix -- is odd under time-reversal, which means that a theory can only be time-reversal invariant in the presence of gravity if the central charge vanishes. In this sense, a theory being invariant in flat spacetime may require a gravitational counterterm to remain invariant when the metric is nontrivial. Noting that $U(1)_{\pm1}$ is essentially trivial (it is an SPT) but has central charge $\pm1$, one may add as many factors of this theory as necessary so that the theory under consideration has vanishing central charge, as required to maintain the time-reversal symmetry when turning on a background metric. If the theory is already spin, tensoring by $U(1)_{\pm1}=\{\boldsymbol1,\psi\}$ has no effect other than changing the central charge; but for a bosonic system, this factor turns the theory into a spin TQFT.

\subsection{Symmetries of $U(1)^n$}

The analysis of the symmetries of a system described by a matrix $K$ is essentially identical to that of $U(1)_k$: the symmetries are those automorphisms of the fusion algebra that respect the spin of the lines. The most general endomorphism of $\mathcal A\cong \mathbb Z^n/\!\sim$ is
\begin{equation}
g\colon\vec\alpha\mapsto Q\vec \alpha
\end{equation}
for some matrix $Q$, its $i$-th column being $g(\vec e_i)$, with $\vec e_i$ the $i$-th unit vector. This map is an automorphism if the action of $Q$ is invertible modulo $\sim$, i.e., if it is a permutation of $\mathcal A$. Finally, this permutation shall be a symmetry if it conserves the spin of all the lines, up to complex conjugation in the anti-unitary case. We discuss this in some more detail below.

\subsubsection{Anti-unitary symmetries}

A natural generalisation of proposition~\ref{th:time_reversal} reads
\begin{proposition}\label{lm:necessary_K}
A necessary condition for the Chern-Simons theory with matrix $K$ to admit an anti-unitary symmetry is that there exists a pair of matrices $(Q,P)\in\mathbb Z^{n\times n}\times \mathbb Z^{n\times n}$ where $P$ has even diagonal elements, and such that
\begin{equation}
PK-Q^tK^{-1}QK=1_n\,.
\end{equation}
\end{proposition}

\emph{Proof}. We shall look for the most general permutation that satisfies the conditions~\eqref{eq:mtc_anti}. 

As in the case of a single $U(1)$ factor, any putative time-reversal operation is fixed once we know how the generators transform. The most general fusion endomorphism reads
\begin{equation}
\mathsf T(\vec\alpha)=Q\vec\alpha
\end{equation}
for some matrix $Q$, the $i$-th column of which represents the action of $\mathsf T$ on the unit vector in the $i$-th direction $\vec e_i$.

Imposing that the spin of $\vec e_i$ is the opposite of that of $\mathsf T(\vec e_i)$, we get
\begin{equation}
\frac12 \vec e_i{}^tK^{-1}\vec e_i=-\frac12 \vec e_i{}^tQ^tK^{-1}Q\vec e_i+P_{ii}
\end{equation}
for some integer $P_{ii}$. Similarly, imposing that $\mathsf T$ commutes with braiding, $B(\vec e_i,\vec e_j)=B(\mathsf T(\vec e_i),\mathsf T(\vec e_j))^*$, we get
\begin{equation}
\vec e_i{}^tK^{-1}\vec e_j=-\vec e_i{}^tQ^tK^{-1}Q\vec e_j+P_{ij}
\end{equation}
for some integer $P_{ij}$. These two equations, in matrix form, take the form quoted in the proposition, as claimed. Note that if this equation is satisfied, then the spin of all the lines behaves as expected, and not only that of the generators:
\begin{equation}
\begin{aligned}
h_{\mathsf T(\vec\alpha)}&=\frac12\vec \alpha^t\left(Q^tK^{-1}Q\right)\vec \alpha\\
&=\frac12\vec \alpha^t\left(-K^{-1}+P\right)\vec \alpha,
\end{aligned}
\end{equation}
which indeed equals $-h_{\vec\alpha}$ modulo $1$. \hfill$\square$

We stress that, unlike in the case of a single $U(1)$ factor, the argument in proposition~\ref{lm:necessary_K} does not prove that any map $\vec\alpha\mapsto Q\vec\alpha$ with $PK-Q^tK^{-1}QK=1_n$ represents a time-reversal operation, even though the conditions~\eqref{eq:mtc_anti} are satisfied. One must also require $Q$ to be a permutation, that is, invertible modulo $ K$ over the integers. This is a non-trivial condition that is not satisfied for every solution of $PK-Q^tK^{-1}QK=1_n$. (In the $1\times 1$ case, the equation $pk-q^2=1$ implies that $\gcd(k,q)=1$, and so any solution is invertible; this is no longer necessarily true in the $n\times n$ case: some solutions may fail to be invertible).

As in proposition~\ref{lm:suff_lagran}, one can also examine the time-reversal invariance of $U(1)^n$ through a Lagrangian argument:
\begin{proposition}\label{lm:suff_K}
A sufficient condition for the Chern-Simons theory described by the matrix $K$ to admit an anti-unitary symmetry is that there exists a pair of matrices $(Q,P)\in\mathbb Z^{n\times n}\times \mathbb Z^{n\times n}$ where $P$ has even diagonal elements, and such that
\begin{equation}\label{eq:suff_K}
PK-Q^tK^{-1}QK=1_n
\end{equation}
subject to the conditions
\begin{equation}\label{eq:extra_cond}
%Q^tK^{-1}QK=Q^tKQK^{-1},\qquad P^{-1}QPQ^t=KQ^tK^{-1}Q,\qquad [Q,PK]=0
%K^{-1}QK=KQK^{-1},\qquad P^{-1}QPQ^t=KQ^tK^{-1}Q,\qquad [Q,PK]=0
%[K^{-1}QK,Q^t]=0,\qquad Q^tPQ=PKQ^tK^{-1}Q,\qquad [KP,Q]=0
[K^{-1}QK,Q^t]=0,
%\qquad Q^tP=PKQ^tK^{-1},
\qquad [KP,Q]=0\,.
\end{equation}
\end{proposition}

\noindent (Note that if $Q$ is normal and commutes with $K$, then these equations are automatically satisfied).

\emph{Proof}. By solving for $P$ in~\eqref{eq:suff_K}, and taking the transpose, it becomes clear that $P$ is symmetric, and so it defines a (bosonic) abelian Chern-Simons theory. Take the Lagrangian with matrix $K\oplus -P$
\begin{equation}
4\pi\mathcal L=a^t K \mathrm da-b^tP\mathrm db
\end{equation}
and perform the $\mathrm{GL}_{2n}(\mathbb Z)$ transformation
\begin{equation}
\mathsf T\colon\begin{pmatrix}a\\b\end{pmatrix}\mapsto \begin{pmatrix}Q^t&-P\\K&-Q\end{pmatrix}\begin{pmatrix}a\\b\end{pmatrix}
\end{equation}
under which
\begin{equation}
\begin{aligned}
\begin{pmatrix}
K&0\\0&-P
\end{pmatrix}&\mapsto %\begin{pmatrix}
%QKQ^t-KPK&-QKP+KPQ\\
%-PKQ^t+Q^tPK&PKP-Q^tPQ
%\end{pmatrix}\\
%&=
\begin{pmatrix}
-KPK+QKQ^t&[KP,Q]\\
[Q^t,PK]&PKP-Q^tPQ
\end{pmatrix}\\
&\equiv \begin{pmatrix}
-K&0\\0&P
\end{pmatrix}\,.
\end{aligned}
\end{equation}
The off-diagonal entries vanish by virtue of $[KP,Q]=0$ and $P$ being symmetric, and the equality for the diagonal entries follows from the assumptions in the proposition.

This proves that the theory is time-reversal invariant. The mapping of lines reads
\begin{equation}
(\vec\alpha,\vec\beta)\mapsto (Q\vec\alpha+K\vec\beta,-P\vec\alpha-Q^t\vec\beta)\,.
\end{equation}
Finally, and thanks to the evenness of $P$, the action of $\mathsf T$ descends to a well-defined operation on the lines of $U(1)_K$:
\begin{equation}
(\vec\alpha,0)\mapsto (Q\vec\alpha,-P\vec\alpha)\sim(Q\vec\alpha,0)
\end{equation}
as required. \hfill$\square$

\begin{remark}\normalfont
It is easy to argue that the conditions in proposition~\ref{lm:suff_K} are $\mathrm{GL}_n(\mathbb Z)$-invariant. Indeed, if we redefine our gauge fields according to $a:= G a'$ for some $G\in\mathrm{GL}_n(\mathbb Z)$, then the lines transform as $\vec\alpha=(G^{-1})^t\vec\alpha'$, and
\begin{equation}
\begin{aligned}
K&=(G^{-1})^tK' G^{-1}\\
Q&=(G^{-1})^tQ'G^t\\
P&=GP'G^t
\end{aligned}
\end{equation}
which leaves the equations~\eqref{eq:suff_K},~\eqref{eq:extra_cond} invariant. This was to be expected, inasmuch as a Chern-Simons theory depends on $K$ modulo congruences. (Two $K$-matrices in the same congruence class have the same determinant; however, the converse is not true: there can multiple congruence classes with a given determinant. The number of congruence classes depends nontrivially on the value of the determinant.)
\end{remark}

Deciding whether the equation $PK-Q^tK^{-1}QK=1_n$ is solvable for a given $K$ is a rather non-trivial problem, unlike in the case of $U(1)_k$ (where it suffices to scan $q\in[0,k)$ for solutions; moreover, and thanks to proposition~\ref{th:prime_factors}, deciding whether $k\in\mathbb T$ requires at most $\omega(k)\le \frac{2\log k}{\log\log k}$ operations if given the prime divisors of $k$). We shall make no attempt at finding an efficient characterisation of the set of $K$-matrices that solve this equation. We will content ourselves with focusing specifically to the case where $K$ is a $2\times 2$ matrix. In particular, we will consider the following two families of $K$-matrices:
\begin{itemize}

\item Diagonal $U(1)_{k_1}\times U(1)_{k_2}$, with matrix $K=\begin{pmatrix}k_1&0\\0&k_2\end{pmatrix}$, and
\item $\mathbb Z_{k_1}$ twisted gauge theory at level $k_2$, denoted by $(\mathbb Z_{k_1})_{k_2}$, with matrix $K=\begin{pmatrix}0&k_1\\k_1&k_2\end{pmatrix}$.
\end{itemize}

\begin{remark}\normalfont
The theory $(\mathbb Z_{k_1})_{k_2}$ is also known as Dijkgraaf-Witten theory when $k_2$ is even~\cite{Dijkgraaf:1989pz}. It admits a Chern-Simons gauge theory realization~\cite{Maldacena:2001ss,Kapustin:2014gua}. One can show that any $2\times2$ matrix $K$ with $\det(K)=-n^2$ for some integer $n$ can be brought into this form by a $\mathrm{GL}_2(\mathbb Z)$ congruence transformation $G^tKG$ (see e.g.~\cite{Lu:2012dt}). Furthermore, it is easy to show that $(\mathbb Z_{k_1})_{k_2}\sim (\mathbb Z_{k_1})_{k_2+2k_1}$, because the corresponding matrices are congruent\footnote{
Given $(\mathbb Z_a)_b$, not all the theories in $b\in[0,2a)$ are independent. For example, if $a$ is odd, one has the duality of spin TQFTs $(\mathbb Z_a)_b\ \longleftrightarrow\ (\mathbb Z_a)_{b+a}$. This follows from the more general duality
\begin{equation}
(\mathbb Z_a)_b\times U(1)_k\ \longleftrightarrow\ (\mathbb Z_a)_{b+a}\times U(1)_k
\end{equation}
which holds if and only if $a=2^\alpha(2m+1)$ and $k=2^\alpha(2n+1)$ for some integers $\alpha,m,n$. The explicit change of variables is $G^tK_{a,b}G\equiv K_{a,a+b}$, where
\begin{equation}
G:=\begin{pmatrix}-1&m + n + 2 m n&2 n+1\\0&-1&0\\0&2m+1&1\end{pmatrix},\quad K_{a,b}:=\begin{pmatrix}0&a&0\\a&b&0\\0&0&k\end{pmatrix}\,.
\end{equation}\label{footnote:dual_Z}
%Conjecture: there are only $2\tilde\tau(k)$ independent theories in this range, where $\tilde\tau(k):=\tau(k^2/2)$ if $k$ is even, and $\tilde\tau(k):=\tau(k^2)$ if odd, where $\tau(n)$ is the divisor function (equal to the number of divisors of $n$).
}.
\end{remark}

We conjecture the following:
\begin{conjecture}\label{conj:diag}
The diagonal theory $U(1)_{k_1}\times U(1)_{k_2}$:
\begin{itemize}
\item If $k_1k_2>0$,
\begin{itemize}
\item Never time-reversal invariant if $k_1k_2=0\mod 4$,
\item If $k_1k_2= 2\mod 4$, say, $k_1=2\tilde k_1$, then the theory is $\mathsf T$-invariant if and only if $k_2\in \mu(\tilde k_1)\mathbb T$, i.e., if $\mu(\tilde k_1)=\mu(k_2)$,
\item If $k_1k_2$ is odd, then the theory is $\mathsf T$-invariant if and only if $k_2\in \mu(k_1)\mathbb T$, i.e., if $\mu(k_1)=\mu(k_2)$
\end{itemize}

\item If $k_1k_2<0$,
\begin{itemize}
\item If $k_1$ is odd, the theory is $\mathsf T$-invariant if and only if $k_2\in \mu(k_1)\mathbb T$,
\item If $k_1=2\tilde k_1$ is even, the theory is $\mathsf T$-invariant if and only if $k_2\in \mu(\tilde k_1)(\mathbb T\cup 2\mathbb T)$.
\end{itemize}

\end{itemize}
\end{conjecture}

\begin{conjecture}\label{cj:DW_time_rev}
The theory $(\mathbb Z_{k_1})_{k_2}$ is time-reversal invariant if and only if $k_2\in \mu(k_1)\mathbb Z$.
\end{conjecture}

Some of these claims are easy to prove. For example, if $k_1$ and $k_2$ are both even and positive, then the theory $U(1)_{k_1}\times U(1)_{k_2}$ is bosonic and has central charge $+2$, and so it cannot be time-reversal invariant. More generally, the conditions above can be seen to be necessary just by insisting that the generating lines $\vec e_1,\vec e_2$ have a line with opposite spin. Proving that they are also sufficient requires more work, but in principle does not seem out of reach: an approach similar to the one-dimensional case $U(1)_k$ should work. In any case, we checked that the conjecture is correct up to $|k_i|\le 200$ in the diagonal case, and $|k_1|\le 200$ and $k_2\in[0,2k_1)$ in the $(\mathbb Z_{k_1})_{k_2}$ gauge theory case.
We stress that the diagonal theory can be be time-reversal invariant even when neither of the factors by itself is; naturally, this also holds for more general theories: a product may have more symmetries than its individual factors.

%Diagonal: tested up to $|k_i|\le200$. DW: tested up to 192.

%\textbf{Question}: what is the number of permutations as a function of $k_1,k_2$? If $k_1=3^k$ and $k_2$ is positive and odd, it seems that the number of solutions is 
%%If $k_1=3$ and $k_2$ is positive and odd, it seems that the number of solutions is $2^{3+\nu(k_2/3)}$, where
%%\begin{equation}
%%\nu(p_1^{\alpha_1}p_2^{\alpha_2}\cdots p_n^{\alpha_n}q_1^{\beta_1}q_2^{\beta_2}\cdots q_m^{\beta_m})=n+m-\left[\sum_{i=1}^m\beta_i\in 2\mathbb Z+1\right]
%%\end{equation}
%%where $p_i\equiv 1\mod6$ and $q_i\equiv 5\mod 6$.
%$3^{k-1}2^{\omega(k_2)+\frac12(3+\lambda(\mu'(k_2)))}$ where $\mu'$ denote the operation of keeping the prime factors congruent to $5$ modulo $6$ only, and $\lambda$ is the Liouville function $\lambda(\,\cdot\,)=(-1)^{\Omega(\,\cdot\,)}$, with $\Omega(\,\cdot\,)$ the number of prime factors counted with multiplicity.

%Alternatively, if we let $\mu'$ denote the operation of keeping the prime factors congruent to $5$ modulo $6$ only, then the number of solutions can also be expressed as $2^{\omega(k_2)+\frac12(3+\lambda(\mu'(k_2)))}$, with $\lambda$ the Liouville function $\lambda(\,\cdot\,)=(-1)^{\Omega(\,\cdot\,)}$, with $\Omega(\,\cdot\,)$ the number of prime factors counted with multiplicity.

Note that if the conjecture above is true, then any odd non-Pythagorean prime factor of $\det(K)$ must appear an even number of times. In fact, it seems that this is true for any $2\times2$ matrix, whether it is of the forms above or not:
\begin{conjecture}
A necessary condition for the matrix $K\in\mathbb Z^{2\times2}$ to describe a time-reversal invariant theory is that $\lambda(\det(K))\in\mathbb T$, where $\lambda(n)$ denotes the squarefree part of $n$\,.
\end{conjecture}

We recall that a number is said to be squarefree if its prime decomposition contains no repeated factors. We have checked that this conjecture is true for all matrices with $|\det(K)|\le500$. (For completeness, we remark that $\lambda(n)\in\mathbb T$ if and only if $n$ can be expressed as the sum of two perfect squares, not necessarily coprime).

It also appears that all primitive matrices with $\det(K)>0$, if time-reversal invariant, have $\mathsf T^2=\mathsf C$, as in the $1\times1$ case:
\begin{conjecture}
If $K\in\mathbb Z^{2\times2}$ is positive definite and primitive (i.e.~with $\gcd(K_{ij})=1$ for all $i,j$), then $\mathsf T^2=\mathsf C$.
\end{conjecture}

We checked that this is true for all matrices with $\det(K)\le400$.

\subsubsection{Unitary Symmetries.}

An essentially identical philosophy allows us to study unitary symmetries rather than anti-unitary ones. Following an argument equivalent to that of proposition~\ref{lm:necessary_K} it is easy to prove that
\begin{proposition}\label{pr:unitary_necessary_K}
Given some $K\in\mathbb Z^{n\times n}$, the most general unitary symmetry $($i.e., a permutation subject to~\eqref{eq:mtc_unitary}$)$ is of the form
\begin{equation}
\mathsf U\colon\vec\alpha\mapsto Q\vec\alpha
\end{equation}
for some $Q\in\mathbb Z^{n\times n}$, invertible over $\mathcal A$, the $i$-th column of which represents $\mathsf U(\vec e_i)$, the action of the unitary symmetry on the unit vector in the $i$-th direction. Invariance of spin and braiding requires
\begin{equation}\label{eq:unitary_sym}
PK+Q^t K^{-1}QK=1_n
\end{equation}
for some integral matrix $P$ with even diagonal components.
\end{proposition}

There is always the trivial solution $Q=1_n$, which leaves all the lines invariant, and its negative $Q=-1_n$, which corresponds to charge-conjugation $\mathsf C\colon\vec\alpha\mapsto-\vec\alpha$. Any other solution $Q$ (invertible modulo $K$) will correspond to some non-trivial unitary zero-form symmetry of the system.

We can finally write down the general expression for the group of symmetries of a given theory:
\begin{proposition}\label{pr:group_symmetries_K}
Given an arbitrary abelian TQFT realized as a $U(1)^n$ Chern-Simons theory with matrix of levels $K$, the group of (unitary and anti-unitary) zero-form symmetries can be expressed as
\begin{equation}
\Aut(K)\cong\{Q\in\mathbb Z^{n\times n}\,|\, PK\pm Q^t K^{-1}QK=1_n\mathrm{\ for\ some\ }P\in\mathbb Z^{n\times n}\}/\sim
\end{equation}
where $P$ is required to have even diagonal components, $Q$ is required to be invertible modulo $K$, and $\sim$ denotes the equivalence
\begin{equation}
Q\sim Q'\qquad\Longleftrightarrow\qquad Q\vec e_i\sim Q'\vec e_i,\qquad i=1,2,\dots,n
\end{equation}
where the last $\sim$ denotes equivalence in $\mathcal A$ $($cf.~\eqref{eq:equiv_A}$)$. The subgroup of unitary symmetries is given by
\begin{equation}
\Aut_U(K)\cong\{Q\in\mathbb Z^{n\times n}\,|\, PK+ Q^t K^{-1}QK=1_n\mathrm{\ for\ some\ }P\in\mathbb Z^{n\times n}\}/\sim
\end{equation}
with the same restrictions as before. A given symmetry $[Q]$ is quantum if and only if $P\neq0$ for all $Q\in[Q]$.
\end{proposition}

\begin{remark}\normalfont
Here we are making a slight abuse of notation in order to simplify the presentation: strictly speaking, if a given matrix $Q$ satisfies both $PK+ Q^t K^{-1}QK=1_n$ and $PK-Q^t K^{-1}QK=1_n$ (possibly with different $P$'s), they are \emph{different} symmetries, and so distinct elements of $\Aut(K)$. The same permutation on the anyons constitutes both a unitary, and an anti-unitary symmetry of the system. In other words, the group of symmetries is the \emph{disjoint union} of the set of anti-unitary symmetries, and the set of unitary symmetries. In order to implement this, one should think of $\Aut(K)$ as pairs $(Q,\sigma)$, where $\sigma=\pm1$ keeps track of whether a given permutation is unitary or anti-unitary, and one must add the condition $\sigma(Q)=\sigma(Q')$ to the equivalence relation $\sim$.
\end{remark}

We propose the following conjecture:
\begin{conjecture}\label{cj:Z_k_symmetry}
The group of unitary symmetries of $(\mathbb Z_k)_0$ is multiplicative in $k$:
\begin{equation}
\Aut_U((\mathbb Z_{ab})_0)=\Aut_U((\mathbb Z_a)_0)\times \Aut_U((\mathbb Z_b)_0),\qquad \gcd(a,b)=1\,.
\end{equation}
Furthermore, for prime powers, it is given by
\begin{equation}
\Aut_U((\mathbb Z_{\pi^n})_0)=D_{2\phi(\pi^n)},\qquad \Aut_U((\mathbb Z_{2^n})_0)=\mathbb Z_2\times D_{\phi(2^n)}\,,
\end{equation}
where $D_{2n}$ denotes the dihedral group of order $2n$. The full group of symmetries, including anti-unitary transformations, is a $\mathbb Z_2$ extension of the unitary sub-group:
\begin{equation}
\Aut((\mathbb Z_k)_0)=\mathbb Z_2\ltimes\Aut_U((\mathbb Z_k)_0)\,.
\end{equation}
\end{conjecture}

\begin{remark}\normalfont
Note the similarity of this group and $\mathbb Z_k^\times:=\Aut(\mathbb Z_k)$, the multiplicative group of integers modulo $k$. As per a classic result of Gauss, this latter group is also multiplicative, and given by $\Aut(\mathbb Z_{\pi^n})=\mathbb Z_{\phi(\pi^n)}$ and $\Aut(\mathbb Z_{2^n})=\mathbb Z_2\times \mathbb Z_{\phi(2^n)/2}$. For $k=\pi$ a prime, the group $\Aut_U((\mathbb Z_{\pi})_0)=D_{2(\pi-1)}$ has been computed in~\cite{Fuchs:2014ema}.
\end{remark}

%\textbf{Conjecture}: the unitary group of symmetries of $\mathbb Z_k$ is multiplicative in $k$, and is given for primer powers by
%\begin{equation}
%\Aut_U(\mathbb Z_{p^n})=D_{2\phi(p^n)},\qquad \Aut_U(\mathbb Z_{2^n})=\mathbb Z_2\times D_{\phi(2^n)}
%\end{equation}
%Note the similarity with the automorphism group of $\mathbb Z_{p^n}$ (cf.~\url{http://people.sju.edu/~smith/Current_Courses/autG.pdf}).

We next illustrate how to compute $\Aut(\,\star\,)$ step by step, through a couple of examples. More examples are worked out, to a lesser degree of detail, in section~\ref{sec:examples}.

Consider the theory $(\mathbb Z_{k_1})_{k_2}$, whose matrix is
\begin{equation}
K=\begin{pmatrix} 0&k_1\\k_1&k_2\end{pmatrix}
\end{equation}
where we can take without loss of generality $k_1>0$ and $0\le k_2<2k_1$. The theory is bosonic if $k_2$ is even, and spin otherwise. In the first case, the lines are of the form $(\alpha,\beta)\in\mathbb Z_{k_1}\times\mathbb Z_{k_1}$, and in the second case $(\alpha,\beta)\in\mathbb Z_{2k_1}\times\mathbb Z_{k_1}$. The spin of an arbitrary line is
\begin{equation}
h_{\alpha,\beta}=\frac{\alpha\beta}{k_1}-\frac{k_2\alpha^2}{2k_1^2}
\end{equation}

A common notation for the lines of $(\mathbb Z_{k_1})_{k_2}$ is $\mathsf e_i=(i,0)$, called the \emph{electric} lines, and $\mathsf m_j=(0,j)$, called the \emph{magnetic} lines. Their product is $\mathsf e_i\mathsf m_j=(i,j)$. There are $i\in[0,k_1)$ electric lines if $k_2$ is even, and $i\in[0,2k_1)$ lines of odd; and $j\in[0,k_1)$ magnetic lines. (The electric line $\mathsf e_i=(i,0)$ should not be confused with the unit vector $\vec e_i=(0,0,\dots,0,1,0,\dots,0)$ in the $i$-th direction). The line $\mathsf e_{k_1}\equiv\psi$ is the transparent fermion, and so $\mathsf e_{i+k_1}\equiv\mathsf e_i\times\psi$.

Take for example $(\mathbb Z_3)_0$. A basis of lines is
\begin{equation}
\begin{array}{llllllllll}
\mathcal A&=\{(0,0),&(1,0),&(2,0),&(0,1),&(1,1),&(2,1),&(0,2),&(1,2),&(2,2)\}\\
&=\{\,\boldsymbol1,&\,\mathsf e_1,&\,\mathsf e_2,&\,\mathsf m_1,&\,\mathsf e_1\mathsf m_1,&\,\mathsf e_2\mathsf m_1,&\,\mathsf m_2,&\,\mathsf e_1\mathsf m_2,&\,\mathsf e_2\mathsf m_2\}
\end{array}
\end{equation}
with spins
\begin{equation}
h=\{0,\,0,\,0,\,0,\,1/3,\,2/3,\,0,\,2/3,\,1/3\}\,.
\end{equation}

Any endomorphism of the fusion algebra is of the form
\begin{equation}
g\colon \begin{pmatrix}\alpha\\\beta\end{pmatrix}\mapsto Q\begin{pmatrix}\alpha\\\beta\end{pmatrix},\qquad Q=\begin{pmatrix}g(\vec e_1)&g(\vec e_2)\end{pmatrix}\,.
\end{equation}

As $\vec e_i$ both have vanishing spin, the condition $g\colon h\mapsto \pm h$ requires
\begin{equation}
g(\vec e_i)\in\{(1,0),\,(2,0),\,(0,1),\,(0,2)\}
\end{equation}
and so there are $4^2-4=12$ candidates for the matrices $Q$:
\begin{equation}
\begin{array}{c|cccc}
g(\vec e_i)&(1,0)&(2,0)&(0,1)&(0,2)\\ \hline
(1,0)&\cdot&\begin{pmatrix}1&2\\0&0\end{pmatrix}&\begin{pmatrix}1&0\\0&1\end{pmatrix}&\begin{pmatrix}1&0\\0&2\end{pmatrix}\\
(2,0)&\begin{pmatrix}2&1\\0&0\end{pmatrix}&\cdot&\begin{pmatrix}2&0\\0&1\end{pmatrix}&\begin{pmatrix}2&0\\0&2\end{pmatrix}\\
(0,1)&\begin{pmatrix}0&1\\1&0\end{pmatrix}&\begin{pmatrix}0&2\\1&0\end{pmatrix}&\cdot&\begin{pmatrix}0&0\\1&2\end{pmatrix}\\
(0,2)&\begin{pmatrix}0&1\\2&0\end{pmatrix}&\begin{pmatrix}0&2\\2&0\end{pmatrix}&\begin{pmatrix}0&0\\2&1\end{pmatrix}&\cdot
\end{array}
\end{equation}

By explicit computation, one may check that the only endomorphisms that are actually automorphisms (i.e., the only matrices $Q$ that are invertible modulo $K$) are
\begin{equation}
\begin{aligned}
\begin{pmatrix}1&0\\0&1\end{pmatrix},\begin{pmatrix}2&0\\0&2\end{pmatrix},\begin{pmatrix}0&1\\1&0\end{pmatrix},\begin{pmatrix}0&2\\2&0\end{pmatrix}\\
\begin{pmatrix}1&0\\0&2\end{pmatrix},\begin{pmatrix}2&0\\0&1\end{pmatrix},\begin{pmatrix}0&2\\1&0\end{pmatrix},\begin{pmatrix}0&1\\2&0\end{pmatrix}
\end{aligned}
\end{equation}
and that the first line satisfies $K^{-1}-Q^tK^{-1}Q=P$, and the second one $K^{-1}+Q^tK^{-1}Q=P$, for some integral-valued matrix $P$. Therefore, the former generate unitary symmetries, and the latter anti-unitary symmetries.

One may check that the two matrices
\begin{equation}
\mathsf T\colon\begin{pmatrix}0&2\\1&0\end{pmatrix},\qquad\mathsf U\colon\begin{pmatrix}0&1\\1&0\end{pmatrix}
\end{equation}
generate the whole group of symmetries, and they satisfy
\begin{equation}
\mathsf T^4=\mathsf U^2=(\mathsf{TU})^2=1
\end{equation}
and so the group of symmetries is dihedral:
\begin{equation}
\Aut((\mathbb Z_3)_0)=D_8=\langle \mathsf T,\mathsf U\rangle\,.
\end{equation}
Similarly, the pair of matrices $\mathsf C:=\mathsf T^2$ and $\mathsf U$ generate the subgroup of unitary symmetries, and they satisfy
\begin{equation}
\mathsf C^2=\mathsf U^2=1
\end{equation}
and so the latter is cyclic:
\begin{equation}
\Aut_U((\mathbb Z_3)_0)=\mathbb Z_2^2=\langle \mathsf C,\mathsf U\rangle\,.
\end{equation}

Consider now what happens when we turn on a non-trivial twisting, say, $(\mathbb Z_3)_2$. The spin of the lines is modified into
\begin{equation}
h=\{0,\,8/9,\,5/9,\,0,\,2/9,\,2/9,\,0,\,5/9,\,8/9\}\,.
\end{equation}

As we can see, there is no line with spin $-8/9=1/9\mod1$, and so $\vec e_1$ has no partner under time-reversal: the theory does not admit anti-unitary symmetries. Therefore the symmetries, if any, must be unitary, and so they must fix the spin; thus, the condition $h\mapsto +h$ requires
\begin{equation}
\begin{aligned}
\mathsf U(\vec e_1)&\in\{(1,0),(2,2)\}\\
\mathsf U(\vec e_2)&\in\{(0,1),(0,2)\}
\end{aligned}
\end{equation}
from where it follows that all the candidates for $Q$ are
\begin{equation}
\begin{array}{c|cc}
\mathsf U(\vec e_i)&(1,0)&(2,2)\\ \hline
(0,1)&\begin{pmatrix}1&0\\0&1\end{pmatrix}&\begin{pmatrix}2&0\\2&1\end{pmatrix}\\
(0,2)&\begin{pmatrix}1&0\\0&2\end{pmatrix}&\begin{pmatrix}2&0\\2&2\end{pmatrix}
\end{array}
\end{equation}

One may check that all these matrices are invertible, but the only two that satisfy $K^{-1}-Q^tK^{-1}Q=P$ for some integral-valued matrix $P$ are
\begin{equation}
\begin{pmatrix}1&0\\0&1\end{pmatrix},\begin{pmatrix}2&0\\2&2\end{pmatrix}\,.
\end{equation}
Finally, the second matrix is easily seen to implement charge conjugation $\mathsf C$, and so it squares to the identity. In other words, the group of symmetries of the system is
\begin{equation}
\Aut((\mathbb Z_3)_2)=\Aut_U((\mathbb Z_3)_2)=\mathbb Z_2=\langle \mathsf C\rangle\,.
\end{equation}

By an identical argument one may calculate the group of symmetries of an arbitrary abelian theory. In table~\ref{tab:Z_k_symmetry_full} we include the group of symmetries of $(\mathbb Z_{k_1})_{k_2}$ for small values of the levels.

\begin{table}[h!]
\begin{equation*}
\begin{array}{r|c|c|c|c}
k & \Aut((\mathbb Z_k)_0)& \Aut_U((\mathbb Z_k)_0)& \Aut((\mathbb Z_k)_{\mu(k)})& \Aut_U((\mathbb Z_k)_{\mu(k)})\\ \hline
2&\mathbb Z_2^2&\mathbb Z_2&\mathbb Z_2&0\\
3&D_8&\mathbb Z^2_2&D_8&\mathbb Z_2^2\\
4&D_8&\mathbb Z^2_2&D_8&\mathbb Z_2^2\\
5&\mathbb Z_4\circ D_8&D_8&\mathbb Z_4&\mathbb Z_2\\
6&\mathbb Z_2\times D_8&\mathbb Z^3_2&D_8&\mathbb Z_2^2\\
7&\mathbb Z_3\rtimes D_8&D_{12}&\mathbb Z_3\rtimes D_8&D_{12}\\
8&\mathbb Z_2\times D_8&\mathbb Z^3_2&\mathbb Z_2\times D_8&\mathbb Z_2^3\\
9&\mathbb Z_3\rtimes D_8&D_{12}&\mathbb Z_3\rtimes D_8&D_{12}\\
10&\mathbb Z_2\times \mathbb Z_4\circ D_8&\mathbb Z_2 \times D_8&\mathbb Z_4&\mathbb Z_2\\
11&\mathbb Z_5\rtimes D_8&D_{20}&\mathbb Z_5\rtimes D_8&D_{20}\\
12&\mathbb Z_2^2\wr \mathbb Z_2&\mathbb Z^4_2&\mathbb Z_2^2\wr \mathbb Z_2&\mathbb Z_2^4\\
13&\mathbb Z_4\circ D_{24}&D_{24}&\mathbb Z_4&\mathbb Z_2\\
14&\mathbb Z_2\times \mathbb Z_3\rtimes D_8&\mathbb Z^2_2 \times S_3&\mathbb Z_3\rtimes D_8&D_{12}\\
15&D_8\rtimes_5D_8&\mathbb Z^2_2 \times D_8&\mathbb Z_2^2\rtimes \mathbb Z_4&\mathbb Z_2^3\\
16&\mathbb Z_4\rtimes D_8&\mathbb Z_2 \times D_8&\mathbb Z_4\rtimes D_8&\mathbb Z_2\times D_8\\
17&\mathbb Z_4\circ D_{32}&D_{32}&\mathbb Z_4&\mathbb Z_2\\
18&\mathbb Z_2\times \mathbb Z_3\rtimes D_8&\mathbb Z^2_2\times S_3&\mathbb Z_3\rtimes D_8&D_{12}\\
19&\mathbb Z_9\rtimes D_8&D_{36}&\mathbb Z_9\rtimes D_8&D_{36}\\
20&D_8\rtimes_5D_8&\mathbb Z^2_2\times D_8&\mathbb Z_2^2\rtimes \mathbb Z_4&\mathbb Z_2^3\\
21&\mathbb Z_2^3\rtimes_2D_{12}&\mathbb Z^3_2\times S_3&\mathbb Z_2^3\rtimes_2D_{12}&\mathbb Z_2^3\times S_3\\
22&\mathbb Z_2\times \mathbb Z_5\rtimes D_8&\mathbb Z^2_2\times D_{10}&\mathbb Z_5\rtimes D_8&D_{20}\\
23&\mathbb Z_{11}\rtimes D_8&D_{44}&\mathbb Z_{11}\rtimes D_8&D_{44}\\
24&\mathbb Z_2\times \mathbb Z_2^2\wr \mathbb Z_2&\mathbb Z^5_2&\mathbb Z_2\times \mathbb Z_2^2\wr \mathbb Z_2&\mathbb Z_2^5\\
25&\mathbb Z_4\circ D_{40}&D_{40}&\mathbb Z_4&\mathbb Z_2\\
26&\mathbb Z_2\times \mathbb Z_4\circ D_{24}&\mathbb Z_2 \times D_{24}&\mathbb Z_4&\mathbb Z_2\\
27&\mathbb Z_9\rtimes D_8&D_{36}&\mathbb Z_9\rtimes D_8&D_{36}
\end{array}
\end{equation*}
\caption{The group of symmetries of $(\mathbb Z_{k_1})_{k_2}$, denoted by $\Aut(\,\star\,)$, and its unitary subgroup $\Aut_U(\,\star\,)$, for $k_1\in[0,27]$ and $k_2=0,\mu(k_1)$. For $k_2\not\propto \mu(k_1)$ there are no anti-unitary symmetries. (See Appendix~\ref{ap:grupo} for basic definitions).}
\label{tab:Z_k_symmetry_full}
\end{table}

Similarly, in tables~\ref{tab:diag_symmetry_full} and~\ref{tab:diag_symmetry_unitary} we include the group of symmetries of the diagonal theory $U(1)_{k_1}\times U(1)_{k_2}$.

\begin{table}[h!]
\begin{equation*}
%\hspace{-40pt}\footnotesize
\hspace{-30pt}
\arraycolsep=-.2pt%\def\arraystretch{2.2}
\begin{array}{r|cccccccccccccc}
 &\phantom12&\phantom13&\phantom14&\phantom15&\phantom16&\phantom17&\phantom18&\phantom19&10&11&12&13&14&15\\ \hline
2&\,\mathbb Z_2&\mathbb Z_2&\mathbb Z_2&\mathbb Z_4&\mathbb Z_2&\mathbb Z_2&\mathbb Z_2&\mathbb Z_2&\mathbb Z_2^2&\mathbb Z_2&\mathbb Z_2^2&\mathbb Z_4&\mathbb Z_2&\mathbb Z_2^2\\
3&\cdot&SD_{16}&\mathbb Z_2^2&\mathbb Z_2^2&D_8&\mathbb Z_2^2&\mathbb Z_2^3&D_{12}&\mathbb Z_2^2&\mathbb Z_2^2&\mathbb Z_2 \times D_8&\mathbb Z_2^2&\mathbb Z_2^2&\mathbb Z_2^2\rtimes \mathbb Z_4\\
4&\cdot&\cdot&D_8&\mathbb Z_2^2&\mathbb Z_2^2&\mathbb Z_2^2&\mathbb Z_2^3&\mathbb Z_2^2&\mathbb Z_2^2&\mathbb Z_2^2&\mathbb Z_2^3&\mathbb Z_2^2&\mathbb Z_2^2&\mathbb Z_2^3\\
5&\cdot&\cdot&\cdot&\mathbb Z_4\circ D_8&\mathbb Z_2^2&\mathbb Z_2^2&\mathbb Z_2^3&\mathbb Z_2^2&\mathbb Z_4 \times S_3&\mathbb Z_2^2&\mathbb Z_2^3&\mathbb Z_4 \times \mathbb Z_2&\mathbb Z_2^2&\mathbb Z_2^2 \times S_3\\
6&\cdot&\cdot&\cdot&\cdot&\mathbb Z_2\times D_8&\mathbb Z_2^2&\mathbb Z_2^2&D_{12}&\mathbb Z_2^2&\mathbb Z_2^2&\mathbb Z_2^3&\mathbb Z_2^2&\mathbb Z_2^3&D_8\rtimes \mathbb Z_4\\
7&\cdot&\cdot&\cdot&\cdot&\cdot&SD_{32}&\mathbb Z_2^3&\mathbb Z_2^2&\mathbb Z_2^2&\mathbb Z_2^2&\mathbb Z_2^3&\mathbb Z_2^2&QD_{32}&\mathbb Z_2^3\\
8&\cdot&\cdot&\cdot&\cdot&\cdot&\cdot&\mathbb Z_2\times D_8&\mathbb Z_2^3&\mathbb Z_2^2&\mathbb Z_2^3&\mathbb Z_2^4&\mathbb Z_2^3&\mathbb Z_2^2&\mathbb Z_2^4\\
9&\cdot&\cdot&\cdot&\cdot&\cdot&\cdot&\cdot&\mathbb Z_{24}\rtimes \mathbb Z_2&\mathbb Z_2^2&\mathbb Z_2^2&\mathbb Z_2^2 \times S_3&\mathbb Z_2^2&\mathbb Z_2^2&\mathbb Z_2^2 \times S_3\\
10&\cdot&\cdot&\cdot&\cdot&\cdot&\cdot&\cdot&\cdot&\mathbb Z_2\times D_8&\mathbb Z_2^2&\mathbb Z_2^3&\mathbb Z_4 \times \mathbb Z_2&\mathbb Z_2^2&\mathbb Z_2 \times D_8\\
11&\cdot&\cdot&\cdot&\cdot&\cdot&\cdot&\cdot&\cdot&\cdot&\mathbb Z_{24}\rtimes \mathbb Z_2&\mathbb Z_2^3&\mathbb Z_2^2&\mathbb Z_2^2&\mathbb Z_2^3\\
12&\cdot&\cdot&\cdot&\cdot&\cdot&\cdot&\cdot&\cdot&\cdot&\cdot&D_8 \times D_8&\mathbb Z_2^3&\mathbb Z_2^3&\mathbb Z_2^4\\
13&\cdot&\cdot&\cdot&\cdot&\cdot&\cdot&\cdot&\cdot&\cdot&\cdot&\cdot&\mathbb Z_4\circ D_{24}&\mathbb Z_2^2&\mathbb Z_2^3\\
14&\cdot&\cdot&\cdot&\cdot&\cdot&\cdot&\cdot&\cdot&\cdot&\cdot&\cdot&\cdot&\mathbb Z_2 \times D_{16}&\mathbb Z_2^3\\
15&\cdot&\cdot&\cdot&\cdot&\cdot&\cdot&\cdot&\cdot&\cdot&\cdot&\cdot&\cdot&\cdot&D_8\rtimes_7SD_{16}
\end{array}
\end{equation*}
\caption{The group of symmetries of the diagonal theory $K=\diag(k_1,k_2)$, to wit, $\Aut(U(1)_{k_1}\times U(1)_{k_2})$, for small values of $k_i$.}
\label{tab:diag_symmetry_full}
\end{table}

\begin{table}[h!]
\begin{equation*}
\hspace{-30pt}
\arraycolsep=2pt
\begin{array}{r|cccccccccccccc}
&\phantom12&\phantom13&\phantom14&\phantom15&\phantom16&\phantom17&\phantom18&\phantom19&10&11&12&13&14&15\\ \hline
2&\mathbb Z_2&\mathbb Z_2&\mathbb Z_2&\mathbb Z_2&\mathbb Z_2&\mathbb Z_2&\mathbb Z_2&\mathbb Z_2&\mathbb Z^2_2&\mathbb Z_2&\mathbb Z^2_2&\mathbb Z_2&\mathbb Z_2&\mathbb Z^2_2\\
3&\cdot&D_8&\mathbb Z_2^2&\mathbb Z_2^2&\mathbb Z_2^2&\mathbb Z_2^2&\mathbb Z_2^3&D_{12}&\mathbb Z_2^2&\mathbb Z_2^2&\mathbb Z_2\times D_8&\mathbb Z_2^2&\mathbb Z_2^2&\mathbb Z_2^3\\
4&\cdot&\cdot&D_8&\mathbb Z_2^2&\mathbb Z_2^2&\mathbb Z_2^2&\mathbb Z_2^3&\mathbb Z_2^2&\mathbb Z_2^2&\mathbb Z_2^2&\mathbb Z_2^3&\mathbb Z_2^2&\mathbb Z_2^2&\mathbb Z_2^3\\
5&\cdot&\cdot&\cdot&D_8&\mathbb Z_2^2&\mathbb Z_2^2&\mathbb Z_2^3&\mathbb Z_2^2&D_{12}&\mathbb Z_2^2&\mathbb Z_2^3&\mathbb Z_2^2&\mathbb Z_2^2&\mathbb Z_2^2\times S_3\\
6&\cdot&\cdot&\cdot&\cdot&\mathbb Z_2\times D_8&\mathbb Z_2^2&\mathbb Z_2^2&D_{12}&\mathbb Z_2^2&\mathbb Z_2^2&\mathbb Z_2^3&\mathbb Z_2^2&\mathbb Z_2^3&\mathbb Z_2\times D_8\\
7&\cdot&\cdot&\cdot&\cdot&\cdot&D_{16}&\mathbb Z_2^3&\mathbb Z_2^2&\mathbb Z_2^2&\mathbb Z_2\times\mathbb Z_2&\mathbb Z_2^3&\mathbb Z_2^2&D_{16}&\mathbb Z_2^3\\
8&\cdot&\cdot&\cdot&\cdot&\cdot&\cdot&\mathbb Z_2\times D_8&\mathbb Z_2^3&\mathbb Z_2^2&\mathbb Z_2^3&\mathbb Z_2^4&\mathbb Z_2^3&\mathbb Z_2^2&\mathbb Z_2^4\\
9&\cdot&\cdot&\cdot&\cdot&\cdot&\cdot&\cdot&D_{24}&\mathbb Z_2^2&\mathbb Z_2^2&\mathbb Z_2^2\times S_3&\mathbb Z_2^2&\mathbb Z_2^2&\mathbb Z_2^2\times S_3\\
10&\cdot&\cdot&\cdot&\cdot&\cdot&\cdot&\cdot&\cdot&\mathbb Z_2\times D_8&\mathbb Z_2^2&\mathbb Z_2^3&\mathbb Z_2^2&\mathbb Z_2^2&\mathbb Z_2\times D_8\\
11&\cdot&\cdot&\cdot&\cdot&\cdot&\cdot&\cdot&\cdot&\cdot&D_{24}&\mathbb Z_2^3&\mathbb Z_2^2&\mathbb Z_2^2&\mathbb Z_2^3\\
12&\cdot&\cdot&\cdot&\cdot&\cdot&\cdot&\cdot&\cdot&\cdot&\cdot&D_8\times D_8&\mathbb Z_2^3&\mathbb Z_2^3&\mathbb Z_2^4\\
13&\cdot&\cdot&\cdot&\cdot&\cdot&\cdot&\cdot&\cdot&\cdot&\cdot&\cdot&D_{24}&\mathbb Z_2^2&\mathbb Z_2^3\\
14&\cdot&\cdot&\cdot&\cdot&\cdot&\cdot&\cdot&\cdot&\cdot&\cdot&\cdot&\cdot&\mathbb Z_2\times D_{16}& \\
15&\cdot&\cdot&\cdot&\cdot&\cdot&\cdot&\cdot&\cdot&\cdot&\cdot&\cdot&\cdot&\cdot&D_8\times D_8
\end{array}
\end{equation*}
\caption{The group of unitary symmetries of the diagonal theory $K=\diag(k_1,k_2)$, to wit, $\Aut_U(U(1)_{k_1}\times U(1)_{k_2})$, for small values of $k_i$.}
\label{tab:diag_symmetry_unitary}
\end{table}

\subsection{Dualities of $U(1)^n$}\label{sec:dualities}

A straightforward extension of the formalism so far can be used to diagnose dualities between different abelian Chern-Simons theories. Given two systems described by matrices $K_1,K_2$, $n_1\times n_1$ and $n_2\times n_2$, respectively, the theories shall describe the same TQFT if they give rise to isomorphic anyon data $\mathcal A_i,\theta_i$. This corresponds to a bijection $f\colon \mathcal A_1\leftrightarrow\mathcal A_2$ that preserves fusion and spin. If we write the anyons $\mathcal A_i$ as $n_i$-tuples of integral charges $\vec \alpha_i$, then preservation of fusion requires $f$ to act linearly, say, $\vec \alpha_1=Q\vec \alpha_2$, with $Q$ an $n_1\times n_2$ integral matrix that has a left inverse modulo $K_2$ (equivalently, $\vec \alpha_2=\tilde Q\vec \alpha_1$ with $\tilde Q$ an $n_2\times n_1$ integral matrix with left inverse modulo $K_1$). Preservation of spin requires the existence of an $n_2\times n_2$ integral matrix $P$, with even diagonal components, such that
\begin{equation}\label{eq:ab_duality}
Q^tK_1^{-1}Q-K_2^{-1}=P\,.
\end{equation}
The theories described by $K_1,K_2$ are dual if and only if such matrices $Q,P$ exist:
\begin{equation}
K_1\quad\longleftrightarrow\quad K_2\,.
\end{equation}

In light of this discussion, we can summarize the content of the previous sections as follows: a theory with matrix $K$ has a unitary symmetry if and only if there is a self-duality $K\leftrightarrow K$, and an anti-unitary symmetry if and only if there is a duality $K\leftrightarrow -K$.

For example, it is a well-known fact that $U(1)_{-8}$ is level-rank dual (as bosonic TQFTs) to $SU(8)_1$. This latter theory can be represented as an abelian Chern-Simons theory with $K$-matrix equal to the Cartan matrix of $SU(8)$. In our terminology, this duality is implemented, for example, via the matrix $Q=(0,0,1,0,0,0,0)$, which indeed satisfies~\eqref{eq:ab_duality} with $P\equiv(2)$.

Other interesting examples of dual abelian theories can be found in twisted gauge theories $(\mathbb Z_a)_b$. Recall that if $a$ is odd, then $(\mathbb Z_a)_b\leftrightarrow (\mathbb Z_a)_{a+b}$, which is already clear at the Lagrangian level (see footnote~\ref{footnote:dual_Z}). There are extra dualities that go beyond this trivial one, for example $(\mathbb Z_7)_2\leftrightarrow (\mathbb Z_7)_4$ and $(\mathbb Z_7)_3\leftrightarrow (\mathbb Z_7)_5$. One can easily check these dualities by finding a suitable $Q,P$ in~\eqref{eq:ab_duality}.

Many more examples of (trivial and non-trivial) dualities between abelian theories can be exhibited. In contrast to the non-abelian case, abelian TQFTs enjoy infinitely-many dualities. For a given finite abelian group $\mathcal A$ and a quadratic form $\theta$ on it, there are infinitely many integral matrices, of varying dimension, that generate the pair $(\mathcal A,\theta)$, so all these matrices are dual. Trivially dual theories can be obtained by looking directly at the Lagrangian: matrices related by $\mathrm{GL}(\mathbb Z)$-conjugation give rise to the same dynamics, $GK_1G^t=K_2$. Non-trivially dual theories, which are not dual at the Lagrangian level, require the generalized condition $Q^tK_1^{-1}Q-K_2^{-1}=P$, which allows for matrices of different dimension. In any case, fixing $K_1$, one can find infinitely-many matrices $K_2$ that are dual to it, just by varying $G$ or $P,Q$ in these equations.

%\begin{equation}
%\begin{pmatrix}
%\cdot & \cdot & 1 & \cdot & \cdot & \cdot & \cdot
%\end{pmatrix}\begin{pmatrix}
% 2 & -1 & \cdot & \cdot & \cdot & \cdot & \cdot \\
% -1 & 2 & -1 & \cdot & \cdot & \cdot & \cdot \\
% \cdot & -1 & 2 & -1 & \cdot & \cdot & \cdot \\
% \cdot & \cdot & -1 & 2 & -1 & \cdot & \cdot \\
% \cdot & \cdot & \cdot & -1 & 2 & -1 & \cdot \\
% \cdot & \cdot & \cdot & \cdot & -1 & 2 & -1 \\
% \cdot & \cdot & \cdot & \cdot & \cdot & -1 & 2 
%\end{pmatrix}^{-1}\begin{pmatrix}
% \cdot \\ \cdot \\ 1 \\ \cdot \\ \cdot \\ \cdot \\ \cdot 
%\end{pmatrix}-\begin{pmatrix}-8\end{pmatrix}^{-1}\equiv \begin{pmatrix}2\end{pmatrix}
%\end{equation}

\section{Examples}\label{sec:examples}

Finally, we discuss some illustrative examples. To avoid repetition, we typically include a theory only if it incorporates a new feature that was not present in the previous examples. We begin by the case of a single abelian factor, $U(1)_k$.

\begin{example}[$k=2$]\normalfont
%\subsubsection{$U(1)_2$}

We have $\varpi(2)=0$, and so the system has no unitary symmetries. As the system is bosonic, there are no anti-unitary symmetries either.

One may regard the system as a spin TQFT, in which case it is usually known as the \emph{semion-fermion theory}~\cite{PhysRevX.3.041016,10.1093/ptep/ptw083}. The system now admits one anti-unitary symmetry, which can be found by solving $2p-q^2=1$, whose only solution in the range $q\in[0,2)$ is $q=1$. This means that the permutation is $s\leftrightarrow s\times\psi$, as is well-known.

We thus have
\begin{equation}
\begin{aligned}
\Aut(U(1)_2)&=\Aut_U(U(1)_2)=0\\
\Aut(U(1)_2\times\{\boldsymbol1,\psi\})&=\mathbb Z_2=\langle\mathsf T\rangle\\
\Aut_U(U(1)_2\times\{\boldsymbol1,\psi\})&=0\,.
\end{aligned}
\end{equation}

The integer $k=2$ is Pell, and so the time-reversal permutation above is a symmetry of the Lagrangian (provided by $\{\boldsymbol1,\psi\}$ we mean $U(1)_{-1}$ rather than $U(1)_{+1}$).

\end{example}

\begin{example}[$k=3$]\normalfont

We have $\varpi(3)=1$, and so the system only has one unitary symmetry: charge conjugation. This is a Lagrangian symmetry.

Similarly, $3\neq1\mod 4$, and so the system is not time-reversal invariant.

We thus have
\begin{equation}
\Aut(U(1)_3)=\Aut_U(U(1)_3)=\mathbb Z_2=\langle\mathsf C\rangle\,.
\end{equation}

\end{example}

\begin{example}[$k=5$]\normalfont

We have $\varpi(5)=1$, and so the system only has one unitary symmetry: charge conjugation. This is a Lagrangian symmetry.

The level satisfies $5= 1\mod 4$, and so the system is time-reversal invariant. The permutation can be found using equation~\eqref{eq:wilson}: $q=\frac{5-1}{2}!+5=7$ (there is a second solution, which differs by a sign: $q=-7=3\mod 10$). The explicit map of lines is
\begin{equation}
\begin{array}{c|c|c|c|c|c|c|c|c|c|c|}
\alpha & 0 & 1 & 2 & 3 & 4 & 5 & 6 & 7 & 8 & 9\\
\mathsf T(\alpha) & 0& 7& 4& 1& 8& 5& 2& 9& 6& 3
\end{array}
\end{equation}

We thus have
\begin{equation}
\begin{aligned}
\Aut(U(1)_5)&=\mathbb Z_4=\langle\mathsf T\rangle\\
\Aut_U(U(1)_5)&=\mathbb Z_2=\langle\mathsf C\rangle\,.
\end{aligned}
\end{equation}

The integer $k=5$ is Pell, and so the time-reversal permutation above is a symmetry of the Lagrangian once we include the gravitational counterterm (but not without it).

\end{example}

\begin{example}[$k=8$]\normalfont

We have $\varpi(8)=1$, and so the system only has one unitary symmetry: charge conjugation. This is a Lagrangian symmetry.

The system is bosonic, and so it is not time-reversal invariant. One may regard the system as a spin TQFT, but $8=0\mod4$, and so it is not time-reversal invariant either.

As a spin TQFT, one has $\varpi(4)+1=2$, and so the system has three unitary symmetries: charge-conjugation and multiplication by $\pm3$. The latter are not Lagrangian symmetries.

We thus have
\begin{equation}
\begin{aligned}
\Aut(U(1)_8)&=\Aut_U(U(1)_8)=\mathbb Z_2=\langle\mathsf C\rangle\\
\Aut(U(1)_8\times\{\boldsymbol1,\psi\})&=\Aut_U(U(1)_8\times\{\boldsymbol1,\psi\})=\mathbb Z_2\times\mathbb Z_2=\langle\mathsf C,\mathsf U\rangle\,.
\end{aligned}
\end{equation}
where $\mathsf U$ is either of $(\alpha,\beta)\mapsto(\pm 3\alpha,\alpha+\beta)$ (the other sign being $\mathsf{CU}$).

\end{example}

\begin{example}[$k=12$]\normalfont

%Non-trivial unitary symmetries as bosonic.

We have $\varpi(12)=2$, and so the system has three unitary symmetries: charge conjugation and multiplication by $\pm5$. The latter are not Lagrangian symmetries.

The system is bosonic, and so it is not time-reversal invariant. One may regard the system as a spin TQFT, but $12=0\mod4$, and so it is not time-reversal invariant either.

As a spin TQFT, one has $\varpi(6)+1=2$, and so the unitary symmetries are the same as in the bosonic case. They are not Lagrangian symmetries either.

We thus have
\begin{equation}
\begin{aligned}
\Aut(U(1)_{12})&=\Aut_U(U(1)_{12})=\mathbb Z_2\times\mathbb Z_2=\langle\mathsf C,\mathsf U\rangle\\
\Aut(U(1)_{12}\times\{\boldsymbol1,\psi\})&=\Aut_U(U(1)_{12}\times\{\boldsymbol1,\psi\})=\mathbb Z_2\times\mathbb Z_2=\langle\mathsf C,\mathsf U\rangle\,.
\end{aligned}
\end{equation}
where $\mathsf U$ denotes multiplication by either of $\pm5$ (the other sign being $\mathsf{CU}$), while fixing the local fermion, if any.

\end{example}

\begin{example}[$k=15$]\normalfont

%Non-trivial unitary symmetries.

We have $\varpi(15)=2$, and so the system has three unitary symmetries: charge conjugation and multiplication by $\pm11$. The latter are not Lagrangian symmetries.

The level can be factored as $15=3\cdot5$, and $3\neq1\mod4$, and so the system is not time-reversal invariant.

We thus have
\begin{equation}
\Aut(U(1)_{15})=\Aut_U(U(1)_{15})=\mathbb Z_2\times\mathbb Z_2=\langle\mathsf C,\mathsf U\rangle\,.
\end{equation}
where $\mathsf U$ denotes multiplication by either of $\pm11$ (the other sign being $\mathsf{CU}$).

\end{example}

\begin{example}[$k=24$]\normalfont

%Non-trivial unitary symmetries as bosonic.

We have $\varpi(24)=2$, and so the system has three unitary symmetries: charge conjugation and multiplication by $\pm7$. The latter are not Lagrangian symmetries.

The system is bosonic, and so it is not time-reversal invariant. One may regard the system as a spin TQFT, but $24=0\mod4$, and so it is not time-reversal invariant either.

As a spin TQFT, one has $\varpi(12)+1=3$, and so the number of unitary symmetries is doubled. The new symmetries, those that mix the bosonic lines with the transparent fermions, are generated by multiplication by $13$.

We thus have
\begin{equation}
\begin{aligned}
\Aut(U(1)_{24})&=\Aut_U(U(1)_{24})=\mathbb Z_2\times\mathbb Z_2=\langle\mathsf C,\mathsf U\rangle\\
\Aut(U(1)_{24}\times\{\boldsymbol1,\psi\})&=\Aut_U(U(1)_{24}\times\{\boldsymbol1,\psi\})=\mathbb Z^3_2=\langle\mathsf C,\mathsf U,\mathsf U'\rangle\,.
\end{aligned}
\end{equation}
where $\mathsf U$ denotes multiplication by either of $\pm7$ (the other sign being $\mathsf{CU}$) while fixing the local fermion, if any, and $\mathsf U'$ denotes multiplication by $13$, while mixing the local fermion.

%Note that $24$ is the largest integer whose number of totatives is the same as the number of solutions to $q^2\equiv1\mod24$. Therefore, $U(1)_{24}$ is the largest theory whose group of symmetries (as a spin theory) is identical to the automorphism group of its fusion algebra (which means that the topological spin transforms properly for \emph{any} transformation that respects fusion):
%\begin{equation}
%\Aut(U(1)_{24}\times\{\boldsymbol1,\psi\})\equiv \mathbb Z^\times_{24}
%\end{equation}
%The fact that $\phi(24)\equiv8$ is equal to the number of solutions to $q^2\equiv1\mod24$ plays an important role in the study of lattices and the moonshine (reference?).

\end{example}

\begin{example}[$k=25$]\normalfont

%First non-Pell.

We have $\varpi(25)=1$, and so the system only has one unitary symmetry: charge conjugation. This is a Lagrangian symmetry.

The level can be factored as $25=5^2$, and $5= 1\mod 4$, which means that the system is time-reversal invariant. In order to find the solution to $q^2=-1\mod25$ one may use Hensel lifting~\eqref{eq:hensel}: the solutions modulo $5$ are $\pm3$, and so the solutions modulo $5^2$ are $\pm3\mp(3^2+1)=\pm7$. The explicit map of lines is
\begin{equation}
\begin{array}{c|c|c|c|c|c|c|c|c|c|c|}
\alpha & 0 & 1 & 2 & 3 & 4 & \cdots & 46 & 47 & 48 & 49\\
\mathsf T(\alpha) & 0& 7& 14& 21& 28& \cdots& 22& 29& 36& 43
\end{array}
\end{equation}

We thus have
\begin{equation}
\begin{aligned}
\Aut(U(1)_{25})&=\mathbb Z_4=\langle\mathsf T\rangle\\
\Aut_U(U(1)_{25})&=\mathbb Z_2=\langle\mathsf C\rangle\,.
\end{aligned}
\end{equation}

The integer $k=25$ is not Pell, and so the time-reversal permutation above is not a symmetry of the Lagrangian, not even if we include the gravitational counterterm.

\end{example}

\begin{example}[$k=65$]\normalfont

%More than one time-reversal.

We have $\varpi(65)=2$, and so the system has three unitary symmetries: charge conjugation and multiplication by $\pm51$. The latter are not Lagrangian symmetries.

The level can be factored as $65=5\cdot 13$, and $5=13= 1\mod 4$, which means that the system is time-reversal invariant. In order to find the solution to $q^2=-1\mod65$ one may use the Chinese Remainder Theorem (cf.~the discussion below~\eqref{eq:hensel}). The solutions modulo $5$ are $q=\pm3$, and the solutions modulo $13$ are $q=\pm5$. Take for example the solution with $q= 3\mod 5$ and $q=5\mod 13$; then, using the Euclidean algorithm, we find $5\cdot8+13\cdot(-3)=1$, which means that $q=3\cdot13\cdot(-3)+5\cdot 5\cdot8= 83\mod65$. Similarly, taking the solution with $q= -3\mod 5$ and $q=5\mod 13$ leads to $q=57\mod 65$. All in all, the solutions of $q^2=-1\mod 65$ are $q=\pm47,\pm57$. The explicit map of lines is
\begin{equation}
\begin{array}{c|c|c|c|c|c|c|c|c|c|c|}
\alpha & 0 & 1 & 2 & 3 & 4 & \cdots & 126 & 127 & 128 & 129\\
\mathsf T_1(\alpha) & 0& 47& 94& 11& 58& \cdots& 72& 119& 36& 83\\
\mathsf T_2(\alpha) & 0& 57& 114& 41& 98& \cdots& 32& 89& 16& 73
\end{array}
\end{equation}

We thus have
\begin{equation}
\begin{aligned}
\Aut(U(1)_{65})&=\mathbb Z_4\times\mathbb Z_2=\langle\mathsf T,\mathsf U\rangle\\
\Aut_U(U(1)_{65})&=\mathbb Z_2\times\mathbb Z_2=\langle\mathsf C,\mathsf U\rangle\,.
\end{aligned}
\end{equation}
where $\mathsf T$ denotes either of $\mathsf T_1,\mathsf T_2$ (the other one being $\mathsf U\mathsf T$), and $\mathsf U$ denotes multiplication by either of $\pm51$ (the other sign being $\mathsf T^2\mathsf U$).

The integer $k=65$ is Pell, and so the permutation above is a symmetry of the Lagrangian once we include the gravitational counterterm (but not without it).

\end{example}

We now move on to $2\times2$ matrices. We denote by $[a,b,c]$ the equivalence class (with respect to congruence) of all matrices of which $\begin{pmatrix}a&b\\b&c\end{pmatrix}$ is a representative. We begin with positive-definite $K$, and order them by $\det(K)$. (We recall that there can be more than one congruence class with a given value of $\det(K)$).

\begin{example}[$\det(K)=2$]\normalfont
The first non-trivial positive-definite time-reversal invariant theory is $K=[2,0,1]$, where the permutation is $Q=\begin{pmatrix}1&1\\0&1\end{pmatrix}$, which is of order $2$. The system has no unitary symmetries. Therefore,
\begin{equation}
\begin{aligned}
\Aut([2,0,1])&=\mathbb Z_2=\langle\mathsf T\rangle\\
\Aut_U([2,0,1])&=0\,.
\end{aligned}
\end{equation}

The transformation $\mathsf T$ is not a symmetry of the Lagrangian (because the central charge is $2$), but it becomes one once we subtract two units of central charge (i.e., we consider the theory $\diag(K,-1,-1)$, which is dual to $U(1)_2\times U(1)_{-1}$).

There are no other $2\times2$ congruence classes with determinant equal to $2$.

\end{example}

\begin{example}[$\det(K)=5$]\normalfont
The next positive-definite time-reversal invariant theories are $K=[5,0,1]$ and $K=[2,1,3]$, where the permutations are
\begin{equation}
\pm Q=\begin{pmatrix}3&0\\0&1\end{pmatrix}
\end{equation}
and
\begin{equation}
\pm Q=\begin{pmatrix}1&2\\2&3\end{pmatrix}
\end{equation}
respectively. They all satisfy $\mathsf T^2=\mathsf C$ (and thus are of order $4$). The system has no non-trivial unitary symmetries. Therefore,
\begin{equation}
\begin{aligned}
\Aut([5,0,1])&=\Aut([2,1,3])=\mathbb Z_4=\langle\mathsf T\rangle\\
\Aut_U([5,0,1])&=\Aut_U([2,1,3])=\mathbb Z_2=\langle\mathsf C\rangle\,.
\end{aligned}
\end{equation}

These are the only $2\times2$ congruence classes with determinant equal to $5$.

%\textbf{Note}: these are not Lagrangian symmetries (because of the central charge), but may become so if we add $\diag(-1,-1)$. As $5$ is Pell, $[5,0,1]$ is Lagrangian (the theory is dual to $U(1)_5\times U(1)_{-1}$). The other case is more complicated: I haven't been able to find a $\mathrm{GL}_4$ change of variables, nor to prove that it does not exist.

\end{example}

\begin{example}[$\det(K)=9$]\normalfont
The next positive-definite time-reversal invariant theory is $K=[3,0,3]$, where the permutations are
\begin{equation}
\pm Q=\begin{pmatrix}1&4\\2&1\end{pmatrix},\begin{pmatrix}4&2\\1&1\end{pmatrix}
\end{equation}
all of which satisfy $\mathsf T^4=\mathsf C$ (and are thus of order $8$), and
\begin{equation}
\pm Q=\begin{pmatrix}2&4\\1&1\end{pmatrix},\begin{pmatrix}4&1\\1&2\end{pmatrix}
\end{equation}
all of which satisfy $\mathsf T^2=\mathsf C$ (and are thus of order $4$).

Similarly, the non-trivial unitary symmetries are
\begin{equation}
\pm Q=\begin{pmatrix}0&5\\1&0\end{pmatrix},\begin{pmatrix}0&1\\1&0\end{pmatrix},\begin{pmatrix}5&0\\0&1\end{pmatrix}
\end{equation}
the first one of which squares to $\mathsf C$ (and is thus of order $4$), and the other two are of order $2$.

As it turns out, all these symmetries can be generated from just the two matrices
\begin{equation}
\mathsf T\colon\begin{pmatrix}1&4\\2&1\end{pmatrix},\quad\mathsf U_1\colon\begin{pmatrix}0&1\\1&0\end{pmatrix}
\end{equation}
which satisfy $\mathsf T^8=\mathsf U_1^2=1$ and $\mathsf U_1\mathsf T=\mathsf T^3\mathsf U_1$, and so the group of symmetries is semidihedral, $SD_{16}\cong\mathbb Z_8\rtimes\mathbb Z_2$. Similarly, the matrices $\mathsf U_1$ and $\mathsf U_2:=\mathsf T^2$ satisfy $\mathsf U_2^4=\mathsf U_1^2=1$ and $\mathsf U_2\mathsf U_1\mathsf U_2=\mathsf U_1$, and generate the whole group of unitary symmetries, and so the latter is dihedral, $D_8\cong\mathbb Z_4\rtimes\mathbb Z_2$. All in all, the group of symmetries is
\begin{equation}
\begin{aligned}
\Aut([3,0,3])&=SD_{16}=\langle \mathsf T,\mathsf U_1\rangle\\
\Aut_U([3,0,3])&=D_8=\langle \mathsf U_1,\mathsf U_2\rangle\,.
\end{aligned}
\end{equation}

The rest of binary forms with $\det(K)=9$ are $K=[1,0,9]$ and $K=[2,1,5]$, neither of which is time-reversal invariant. They have no non-trivial unitary symmetries either.

\end{example}

\begin{example}[$\det(K)=12$]\normalfont

The matrix $K=[4,2,4]$ has a unitary symmetry with $\mathsf U^6=1$ (and no time-reversal).

The non-trivial permutations are
\begin{equation}
\pm Q=\begin{pmatrix}1&3\\1&2\end{pmatrix},\begin{pmatrix}2&1\\3&1\end{pmatrix},\begin{pmatrix}1&2\\1&3\end{pmatrix},\begin{pmatrix}2&3\\3&2\end{pmatrix},\begin{pmatrix}3&1\\2&1\end{pmatrix}
\end{equation}
which are of order $6,6,2,2,2$, respectively.

The whole group can be generated from one of the order $6$ permutations, and one of the order $2$ ones. They satisfy $\mathsf U_1^6=\mathsf U_2^2=1$, together with $\mathsf U_1\mathsf U_2\mathsf U_1=\mathsf U_2$, and so the group structure is dihedral:
\begin{equation}
\Aut([4,2,4])=\Aut_U([4,2,4])=D_{12}=\langle \mathsf U_1,\mathsf U_2\rangle\,.
\end{equation}

The rest of binary forms of the same determinant are $K=[1,0,12]$, $K=[2,0,6]$, and $K=[3,0,4]$, none of which is time-reversal invariant. One has
\begin{equation}
\begin{aligned}
\Aut([1,0,12])&=\Aut_U([1,0,12])=\mathbb Z_2\times\mathbb Z_2\\
\Aut([2,0,6])&=\Aut_U([2,0,6])=\mathbb Z_2\\
\Aut([3,0,4])&=\Aut_U([3,0,4])=\mathbb Z_2\times\mathbb Z_2\,.
\end{aligned}
\end{equation}

\end{example}

\begin{example}[$\det(K)=18$]\normalfont
The first positive-definite time-reversal invariant theory with a $\mathsf T$ such that $\mathsf T^2\neq\mathsf C$ and $\det(K)$ not a perfect square is $K=[3,0,6]$, where the permutations are
\begin{equation}
\pm Q=\begin{pmatrix}3&1\\2&3\end{pmatrix}
\end{equation}
both of which satisfy $\mathsf T^2=\mathsf C$ (and thus are of order $4$), and
\begin{equation}
\pm Q=\begin{pmatrix}3&1\\4&2\end{pmatrix}
\end{equation}
both of which are of order $2$, i.e. $\mathsf T^2=1$. If one chooses $\mathsf T^2=(-1)^F$, the latter admit a well-defined $\mathbb Z_{16}$ anomaly, which is easily evaluated to be $\pm2$.

The only non-trivial unitary symmetry is
\begin{equation}
\pm Q=\begin{pmatrix}1&0\\0&5\end{pmatrix}
\end{equation}
which is of order $2$.

If we denote by $\mathsf T$ one of the order $4$ time-reversal symmetries, and by $\mathsf U$ one of the unitary ones, then one may check that these two operations generate the whole group of symmetries. One has $\mathsf T^4=\mathsf U^2=1$ and $\mathsf T\mathsf U\mathsf T=\mathsf U$, and so
\begin{equation}
\begin{aligned}
\Aut([3,0,6])&=D_8=\langle \mathsf T,\mathsf U\rangle\\
\Aut_U([3,0,6])&=\mathbb Z_2\times\mathbb Z_2=\langle \mathsf C,\mathsf U\rangle\,.
\end{aligned}
\end{equation}

The rest of binary forms with $\det(K)=18$ are $K=[1,0,18]$ and $K=[2,0,9]$, neither of which is time-reversal invariant. They have no non-trivial unitary symmetries either.

\end{example}

\begin{example}[$\det(K)=49$]\normalfont
The first example of a time-reversal invariant theory where the order of the symmetry is greater than $8$ is $K=[7,0,7]$, where the permutations are
\begin{equation}
\pm Q=\begin{pmatrix}2&11\\3&2\end{pmatrix},\begin{pmatrix}3&12\\2&3\end{pmatrix},\begin{pmatrix}9&3\\4&2\end{pmatrix},\begin{pmatrix}10&2\\5&3\end{pmatrix},
\end{equation}
all of which satisfy $\mathsf T^8=\mathsf C$ (and thus are of order $16$), and
\begin{equation}
\pm Q=\begin{pmatrix}2&10\\3&5\end{pmatrix},\begin{pmatrix}3&9\\2&4\end{pmatrix},\begin{pmatrix}9&4\\4&5\end{pmatrix},\begin{pmatrix}10&5\\5&4\end{pmatrix},
\end{equation}
all of which satisfy $\mathsf T^2=\mathsf C$ (and thus are of order $4$).

The non-trivial unitary symmetries are
\begin{equation}
\pm Q=\begin{pmatrix}0&1\\1&0\end{pmatrix},\begin{pmatrix}13&0\\0&1\end{pmatrix},\begin{pmatrix}5&9\\2&2\end{pmatrix},\begin{pmatrix}9&2\\2&5\end{pmatrix},\begin{pmatrix}0&13\\1&0\end{pmatrix},\begin{pmatrix}9&5\\2&2\end{pmatrix},\begin{pmatrix}2&9\\5&2\end{pmatrix}
\end{equation}
which are of order $2,2,2,2,4,8,8$, respectively.

If we let $\mathsf T$ denote one of the order $16$ time-reversal permutations, and $\mathsf U_1$ one of the order $2$ unitary permutations, then one may check that these two operations generate the whole group. Furthermore, one has $\mathsf T^{16}=\mathsf U_1^2=1$ and $\mathsf U_1\mathsf T\mathsf U_1=\mathsf T^7$, and so the group is the semidihedral group of order $32$. On the other hand, if we let $\mathsf U_2$ be one of the order $8$ unitary symmetries, then one may check that these two operations generate the whole unitary group. One has $\mathsf U_2^8=\mathsf U_1^2=1$ and $\mathsf U_2\mathsf U_1\mathsf U_2=\mathsf U_1$, which is the dihedral group of order $16$. All in all, the group of symmetries is
\begin{equation}
\begin{aligned}
\Aut([7,0,7])&=SD_{32}=\langle \mathsf T,\mathsf U_1\rangle\\
\Aut_U([7,0,7])&=D_{16}=\langle \mathsf U_1,\mathsf U_2\rangle\,.
\end{aligned}
\end{equation}

The rest of binary forms with $\det(K)=49$ are $K=[1,0,49]$, $K=[2,1,25]$, and $K=[5,\pm 1,10]$, neither of which is time-reversal invariant. They have no non-trivial unitary symmetries either.

\end{example}

\begin{example}[$\det(K)=50$]\normalfont

Take for example $K=[5,0,10]$. The anti-unitary symmetries are
\begin{equation}
\pm Q=\begin{pmatrix}1&3\\4&1\end{pmatrix},\begin{pmatrix}1&7\\6&1\end{pmatrix},\begin{pmatrix}1&7\\4&9\end{pmatrix},\begin{pmatrix}1&3\\6&9\end{pmatrix},\begin{pmatrix}3&5\\0&3\end{pmatrix},\begin{pmatrix}3&5\\0&7\end{pmatrix}
\end{equation}
the first two of which satisfy $\mathsf T^6=\mathsf C$ (and are thus of order $12$), and the rest of which satisfy $\mathsf T^2=\mathsf C$ (and are thus of order $4$).

The non-trivial unitary symmetries are
\begin{equation}
\pm Q=\begin{pmatrix}3&4\\2&3\end{pmatrix},\begin{pmatrix}3&6\\8&3\end{pmatrix},\begin{pmatrix}3&6\\2&7\end{pmatrix},\begin{pmatrix}3&4\\8&7\end{pmatrix},\begin{pmatrix}1&0\\0&9\end{pmatrix}
\end{equation}
the first two of which satisfy $\mathsf U^3=\mathsf C$ (and are thus of order $6$), and the rest of which are of order $2$.

One may check that the three matrices
\begin{equation}
\mathsf T\colon\begin{pmatrix}3&5\\0&3 \end{pmatrix},\quad\mathsf U_1\colon\begin{pmatrix}7&4\\2&7 \end{pmatrix},\quad\mathsf U_2\colon\begin{pmatrix}1&0\\0&9 \end{pmatrix}
\end{equation}
generate the whole group, and satisfy $\mathsf T^4=\mathsf U_1^3=\mathsf U_2^2=[\mathsf T,\mathsf U_i]=(\mathsf U_1\mathsf U_2)^2=1$, and therefore
\begin{equation}
\begin{aligned}
\Aut([5,0,10])&=\mathbb Z_4\times D_6=\langle\mathsf T,\mathsf U_1,\mathsf U_2\rangle\\
\Aut_U([5,0,10])&=D_{12}=\langle\mathsf T^2\mathsf U_1,\mathsf U_2\rangle\,.
\end{aligned}
\end{equation}

The rest of binary forms with $\det(K)=50$ are $[6,\pm2,9]$, $[3,\pm1,17]$, $[1,0,50]$, and $[2,0,25]$, and they are all time-reversal invariant with symmetry group $\Aut(\,\star\,)=\mathbb Z_4=\langle\mathsf T\rangle$ and $\Aut_U(\,\star\,)=\mathbb Z_2=\langle\mathsf C\rangle$.

\end{example}

We now move on to $2\times2$ indefinite matrices.

\begin{example}[$\det(K)=-2$]\normalfont

The only binary form with $\det(K)=-2$ is $[1,1,-1]$, which contains four lines. The theory has no non-trivial unitary permutations, and one anti-unitary one, effected by
\begin{equation}
Q=\begin{pmatrix}2&1\\1&0\end{pmatrix}
\end{equation}
which squares to the identity. Therefore,
\begin{equation}
\begin{aligned}
\Aut([1,1,-1])&=\mathbb Z_2=\langle\mathsf T\rangle\\
\Aut_U([1,1,-1])&=0\,.
\end{aligned}
\end{equation}

When $\mathsf T^2=(-1)^F$, this symmetry admits a well-defined $\mathbb Z_{16}$ anomaly, which is easily evaluated to be $\nu=\pm2$.

\end{example}

\begin{example}[$\det(K)=-3$]\normalfont

The two binary forms are $K=[1,1,-2]$ and $K=[2,1,-1]$, neither of which admits an anti-unitary permutation. The unitary permutations are the trivial one, i.e.,
\begin{equation}
\begin{aligned}
\Aut([1,1,-2])&=\Aut_U([1,1,-2])=\mathbb Z_2=\langle \mathsf C\rangle\\
\Aut([2,1,-1])&=\Aut_U([2,1,-1])=\mathbb Z_2=\langle \mathsf C\rangle\,.
\end{aligned}
\end{equation}

\end{example}

\begin{example}[$\det(K)=-4$]\normalfont

All the matrices are of the twisted gauge theory type, $K=[0,2,k]$, with $k=0,1,2,3$. For $k$ odd there are no anti-unitary symmetries, while the unitary ones are trivial:
\begin{equation}
\Aut([0,2,k])=\Aut_U([0,2,k])=\mathbb Z_2=\langle\mathsf C\rangle,\qquad k=1,3\,.
\end{equation}

For $k$ even, there are anti-unitary symmetries. In particular, for $k=0$ we have the trivial permutation and the electric-magnetic duality $\mathsf e\leftrightarrow \mathsf m$, as is well known. There is also the unitary symmetry $\mathsf e\leftrightarrow \mathsf m$, which can be obtained from composing the two anti-unitary symmetries. Similarly, for $k=2$, the anti-unitary permutation is $\mathsf m\leftrightarrow \mathsf{em}$, and there are no unitary symmetries. In short,
\begin{equation}
\begin{aligned}
\Aut([0,2,0])&=\mathbb Z_2\times\mathbb Z_2=\langle \mathsf T,\mathsf T'\rangle\\
\Aut_U([0,2,0])&=\mathbb Z_2=\langle\mathsf T\mathsf T'\rangle\\
\Aut([0,2,2])&=\mathbb Z_2=\langle\mathsf T\rangle\\
\Aut_U([0,2,2])&=0\,.
\end{aligned}
\end{equation}

%\textbf{Question}: what are the time-reversal anomalies? In the $\mathsf T^2=1$ case, we have a $\mathbb Z_2\times\mathbb Z_2$-valued anomaly, which is easily shown to be $(1,\pm1)$. What about the other case?

\end{example}

\begin{example}[$\det(K)=-5$]\normalfont

The representatives are $K=[2,1,-2]$ and $K=[1,2,-1]$. They both have a $\mathsf T^2=\mathsf C$ permutation, and no non-trivial unitary symmetries. In other words,
\begin{equation}
\begin{aligned}
\Aut([2,1,-2])&=\Aut([1,2,-1])=\mathbb Z_4=\langle\mathsf T\rangle\\
\Aut_U([2,1,-2])&=\Aut_U([1,2,-1])=\mathbb Z_2=\langle\mathsf C\rangle\,.
\end{aligned}
\end{equation}

\end{example}

\begin{example}[$\det(K)=-9$]\normalfont

All the matrices are of the twisted gauge theory type, $K=[0,3,k]$, with $k=0,\dots,5$. There are anti-unitary symmetries only for $k=0,3$:
\begin{equation}
\begin{aligned}
\Aut([0,3,k])&=D_8\\
\Aut_U([0,3,k])&=\mathbb Z_2\times\mathbb Z_2\\
\end{aligned},\qquad k=0,3
\end{equation}
while for the rest of levels the only symmetry is charge conjugation:
\begin{equation}
\Aut([0,3,k])=\Aut_U([0,3,k])=\mathbb Z_2,\qquad k=1,2,4,5\,.
\end{equation}

\end{example}

\begin{example}[$\det(K)=-18$]\normalfont

The first example with time-reversal with order greater than $4$ is $K=[3,3,-3]$, whose anti-unitary permutations read
\begin{equation}
\pm Q=\begin{pmatrix}2&5\\-1&0\end{pmatrix},\begin{pmatrix}6&7\\1&2\end{pmatrix},\begin{pmatrix}2&5\\-1&4\end{pmatrix},\begin{pmatrix}6&5\\1&0\end{pmatrix}
\end{equation}
(which are of order $8,8,4,4$), and whose non-trivial unitary permutations read
\begin{equation}
\pm Q=\begin{pmatrix}5&4\\4&1\end{pmatrix},\begin{pmatrix}1&2\\0&-1\end{pmatrix},\begin{pmatrix}5&6\\4&1\end{pmatrix}
\end{equation}
(which are of order $4,2,2$). It is a simple exercise to check that
\begin{equation}
\begin{aligned}
\Aut([3,3,-3])&=SD_{16}\\
\Aut_U([3,3,-3])&=D_8\,.
\end{aligned}
\end{equation}

The rest of the binary forms with the same determinant are $[1,4,-2]$ and $[2,4,-1]$, which have $\Aut(\,\star\,)=\Aut_U(\,\star\,)=\mathbb Z_2=\langle\mathsf C\rangle$.

\end{example}

\begin{example}[$\det(K)=-20$]\normalfont

The next interesting example is $K=[4,2,-4]$, which has 
\begin{equation}
\begin{aligned}
\Aut([4,2,-4])&=\mathbb Z_4\times D_6\\
\Aut_U([4,2,-4])&=D_{12}\,.
\end{aligned}
\end{equation}

The rest of binary forms with the same determinant are $[2,4,-2]$, which has $\Aut(\,\star\,)=\mathbb Z_4$ and $\Aut_U(\,\star\,)=\mathbb Z_2$, and $[1,4,-4]$ and $[4,4,-1]$, which have $\Aut(\,\star\,)=\Aut_U(\,\star\,)=\mathbb Z_2\times\mathbb Z_2$.

\end{example}

\begin{example}[$\det(K)=-27$]\normalfont

Another interesting example is the pair $K=[3,3,-6]$, $K=[6,3,-3]$, which has 
\begin{equation}
\Aut(\,\star\,)=\Aut_U(\,\star\,)=D_{12}\,.
\end{equation}

The rest of binary forms with the same determinant are $[1,5,-2]$ and $[2,5,-1]$, which have $\Aut(\,\star\,)=\Aut_U(\,\star\,)=\mathbb Z_2$.

\end{example}

\begin{example}[$\det(K)=-49$]\normalfont

As $49$ is a perfect square, these matrices are of the twisted gauge theory type. One has
\begin{equation}
\begin{aligned}
\Aut([0,7,k])&=\mathbb Z_3\rtimes D_8\\
\Aut_U([0,7,k])&=D_{12}
\end{aligned}
\end{equation}
if $k\propto 7$, and
\begin{equation}
\Aut([0,7,k])=\Aut_U([0,7,k])=\mathbb Z_2
\end{equation}
otherwise.

\end{example}

\begin{example}[$\det(K)=-121$]\normalfont

The next interesting example is, again, of the twisted gauge theory type. One has
\begin{equation}
\begin{aligned}
\Aut([0,11,k])&=\mathbb Z_5\rtimes D_{8}\\
\Aut_U([0,11,k])&=D_{20}
\end{aligned}
\end{equation}
if $k\propto 11$, and
\begin{equation}
\Aut([0,11,k])=\Aut_U([0,11,k])=\mathbb Z_2
\end{equation}
otherwise.

\end{example}

Finally, we consider a few higher-dimensional examples, chosen at random:

\begin{example}[$\det(K)=16$]\normalfont

The theory with matrix
\begin{equation}
K=\begin{pmatrix}3&-1&-1\\-1&3&-1\\-1&-1&3\end{pmatrix}
\end{equation}
has
\begin{equation}
\begin{aligned}
\Aut(\,\star\,)&=A_4\rtimes D_8\\
\Aut_U(\,\star\,)&=\mathbb Z_2\times S_4\,.
\end{aligned}
\end{equation}

\end{example}

\begin{example}[$\det(K)=36$]\normalfont

The theory with matrix
\begin{equation}
K=\begin{pmatrix}3&0&0\\0&4&2\\0&2&4\end{pmatrix}
\end{equation}
has
\begin{equation}
\begin{aligned}
\Aut(\,\star\,)&=S_3\times SD_{32}\\
\Aut_U(\,\star\,)&=S_3\times D_8\,.
\end{aligned}
\end{equation}

\end{example}

\begin{example}[$\det(K)=48$]\normalfont

The theory with matrix
\begin{equation}
K=\begin{pmatrix}1&0&0\\0&8&4\\0&4&8\end{pmatrix}
\end{equation}
has
\begin{equation}
\Aut(\,\star\,)=\Aut_U(\,\star\,)=\mathbb Z_2^2\times S_4\,.
\end{equation}

\end{example}

\vfill

\section*{Acknowledgments}

We would like to thank Clay C\'ordova, Dan Freed, Davide Gaiotto, Po-Shen Hsin, Theo Johnson-Freyd, Anton Kapustin, Zohar Komargodski, Nathan Seiberg, Ryan Thorngren, Senthil Todadri, Chong Wang and Jon Yard for useful discussions. The research of D.D.~and J.G.~was supported by the Perimeter Institute for Theoretical Physics. Research at Perimeter Institute is supported by the Government of Canada through Industry Canada and by the Province of Ontario through the Ministry of Economic Development and Innovation. Any opinions, findings, and conclusions or recommendations expressed in this material are those of the authors and do not necessarily reflect the views of the funding agencies.

\clearpage
\appendix

\section{Notation and definitions.}\label{ap:grupo}
\label{app:not}

For the convenience of the reader, we gather here some common definitions we use throughout the text.

We denote by $\mathbb Z:=\{0,\pm1,\pm2,\dots\}$ the set of all integers, and by $\mathbb T,\mathbb P$ the two subsets
\begin{equation}
\begin{aligned}
\mathbb T&:=\{k\in\mathbb Z\,\mid\, kp-q^2=1\text{ for some $p,q\in\mathbb Z$}\}\\
\mathbb P&:=\{k\in\mathbb Z\,\mid\, kp^2-q^2=1\text{ for some $p,q\in\mathbb Z$}\}\,.
\end{aligned}
\end{equation}
One has $\mathbb P\subset\mathbb T\subset\mathbb Z$.

All primes greater than $2$ are odd, and so they can be written as $4n\pm1$ for some integer $n$. Those of the form $4n+1$ are called \emph{Pythagorean} (because they can be written as the sum of two squares, unlike those of the form $4n-1$, as per Fermat's theorem).

The function $\phi\colon\mathbb Z\to\mathbb Z$ denotes the Euler totient function: $\phi(k)$ is the number of integers $q$ such that $0<q<k$ and $\gcd(q,k)=1$, where $\gcd$ denotes the greatest common divisor. In other words, there are $\phi(k)$ integers smaller than $k$ that are coprime to it. This function is multiplicative, $\phi(ab)=\phi(a)\phi(b)$ for any $a,b\in\mathbb Z$ with $\gcd(a,b)=1$, and is given by $\phi(\pi^n)=\pi^{n-1}(\pi-1)$ for prime $\pi$ and integer $n$.

The function $\omega\colon\mathbb Z\to\mathbb Z$ counts the number of distinct prime factors, i.e.~the prime decomposition of a given $k\in\mathbb Z$ reads
\begin{equation}
k\equiv\prod_{i=1}^{\omega(k)}\pi_i^{n_i}\,.
\end{equation}
We also denote $\varpi(k):=\omega(k)$ if $k$ is odd, and $\varpi(k):=\omega(k/2)$ if even. For example,
\begin{equation}
\begin{aligned}
\omega(1)&=0,\quad\omega(2)=\omega(3)=\omega(4)=\omega(5)=1,\quad \omega(6)=2,\dots\\
\varpi(1)&=\varpi(2)=0,\quad\varpi(3)=\varpi(4)=\dots=\varpi(11)=1,\quad\varpi(12)=2,\dots
\end{aligned}
\end{equation}
The function $\mu\colon\mathbb Z\to\mathbb Z$ denotes the operation of removing the Pythagorean prime factors:
\begin{equation}
\mu(1)=1,\quad\mu(2)=2,\quad\mu(3)=3,\quad\mu(4)=4,\quad\mu(5)=1,\quad\mu(6)=6,\dots
\end{equation}
One has $k\in\mathbb T$ if and only if $\mu(k)=1$ or $\mu(k)=2$. The function $\lambda\colon\mathbb Z\to\mathbb Z$ denotes the squarefree part (i.e., $\lambda(k)$ is the smallest divisor of $k$ such that $k/\lambda(k)$ is a perfect square):
\begin{equation}
\lambda(1)=1,\quad\lambda(2)=2,\quad\lambda(3)=3,\quad\lambda(4)=1,\quad\lambda(5)=5,\dots,\lambda(8)=2,\dots
\end{equation}

We denote by $\mathbb Z^{n\times n}$ the set of all integral $n\times n$ matrices, and by $\mathrm{GL}_n(\mathbb Z)\subset \mathbb Z^{n\times n}$ the subset of invertible matrices over $\mathbb Z$. A given matrix is invertible over $\mathbb Z$ if and only if its determinant is $\pm1$, and so the elements of $\mathrm{GL}_n(\mathbb Z)$ are known as \emph{unimodular matrices}.

Given some set $A$ with some extra structure $\sigma$, we denote by $\Aut(A,\sigma)\subseteq S_A$ the set of all permutations of $A$ that ``respect'' the structure $\sigma$, and whose group operation is that inherited from $S_A$ (i.e., composition). For example, if $\times\colon A\times A\to A$ is a binary product such that $(A,\times)$ is a group, then $\Aut(A,\times)$ is the set of permutations that are group homomorphisms. Similarly, if $A$ is a group and $\theta\colon A\to U(1)$ is a quadratic form on it, $\Aut(A,\theta)$ denotes the set of automorphisms of $A$ that leave $\theta$ invariant, perhaps up to complex conjugation: $\theta(\pi(a))=\theta(a)^{\pm1}$ for all $a\in A$ and $\pi\in\Aut(A)$. If the data $(A,\theta)$ comes from a Chern-Simons theory with matrix $K$, we also use the notation $\Aut(K)\equiv\Aut(A,\theta)$, or even $\Aut(U(1)_k)$ in the $1\times1$ case.

Given some unital ring $A$, we denote by $A^\times$ the group of units of $A$ -- the set of its invertible elements. For example, one has $\mathrm{GL}_n(\mathbb Z)\equiv (\mathbb Z^{n\times n})^\times$.

The group $\mathbb Z_k$ denotes the cyclic group of order $k$, which consists of the set $\{0,1,\dots,k-1\}$, where the product operation is just addition, followed by reduction modulo $k$. One can also endow $\mathbb Z_k$ with integer product, which makes it into a ring (integer product is not usually invertible); the group of units is denoted by $\mathbb Z_k^\times$, and its order is $\phi(k)$. %see lemma 2.1.9 in McKay, page 39

We also recall some basic definitions from group theory, following~\cite{leedham2002structure}.

\paragraph{Definition 2.1.3} Let $N$ and $G$ be groups. Then an \emph{action} of $G$ on $N$ is a homomorphism $\theta\colon G\to \Aut(N)$. This is described by saying that $G$ \emph{acts} on $N$ or that $N$ is a $G$-\emph{group}.

\paragraph{Definition 2.1.4} Let $G$ and $N$ be groups such that $G$ acts on $N$ with action given by $\theta$. Then the \emph{semi-direct product $N\rtimes_\theta G$ of $N$ by $G$ with this action} is defined as follows. The underlying set of $N\rtimes_\theta G$ is $G\times N$ and the multiplication is defined by $(g_1,n_2)(g_2,n_2)=(g_1g_2,(n_1^{g_2\theta})n_2)$.

%(Given some $x,y\in G$, we denote by $y^x$ the element $x^{-1}yx\in G$, cf.~definition 1.1.1)

\paragraph{Definition 2.2.6} [...] A group $G$ is an \emph{external central product} $H\circ K$ of two groups $H$ and $K$ if there exists an isomorphism $\theta\colon Z(H)\to Z(G)$ such that $G$ is $(H\times K)/N$ where $N=\{(h,h^{-1}\theta)\,\mid\,h\in Z(H)\}$.

\paragraph{Definition 2.3.1} Let $G$ be a group and $\Omega$ a non-empty finite set. Then $G$ \emph{acts on} $\Omega$ if, to each $\omega\in\Omega$ and $g\in G$, there corresponds a unique element $\omega^g\in \Omega$ such that, if $g_1$ and $g_2\in G$ then $(\omega^{g_1})^{g_2}=\omega^{g_1g_2}$; and $\omega^1=\omega$. If $G$ acts on $\Omega$ then the \emph{permutation representation} of $G$ corresponding to the action is the homomorphism $\rho\colon G\to \Sigma_\Omega$, the symmetric group on $\Omega$, defined by $\omega(g\rho)=\omega^g$ for all $\omega\in\Omega$ and all $g\in G$.

\paragraph{Definition 2.3.2} Let $H$ be a group and $\Omega$ a non-empty finite set. Then $H^\Omega$ denotes the set of all maps from $\Omega$ to $H$. For $f_1,f_2\in H$, define $f_1f_2\in H^\Omega$ by $\omega(f_1f_2)=(\omega f_1)(\omega f_2)$ for all $\omega\in\Omega$.

\paragraph{Definition 2.3.3} Let $H$ be a group, and $G$ be a finite group acting on a non-empty finite set $\Omega$. Then an action of $G$ on the group $H^\Omega$ is defined as follows. For each $g\in G$ and $f\in H^\Omega$, define $f^g\in H^\Omega$ by $\omega f^g=\omega^{g^{-1}}f$ for all $\omega\in\Omega$. The \emph{(permutational) wreath product} $H\wr G$ of $H$ with $G$ corresponding to this action of $G$ on $\Omega$ is the split extension $H^\Omega\rtimes G$ with this action of $G$ on $H^\Omega$.

%Given two groups $A,B$, the group $A\times B$ denotes the group whose underlying set is $A\times B$ (the Cartesian product), and whose binary operation is $(a,b)\cdot (a',b'):=(a\cdot a',b\cdot b')$. If $A$ is a normal subgroup of $G$, and $B:=G/A$, then we write $G\equiv A\rtimes B$, called the semidirect product of $A,B$, or the (split) extension of $B$ by $A$. If $A,B$ do not specify $G$ uniquely (i.e., if there is more than one extension), one usually indicates which of the groups one refers to via a subindex: $A\ltimes i B$, where $i=1,2,\dots$ labels the different possible extensions (following e.g.~the ordering in GAP, reference ??).
%
%
%%From H. Neumann, "Varieties of groups" , Springer (1967):
%Similarly, if $A,B$ are two groups, we denote by $A\wr B$ (the \emph{wreath product} of $A$ by $B$) the semidirect product of $A^B$ and $B$, where $A^B$ denotes the group of all maps $A\to B$ under composition, and where $B$ acts on $A^B$ as follows:
%\begin{equation}
%b\colon \varphi\mapsto \varphi^b\qquad \text{such that}\quad\varphi^b(y):=\varphi(yb^{-1})\quad\forall y\in B
%\end{equation}
%where $b\in B$ and $\varphi\in A^B$. %The map $b\colon\varphi\mapsto\varphi^b$ defines an automorphism of $A^B$, and the set of all such maps is isomorphic to $B$.
%The underlying set of $A\wr B$ is $A^B\times B$, and the group product is given by
%\begin{equation}
%(\varphi,b)\cdot (\varphi',b'):=(\varphi^{b'}\circ\varphi',b\cdot b')
%\end{equation}

Finally, we define a few important finite groups (see e.g.~Definition 2.1.11 in~\cite{leedham2002structure}):
\begin{itemize}
\item The \emph{dihedral group} $D_{2n}$ of order $2n$ is defined by
\begin{equation}
D_{2n}=\langle x,y\,\mid\, y^n=x^2=(xy)^2=1\rangle\cong \mathbb Z_n\rtimes\mathbb Z_2
\end{equation}
\item The \emph{semidihedral group} $SD_{2^{n+1}}$ of order $2^{n+1}$ is defined by
\begin{equation}
SD_{2^{n+1}}=\langle x,y\,\mid\, y^{2^n}=x^2=(xy)^2y^{2^{n-1}}=1\rangle
\end{equation}
\item The \emph{symmetric group} $S_n$ of order $n!$, corresponding to all the permutations of $n$ objects, and its commutator subgroup $A_n$, of order $n!/2$, known as the \emph{alternating group} and given by the even permutations of $S_n$. One has $S_n=A_n\rtimes\mathbb Z_2$ for $n\ge5$.

\end{itemize}

\section{Further results.}\label{app:further}

In this appendix we collect some further results concerning the theory $U(1)_k$ which may prove useful in subsequent studies of this system. We begin by making some remarks concerning the set $\mathbb T$, defined as those integers $k$ such that $-1$ is a quadratic residue modulo $k$, i.e., those integers for which the equation $q^2=-1+pk$ is solvable for some integers $p,q$.

It is straightforward to show that any solution $(p,q)$ is such that $q$ is congruent to $q_0$ modulo $k$, where $(p_0,q_0)$ is a solution with $q_0\in[0,k)$. More precisely, if $(p_0,q_0)$ is a solution, then so is $(P(n),Q(n))$ for any $n\in \mathbb Z$, where
\begin{equation}\label{eq:poly_PQ}
\begin{aligned}
P(n)&:=p_0+2q_0n+kn^2\\
Q(n)&:=q_0+kn
\end{aligned}
\end{equation}
as is easily checked. This is not particular to our problem; the solutions to congruences of the form $f(q)=0\mod k$, for some polynomial $f\colon\mathbb Z\to\mathbb Z$, are always defined modulo $k$.

Generically speaking, this type of congruences are solved by first solving them modulo the prime divisors of $k$. Indeed, if $k$ is to divide $f(q)$, then so must its divisors. This means that the prime divisors of $k$ are essential in deciding whether $q^2+1=0\mod k$ is solvable or not. To be precise, one of the key results concerning the set $\mathbb T$ is the following:
\begin{proposition}\label{th:prime_factors}
A given $k$ is in $\mathbb T$ if and only if all its prime factors are Pythagorean (that is, congruent to $1$ modulo $4$), perhaps up to a single factor of $2$.
\end{proposition}
 
\emph{Proof}. By reducing $kp=1+q^2$ modulo $4$, and considering the odd $q$ and even $q$ cases separately, it becomes clear that $k$ cannot be a multiple of $4$. Similarly, by Gaussian reciprocity, $-1$ is a quadratic residue modulo a prime $\pi$ if and only if $\pi$ is Pythagorean, and so $k$ cannot be a multiple of a non-Pythagorean prime either. This proves that the conditions above are necessary; proving that they are also sufficient can be done by explicitly constructing a solution $q$. We now sketch how this can be done.

First off, if $k$ is a Pythagorean prime, we can use Wilson's theorem to obtain an explicit expression for $q$. Indeed,
\begin{equation}\label{eq:wilson}
q=\left(\frac{k-1}{2}\right)!
\end{equation}
satisfies $q^2=-1\mod k$. One can also take
\begin{equation}
q=(k-a)!!
\end{equation}
where $a$ is any of $\{\pm1,2,3\}$.

Lifting the solution to a prime power $k=\pi^n$ can be done using the Hensel lemma. If we let $q_1$ be the solution for $n=1$, then the general solution can be obtained via the quadratic map
\begin{equation}\label{eq:hensel}
q_n=q_{n-1}-a(q_{n-1}^2+1)
\end{equation}
where $a$ is a solution to $2q_1a=1\mod \pi$ (e.g., $a=(2q_1)^{\pi-2}$, as per Fermat's little theorem).

Finally, finding a solution for arbitrary $k$ requires the use of the Chinese Remainder Theorem. %Note that $q+nk$ also satisfies $q^2\equiv-1\mod k$, for any $n\in\mathbb Z$ (cf.~\ref{lm:range_0k}). If $k$ is odd, one may need to adjust $n$ so that the solution lies in the range $[0,2k)$ and is odd.
For example, let $k=a_1a_2$ with $a_1,a_2$ two prime powers. Then $q^2=-1\mod k$ requires $q^2=-1\mod a_i$, which by the previous paragraph has a solution $q_i$. With this, the solution of $q^2=-1\mod k$ is $q= q_1\alpha_1a_2+q_2\alpha_2a_1\mod k$, where $\alpha_1,\alpha_2$ are the B\'ezout coefficients for $a_1,a_2$ (i.e., a pair of integers such that $a_1\alpha_1+a_2\alpha_2=1$, which can be computed using the Euclidean algorithm). By iteration we can easily find the solutions for an arbitrary integer $k=a_1a_2\dots a_n$, and so the conditions in proposition~\ref{th:prime_factors} are also sufficient.\hfill$\square$

\medskip

The integers $q$ that solve $q^2= -1\mod k$ implement the time-reversal permutations on the anyons of $U(1)_k$. The lines $a\in\mathcal A$ that are fixed under time-reversal (modulo local operators) play a special role in analysing the time-reversal symmetry of a system (and its anomalies), see e.g.~\cite{Chan_2016,Barkeshli_2019}. We have the following:
\begin{proposition}
The only lines that satisfy $\mathsf T(a)\equiv a$ are the identity and the transparent fermion. If $k$ is odd, no line satisfies $\mathsf T(a)=a\times\psi$, while if $k$ is even, the only lines satisfying $\mathsf T(a)=a\times\psi$ are $a=k/2\times\boldsymbol 1$ and $a=k/2\times\psi$.
\end{proposition}

\emph{Proof}. Any line fixed by $\mathsf T$ (perhaps up to $\psi$) has $a=\mathsf T^2(a)=\mathsf C(a)$. Let $k$ be odd; then lines fixed by $\mathsf C$ satisfy $2\alpha=0\mod 2k$, that is, $\alpha\propto k$. Both lines $\alpha=0,k$ have $\mathsf T(\alpha)=\alpha$, and so there are no lines with $\mathsf T(a)=a\times\psi$.

Now let $k$ be even; then lines fixed by $\mathsf C$ satisfy $2\alpha=0\mod k$, that is, $\alpha\propto k/2$. One may check that $a=(0,\beta)$ satisfies $\mathsf T(a)=a$, and $a=(k/2,\beta)$ satisfies $\mathsf T(a)=a\times\psi$. \hfill$\square$

We thus see that the property $\mathsf T^2=\mathsf C$ implies that the set of lines that are fixed by time-reversal is very small. More generally, it is possible to argue that, due to $\theta(\mathsf T(a))=\theta(a)^*$, an anyon can only be fixed by $\mathsf T$ (perhaps up to $\psi$) if its spin is either $\theta(a)=\pm1$ or $\theta(a)=\pm i$, i.e., if $h\in\{0,\tfrac14,\tfrac12,\tfrac34\}$. These are the bosons, fermions, semions, and anti-semions of the theory. For some purposes, it may be useful to know how many of these lines the theory supports. We have the following:
\begin{proposition}\label{lm:count}
Let $k\in \mathbb Z$, and denote by $N_h$ the number of lines of spin $h$ in $U(1)_k$ (as a spin TQFT), and by $\lambda(k)$ the squarefree part of $k$. Then we have $N_0=N_{1/2}= \sqrt{k/\lambda(k)}$. Furthermore, if we write $k=2^e\tilde k$, with $\tilde k$ odd, then $N_{1/4}=N_{3/4}=0$ if $e$ is even, and $N_{1/4}=N_{3/4}=\sqrt{k/\lambda(k)}$ if odd.
\end{proposition}

\emph{Proof}. We shall need the following trivial fact: given some integer $k\in\mathbb Z$, all solutions to the equation %http://oeis.org/A007913
\begin{equation}
\alpha^2=k\beta,\qquad \alpha,\beta\in\mathbb Z
\end{equation}
are of the form $(\alpha_n,\beta_n)=(n\sqrt{k\lambda(k)},n^2\lambda(k))$ for some integer $n$. Indeed, if $k\beta$ is to be a perfect square, then $\beta$ must be proportional to $\lambda(k)$; and the constant of proportionality must itself be a perfect square.

We next count the bosons and fermions of $U(1)_k$.

We begin with the $k$ odd case. An anyon $\alpha\in[0,2k)$ has vanishing spin iff $\alpha^2=2k\beta$ for some integer $\beta$. All the solutions to this equation are of the form $\alpha=n\sqrt{2k\lambda(2k)}$ for some integer $n=0,1,\dots,\lfloor \frac{2k-1}{\sqrt{2k\lambda(2k)}}\rfloor$. Therefore, there are
\begin{equation}
\left\lfloor \frac{2k-1}{\sqrt{2k\lambda(2k)}}\right\rfloor+1\equiv \sqrt{\frac{k}{\lambda(k)}}
\end{equation}
bosons. Similarly, the fermions are given by the solutions to $\alpha^2=k(2\beta+1)$, that is, $\alpha=n\sqrt{k\lambda(k)}$ with $n=1,3,\dots,\lfloor \frac{2k-1}{\sqrt{k\lambda(k)}}\rfloor$. Therefore, there are
\begin{equation}
\frac12\left(\left\lfloor \frac{2k-1}{\sqrt{k\lambda(k)}}\right\rfloor+1\right)\equiv \sqrt{\frac{k}{\lambda(k)}}
\end{equation}
fermions.

We now move on the the $k$ even case. The bosons in the spin theory come from the bosons and fermions in the non-spin theory. The former solve $\alpha^2=2k\beta$ and the latter solve $\alpha^2=k(2\beta+1)$. Together, they solve $\alpha^2=k\beta$, that is, $\alpha=n\sqrt{k\lambda(k)}$, with $n=0,1,\dots,\lfloor\frac{k-1}{\sqrt{k\lambda(k)}}\rfloor$. Therefore, there are
\begin{equation}
\left\lfloor \frac{k-1}{\sqrt{k\lambda(k)}}\right\rfloor+1\equiv \sqrt{\frac{k}{\lambda(k)}}
\end{equation}
bosons. The counting of the fermions is identical.

A very similar argument proves the claim for the semions. For $k$ odd, the counting is straightforward. For $k$ even, one is to count the spin $1/4$ and $3/4$ lines in the bosonic theory, which solve $2\alpha^2=k(2p+1)$. Writing $k=2^e\tilde k$, with $\tilde k$ odd, it is clear that no solutions exist for $e$ even (because $\sqrt 2$ is not integral). For $e$ odd, the solution is $\alpha=2^{(e-1)/2}n\sqrt{\tilde k\lambda(\tilde k)}$, with $n=1,3,\dots,\lfloor\frac{2^e\tilde k-1}{2^{(e-1)/2}\sqrt{\tilde k\lambda(\tilde k)}}\rfloor$. Thus, there are
\begin{equation}
\frac12\left(\left\lfloor\frac{2^e\tilde k-1}{2^{(e-1)/2}\sqrt{\tilde k\lambda(\tilde k)}}\right\rfloor+1\right)\equiv \sqrt{\frac{k}{\lambda(k)}}
\end{equation}
spin $h=1/4$ lines in the spin theory, and as many spin $3/4$ lines. \hfill$\square$

A similar technique can be applied to counting other lines $N_h$. %For example, $N_{-1/2k}$ counts the anti-unitary symmetries, and $N_{1/2k}$ the unitary ones; these numbers are given by $2$ to a power of $\omega(k)$, as discussed above.
%$N_{\frac{2k-1}{2k}}=2^{\varpi(k)}$ if $k\in\mathbb T$, and zero otherwise. \textbf{Question}: what about other values of $h$? what is $N_{1/2k}$? Are other cases where interesting number-theoretic functions appear?

%\begin{remark}\normalfont
%If we blindly apply the anomaly indicator formula of~\cite{PhysRevLett.119.136801,Tachikawa:2016nmo} to the spin TQFT $U(1)_k$, we get $\mathrm e^{2\pi i \nu/16}=1/\sqrt{\lambda(k)}$ for $k$ odd, and $\mathrm e^{2\pi i \nu/16}=(1\pm i)/\sqrt{\lambda(k)}$ for $k$ even. There is no reason to expect this formula is in any way valid for a $\mathbb Z_4^{\mathsf T}$ symmetry; but it is interesting to note that, in the case of $k$ a perfect square, it predicts that $\nu\equiv0$, which is consistent with the fact that no anomalies are to be expected in this case. It also predicts that, if $k$ is twice a perfect square, then $\nu=\pm2$, although it is not clear whether this result is meaningful at all. In any case, it is not unreasonable to expect that a similar indicator formula should hold for $\mathbb Z_4^{\mathsf T}$ anomalies, in which case we hope proposition~\ref{lm:count} will prove useful in computing them.
%\end{remark}

We now move on to the so-called Pell numbers:
\begin{definition}\label{def:pell}
\normalfont An integer $k$ is said to be \emph{Pell} if there exists a pair of integers $p,q$ such that $kp^2-q^2=1$. The set of Pell numbers is denoted by $\mathbb P$.
\end{definition}

We include here some known facts about Pell numbers, the first few of which are $k=1, 2, 5, 10, 13, 17, 26, 29,\dots$:
\begin{itemize}

\item No perfect square other than $1$ is ever Pell. (Indeed, $n^2-m^2>2m$ for $n>m>0$, and so this expression cannot equal $1$).

\item All Pell numbers are in $\mathbb T$ (but the converse is not true; the first few exceptions are $\mathbb T\setminus\mathbb P=\{25, 34, 146, 169, 178, 194,\dots\}$).

\item A squarefree integer $k$ is Pell iff the fundamental unit $\sigma$ of $\mathbb Q(\sqrt k)$ has norm $-1$. The rest of units are of the form $\pm\sigma^n$ for some integer $n$ (see e.g.~\cite{alaca_williams_2003}, theorem 11.4.1).

\item $k$ is Pell iff the convergents of $\sqrt k$ have odd period. If $(p_0,q_0)$ denotes the fundamental solution, then the rest of solutions are $q_n+p_n\sqrt k=(q_0+p_0\sqrt k)^{2n+1}$ (see e.g.~\cite{Mollin:2008:FNT:1628707}, theorems 5.15 and 5.16). Equivalently,
\begin{equation}
\begin{pmatrix}p_n\\q_n\end{pmatrix}=\begin{pmatrix}q_0&p_0\\kp_0&q_0\end{pmatrix}^{\!2n}\begin{pmatrix}p_0\\q_0\end{pmatrix}
\end{equation}
(Note that the determinant of this matrix is $-1$, and so its odd powers generate \emph{positive} norm units).

\item $k$ is Pell iff it can be written as $k=a^2+b^2$ for relatively prime $a,b\in\mathbb Z$, with $b$ odd, and such that the Gauss-type Diophantine equation $b(V^2-W^2)-2aVW=1$ is solvable with $V,W\in\mathbb Z$~\cite{pub:11718}.

\item Let $\pi$ denote a prime not congruent to $3$ mod $4$. Then any integer of the form $k=\pi$, or $k=\pi_1\pi_2$ with $(\pi_1,\pi_2)=-1$, is Pell (where $(\cdot,\,\cdot)$ is the Legendre symbol; see e.g.~\cite{alaca_williams_2003}, theorem 11.5.7). Furthermore, any odd integer of the form $k=\pi_1\pi_2\cdots\pi_{2n+1}$ such that there is no triplet $(a,b,c)$ with $(\pi_a,\pi_b)=(\pi_b,\pi_c)=+1$, is Pell~\cite{10.2307/2036951}.

\end{itemize}

Pell numbers appear naturally in the study of the time-reversal properties of $U(1)_k$. For example, one has the following:
\begin{proposition}\label{lm:double_pell}
If $kk'$ satisfies the Pell equation
%\begin{equation}
%kk'p^2-q^2=1,\qquad p,q\in\mathbb Z
%\end{equation} then
the theory $U(1)_k\times U(1)_{-k'}$ is time-reversal invariant.
\end{proposition}

\emph{Proof}. Assume that
\begin{equation}
kk'p^2-q^2=1,\qquad p,q\in\mathbb Z
\end{equation}
Let
\begin{equation}
4\pi\mathcal L=ka\,\mathrm da-k'b\,\mathrm db
\end{equation}
and introduce the $\mathrm{GL}_2(\mathbb Z)$ transformation
\begin{equation}
\mathsf T\colon\begin{pmatrix}a\\b\end{pmatrix}\mapsto \begin{pmatrix}q&k'p\\kp&q\end{pmatrix}\begin{pmatrix}a\\b\end{pmatrix}
\end{equation}

The Lagrangian becomes
\begin{equation}
\mathsf T\colon4\pi\mathcal L\mapsto-k\,a\,\mathrm da+k'\,b\,\mathrm db
\end{equation}
as required.\hfill$\square$

\medskip

%\begin{remark}\normalfont This duality only makes sense as a duality of spin TQFTs. Indeed, if $kk'$ is Pell, then it cannot be a multiple of $4$, and so $k$ and $k'$ cannot be both even. Therefore, at least one factor in $U(1)_k\times U(1)_{-k'}$ has odd level, and the theory is necessarily spin.
%\end{remark}

Taking $k'=1$ leads to the invariance of $U(1)_k\times U(1)_{-1}$ (cf.~proposition~\ref{lm:lagrangian_pell}). Moreover, this result, together with conjecture~\ref{conj:diag}, leads to the following interesting purely number-theoretic conjecture:
\begin{conjecture}\label{cj:pell}
An integer $k$ satisfies $q^2=-1\mod k$ for some $q\in\mathbb Z$ if and only if there exists some Pell integer $k'$ such that $kk'$ is also Pell.
\end{conjecture}

Recall that any solution of $q^2=-1+pk$ is of the form $p=p_0+2q_0n+kn^2$ (cf.~\eqref{eq:poly_PQ}). If $p$ is Pell for some $n$, then it suffices to take $k'=p$, from where the conjecture would follow (because $kp=q^2+1$ is automatically Pell). Noting that whenever this polynomial is prime, it is also Pell, our conjecture actually follows from the so-called Hardy-Littlewood ``conjecture F''~\cite{hardy1923}, which states that $ax^2+bx+c$ is prime infinitely often unless $b^2-4ac$ is a perfect square or $a+b$ and $c$ are both even (neither condition being satisfied by our polynomials). It is widely believed that the Hardy-Littlewood conjecture is true, which implies that our conjecture -- being much weaker -- should be true as well.

There is a more specific result due to Lemke Oliver and Iwaniec~\cite{RobertJ2012,Iwaniec1978} that states that a polynomial of the type above represent primes or semiprimes infinitely often. But any prime, or any semiprime $\pi_1\pi_2$ with $(\pi_1,\pi_2)=-1$ is Pell. Having no reason to expect otherwise, one is lead to conjecture that both options $(\pi_1,\pi_2)=\pm1$ appear with the same probability -- which is confirmed by numerical analysis -- from where it would follow that $p_0+2q_0n+kn^2$ generates infinitely many Pell numbers. In fact, the only possibility for a failure of our conjecture is that this polynomial never represents a prime (disproving the Hardy-Littlewood conjecture), and that all the semiprimes it represents have $(\pi_1,\pi_2)=+1$. This is extremely unlikely, but we have no proof that it cannot happen.

In any event, we checked that the conjecture is true for $k$ up to $10^9$. For now it remains an interesting open question.

If the conjecture is true, we can in fact invert the logic and use the time-reversal invariance of $U(1)_k\times U(1)_{-k'}$ to argue that of $U(1)_k$, for any $k\in\mathbb T$, by mimicking the argument of proposition~\ref{lm:suff_lagran}.

\paragraph{Added note:} An unconditional proof of conjecture~\ref{cj:pell} has been discussed in~\href{mathoverflow.net/q/364415}{MathOverflow}.

\clearpage

\clearpage
\printbibliography
\end{document}